\documentstyle[aps,preprint,tighten,floats,eqsecnum,epsf,rotate,prd]{revtex}

\newbox\rotbox

\begin{document}
\preprint{\vbox{\noindent Accepted for Publication in
                          {\it Annals of Physics}
\hfill DOE/ER/40427-17-N95\\}} 


\title{\huge QCD Sum Rules\\ for\\ Skeptics}

\author{\sc Derek
B. Leinweber\footnote{E-mail:~derek@phys.washington.edu
~$\bullet$~ Telephone: (206)~616--1447 ~$\bullet$~ Fax:
(206)~685--0635 
WWW:~http://www.phys.washington.edu/$\sim$derek/Welcome.html}}

\address{
Department of Physics, Box 351560, University of Washington, Seattle,
WA 98195}

\address{and}

\address{
TRIUMF, 4004 Wesbrook Mall, Vancouver, BC, V6T 2A3, Canada}

\date{October 19, 1995)\\ (Revised July 10, 1996}

\maketitle

\begin{abstract} 

   A new Monte-Carlo based uncertainty analysis is introduced to
quantitatively determine the predictive ability of QCD sum rules.  A
comprehensive analysis of ground state $\rho$-meson and nucleon
spectral properties is performed.  Many of the findings contradict the
conventional wisdom of both practitioners and skeptics alike.
Associations between the phenomenological fit parameters are
particularly interesting as they reveal how the sum rules resolve the
spectral properties.  The use of derivative sum rules for the
determination of $\rho$-meson spectral properties is shown to be a
very unfavorable approach.  Most prior nucleon sum rule analyses are
based on a sum rule which is found to be invalid; the results are
suspect, and should be reevaluated.  The ``Ioffe formula'', argued by
many to qualitatively encapsulate a description of the nucleon mass in
terms of the chiral symmetry breaking order parameter $\left <
\overline q q \right >$ is misleading at best.  QCD Sum Rules are
found to be self-consistent without contributions from direct
instantons.  This implies that instanton effects are adequately
accounted for in the nonperturbative vacuum condensates.  This
in-depth examination of QCD sum rule self consistency paints a
favorable picture for further quantitative refinements of the QCD sum
rule approach.

\end{abstract}

\newpage

\narrowtext

\section{INTRODUCTION}
\label{intro}

\subsection{Prologue}

   The QCD Sum Rule (QCD-SR) approach to QCD continues to be a highly
active field.  The influence of the field is reflected in over 1500
references to the seminal paper of Shifman, Vainshtein and Zakharov
(SVZ) \cite{shifman79}, with over 380 of these references following
the beginning of 1992 \cite{slacPPlib}.  While the approach has been
applied to a variety of hadronic observables in both vacuum and finite
density nuclear matter, a comprehensive review of the systematic
errors associated with the approach has not been considered.  

   One of the key assumptions of the approach is the use of the so
called ``continuum model'' to remove excited state contaminations from
the hadron correlator under investigation.  Recently, the continuum
model for the nucleon was tested by using correlation functions
calculated via lattice regularized QCD
\cite{leinweber95a,leinweber94e}.  The results suggest that the
continuum model adequately removes the excited state contaminations
and allows the isolation of the ground state.  The success of the
continuum model warrants a more careful investigation of other aspects
of the QCD-SR approach so that uncertainties in the predictions may
be reliably and quantitatively determined.

   The focus of this paper is to establish a rigorous procedure for
extracting quantities of phenomenological interest from QCD Sum Rules.
The main conclusion of this investigation is that QCD Sum Rules work
and they are predictive when the analysis is done rigorously.  The
ground state masses of the $\rho$-meson and the nucleon are used to
introduce the concepts.  However, the procedures introduced here are
more generally applicable to quantities of current experimental
interest.  The goal is to shed light on the stability of QCD Sum Rule
analyses and thus the accuracy to which observables may be reliably
determined.

   Implementation of the procedure introduced here does not lead to
slight adjustments in the extracted parameters, but rather leads to
qualitatively different conclusions.  It will become apparent that
most existing QCD-SR investigations of nucleon properties are in fact
unreliable.  Previous calculations are based on sum rules in which
neither the operator product expansion (OPE) nor the phenomenological
description are under acceptable control.

   The QCD-SR method is presently the best fundamentally based
approach for investigating the properties of hadrons in nuclear
matter, as the lattice approach is challenged by a number of
formidable obstacles \cite{kogut95}.  Issues surrounding the finite
width of the rho-meson make reliable lattice calculations of
$\rho-\omega$ mixing very difficult.  These are two of the many topics
of current interest to nuclear physicists where the best fundamentally
based approach is that of QCD-SRs.  Moreover, the approach has minimal
model dependence.  Hence it is imperative to probe the predictive
ability of the approach and the ideas presented here are currently
being applied to these topics mentioned above
\cite{leinweber95f,leinweber95e,leinweber95g}.

\subsection{Systematic Uncertainties}

   The most significant source of uncertainty in the QCD Sum Rule
approach is an imprecise knowledge of the vacuum condensates appearing
in the operator product expansion (OPE).  This is particularly
problematic for the higher dimension operators, where one usually
invokes the vacuum saturation hypothesis (factorization) to replace
higher-dimension-operator vacuum-expectation values by products of
lower-dimension operators.  A list of key sources of uncertainty in
the QCD-SR approach should include:
\begin{itemize}
\item the unfactorized condensate values, 
\item factorization of higher dimensional operators, 
\item small but neglected $\alpha_s$ corrections, 
\item the truncation of the operator product expansion (OPE),
\item the selection of the regime for matching the QCD and phenomenological
      sides of the sum rules,
\item uncertainties associated with the summation of perturbation
      theory in large orders, and
\item the possibility of significant direct instanton contributions.
\end{itemize}
With the exception of direct instanton contributions, the effects of
these uncertainties will be estimated via a Monte-Carlo error
analysis.

   Gaussian distributions for the condensate values are generated via
Monte Carlo.  These distributions are selected to reflect the spread
of values assumed in previously published QCD-SR analyses, and the
uncertainties such as operator factorization listed above.  These
distributions provide a distribution for the OPE and thus uncertainty
estimates for the OPE which will be used in a $\chi^2$ fit.  In turn,
the OPE distribution provides distributions for the phenomenological
fit parameters, thus establishing the predictive ability of the QCD-SR
approach.

\subsection{Conventional Analyses}

   In the field of QCD inspired models, the focus is often on whether
or not the model is able to encompass the known experimental data, as
opposed to assessing the predictive potential of the method.  A review
of the QCD-SR literature leaves no doubt as to whether or not the
QCD-SR method is able to encompass the established data.  The poorly
determined condensate values and arbitrary aspect of the matching
Borel regime provides tremendous freedom in devising ways to achieve
the desired result.  In fact, it is difficult to find QCD-SR-based
predictions of hadronic observables which fail to agree with
experiment.  This is somewhat surprising, since the physics
responsible for some observables may be poorly represented in a
truncated OPE commonly limited to distances of less than 0.3 fm.

   Most QCD-SR calculations of nucleon properties in the literature
fall short in the analysis stage where the phenomenological parameters
are determined by matching the QCD and phenomenological sides of the
sum rules.  Typical analysis short-comings include:
\begin{itemize}
\item selecting a single value for the Borel parameter which gives
``nice'' results.  A Borel {\it regime} should be selected in order to
evaluate the stability and reliability of the results.
\item a selection of the Borel regime without careful regard to OPE 
convergence\footnote{Here and in the following, ``convergence'' of the
OPE simply means that the highest dimension terms considered in the
OPE, with their Wilson coefficients calculated to leading order in
perturbation theory, are small relative to the leading terms of the
OPE.} nor maintaining ground state dominance of the phenomenological
side of the sum rules.  
\item the fixing of search parameters (such as the continuum
threshold) to preferred values.  It will become apparent that this
introduces a strong bias to the remaining fit parameters which may not
reflect the properties of QCD.
\item claiming an accuracy for QCD-SR predictions without supporting
calculations.  Occasionally a ``stability analysis''
\cite{leinweber90,chung84} is considered, in
which fit parameters are monitored as a single condensate value is
varied.  However, such analyses explore a relatively small corner of
the condensate parameter space.
\end{itemize}
In general, former procedures adopted for matching the two sides of
the sum rules lack the level of rigor which will be presented here.

   This general lack of rigor can be traced back to the original paper
of SVZ \cite{shifman79}.  In their conclusions, they comment ``we
prefer not to deepen into the debris of computer calculations,
sophisticated fit programs, arranging error bars here and
there$\ldots$'' While this may be appropriate for a seminal paper on
the QCD-SR approach, continuation of this philosophy has lead to
strong negative criticism of the QCD-SR field.  It is not uncommon to
hear remarks such as ``QCD Sum Rules are useful only if you know the
answer.'' or ``You can get anything you want from QCD Sum Rules.'' and
``10\% here, 10\% there\ldots Pretty soon you're talking about real
numbers!''  These remarks are unfortunate, as a rigorous and
proficuous analysis is possible.

\subsection{Outline}

   In this paper, a method for the quantitative determination of
phenomenological quantities with uncertainties will be presented.
Since this paper is directed to both QCD-Sum-Rule practitioners and
skeptics alike, we begin by briefly outlining the QCD-SR approach in
Section \ref{qcdsr}.  A more pedagogical review of the approach may be
found in the recent review article of Ref.\ \cite{cohen95}.  This
section also introduces the ``Ioffe formula'' \cite{ioffe81}, argued
by many to qualitatively encapsulate a description of the nucleon mass
in terms of the chiral symmetry breaking order parameter $\left <
\overline q q \right >$. 

   Uncertainties associated with the evaluation of the OPE will be
addressed via a new Monte-Carlo based procedure.  Estimates for these
uncertainties are presented in Section \ref{uncert}.  Section
\ref{analysis} details the Monte-Carlo uncertainty analysis and
discusses the optimization algorithm and cautious convergence criteria
used in the following.

   Section \ref{rho} presents an analysis of $\rho$-meson sum rules.
Here the criteria for the selection of the regime for matching the QCD
and phenomenological sides of the sum rules is reviewed.  Reasonable
alternatives are considered via Monte Carlo.

   These considerations lead to the selection of the optimal
interpolating field for nucleon sum rules.  New insights into
unconventional nucleon interpolators obtained from the lattice QCD
investigation of Ref. \cite{leinweber95b} play a significant role
here.  Section \ref{nucleon11} reviews the selection of the optimal
nucleon interpolating field for correlators obtained from spin-1/2
interpolating fields.  The effects of choosing non-optimal
interpolators is demonstrated.  Section \ref{nucleon13} explores
additional sum rules obtained from the overlap of the generalized
spin-1/2 and a spin-3/2 interpolator.

   In Section \ref{correl} we explore the correlations between
condensate values, fit parameters and sum rule consistency.  These
correlations serve to identify the roles of the leading terms of the
OPE in hadronic physics.  The validity of the ``Ioffe formula'' is
evaluated here, and these results may be of interest to those modeling
the QCD vacuum.  The contingency table analysis reveals that the
intimate relationship between the quark condensate and the nucleon
mass suggested in the ``Ioffe formula'' is invalid.

   The necessity of direct instanton contributions to the QCD Sum
Rules is determined in Section \ref{instantons}.  We demonstrate that
there is no evidence indicating that direct instanton contributions
are required to maintain sum rule consistency.

   The implications of the results of this analysis on coordinate
space correlators are discussed in Section \ref{CoordImpl}.  Finally,
Section \ref{conclusions} summarizes the findings of this in-depth
investigation.

\section{QCD SUM RULE FORMALISM}
\label{qcdsr}

   The underlying principle of field theoretic approaches to hadron
phenomenology is the Ansatz of duality.  That is, it is possible
to simultaneously describe a hadron as quarks propagating in the QCD
vacuum, and as a phenomenological field with the appropriate quantum
numbers. 

   Hadron masses are extracted from an analysis of the two-point
function
\begin{equation}
\Pi(p) = i \, \int d^4 x\, e^{i p \cdot x}\,
\langle \Omega \bigm | T \{ \chi (x) \; 
\overline \chi (0) \} \bigm | \Omega \rangle \, .
\label{twopt}
\end{equation}
The interpolating field $\chi$ is typically constructed from quark
field operators combined to give the quantum numbers of the hadron
under investigation.  The $\rho$-meson interpolator related to the
decay constant is
\begin{equation}
\chi_\mu =  {1 \over 2} \left ( \overline u \, \gamma_\mu \, u -
  \overline d \, \gamma_\mu \, d \right ) \, .
\label{chirho}
\end{equation}
To maintain maximal overlap with the ground state relative to excited
states, only interpolators without derivatives are considered.  For
the nucleon we begin by considering the most general spin-1/2 nucleon
interpolator
\begin{equation}
\chi_{\cal O} = \chi_1 + \beta \chi_2 \, ,
\label{chiO}
\end{equation}
where
\begin{mathletters}
\begin{equation}
\chi_1(x) = \epsilon^{abc}
                 \left ( u^{Ta}(x) C \gamma_5 d^b(x) \right ) u^c(x)
\, ,
\label{chiN1}
\end{equation}
and
\begin{equation}
\chi_2(x) = \epsilon^{abc}
                 \left ( u^{Ta}(x) C d^b(x) \right ) \gamma_5 u^c(x)
\, .
\label{chiN2}
\end{equation}
\end{mathletters}%
$\chi_1(x)$ is the interpolator typically used in lattice QCD
analyses.  $\chi_2(x)$ vanishes in the nonrelativistic limit.  However
in a theory with light relativistic current quarks, there is no reason
to exclude such an interpolating field {\it a priori} \cite{chung82}.
With the use of the Fierz relations, the combination of the above two
interpolating fields for $\beta = -1$ may be written
\begin{mathletters}
\begin{eqnarray}
\chi_{\rm SR}(x) &=& \epsilon^{abc}
                 \left ( u^{Ta}(x) C \gamma_\mu u^b(x) \right )
                 \gamma_5 \gamma^\mu d^c(x) \, , \\
                 &=& 2 \, \left ( \chi_2 - \chi_1 \right ) \, ,
\end{eqnarray}%
\label{chiSR}%
\end{mathletters}%
giving the proton interpolating field advocated by Ioffe
\cite{ioffe83} and often found in QCD-SR calculations.

\subsection{Phenomenology}

   At the phenomenological level one proceeds by inserting a complete
set of eigenstates with the quantum numbers of the interpolator.  The
ability of the interpolator to annihilate a positive parity baryon to
the QCD vacuum is described by the parameter $\lambda$ in
\begin{equation} 
\langle \Omega \bigm | \chi_{\cal O} (0) \bigm | i,p,s \rangle \, =
\lambda_{{\cal O}i} \, u(p,s) \, ,
\label{LambdaDef}
\end{equation}
for state $i$ of momentum $p$ and spin $s$.  $u(p,s)$ is a Dirac
spinor.  Negative parity states require a $\gamma_5$ on the right-hand
side of (\ref{LambdaDef}).  For the nucleon, the two point function
has the form
\begin{equation}
\Pi(p) = \sum_i \, \lambda_{{\cal O}i}^2\, 
{\gamma \cdot p \pm M_i \over p^2 - M_i^2 - i \epsilon'} \, ,
\end{equation}
where $+/-$ corresponds to positive/negative parity baryons.
While there is some suppression of excited state contributions to the
two-point function, there is little hope of isolating ground state
properties from such a function.  Hence one uses the celebrated
Borel Transform
\begin{equation}
\widehat B [ f(p^2) ] = \lim_{{\scriptstyle -p^2 \to \infty \atop
                        \scriptstyle    n \to \infty } \atop
                        \scriptstyle -p^2/n = M^2 }
{1\over n !} (-p^2)^{n+1} \left ( {d \over d p^2} \right )^n
f (p^2) \, ,
\end{equation}
at each Dirac-$\gamma$ structure to obtain exponential suppression of
excited states.  The parameter $M$ is commonly referred to as the
Borel mass.  The Borel transform is easily applied to a dispersion
relation for $\Pi(p)$ at each Dirac-$\gamma$ structure
\begin{equation}
\Pi(p^2)=\frac{1}{\pi}\int_0^\infty ds\,
\frac{{\rm Im}\,\Pi(s)}
{s - p^2}+{\mbox{subtractions}} \, ,
\label{dispursion}
\end{equation}
and the polynomial subtraction terms are eliminated by the Borel
transform.  At the structure ${\gamma \cdot p}$ one has
\begin{mathletters}
\begin{equation}
\Pi_{\gamma \cdot p}(M) = \int \rho_{\gamma \cdot p}(s) \, e^{-s/M^2} \,
ds  \, , 
\end{equation}
where we have introduced the spectral density $\rho(s) = {\rm Im} \,
\Pi(s) / \pi$,
\begin{equation}
\rho_{\gamma \cdot p}(s) = \lambda_{\cal O}^2 \, \delta( s - M_N^2 ) +
\xi_{\gamma \cdot p}(s) \, .
\label{rhop}%
\end{equation}%
\label{Gp}%
\end{mathletters}%
At the structure $1$
\begin{mathletters}
\begin{equation}
\Pi_{1}(M) = \int \rho_{1}(s) \, e^{-s/M^2} \, ds  \, ,
\label{G1}
\end{equation}
where
\begin{equation}
\rho_{1}(s) = \left (\pm M_N \right ) \, \lambda_{\cal O}^2 \, 
\delta( s - M_N^2 ) + \xi_1(s) \, .
\label{rho1}%
\end{equation}%
\end{mathletters}%
Here, $\xi_{\gamma \cdot p}(s)$ and $\xi_1(s)$ account for excited
state contributions, and $+/-$ in (\ref{rho1}) corresponds to
positive/negative parity states.

   For the vector meson one has
\begin{mathletters}
\begin{equation}
\Pi_V(M) = \int \rho_V(s) \, e^{-s/M^2} \, ds  \, ,
\label{rhoVa} 
\end{equation}
\begin{equation}
\rho_V(s) = f_\rho^2 \, \delta( s - M_\rho^2 ) +
\xi_V(s) \, ,
\label{rhoVb}%
\end{equation}%
\label{rhoV}%
\end{mathletters}%
at the structure $g_{\mu \nu} - p_\mu p_\nu / p^2$.

\subsection{QCD}

At the quark level, one exploits the operator product expansion (OPE)
to describe the short distance behavior of the two-point function
\begin{mathletters}
\begin{eqnarray}
T \left \{ \chi (x),\ \overline \chi (0) \right \}
&=& \sum_n C_n (x,\mu) \ {\cal O}_n (0,\mu) \, ,  
\label{OPEmu} \\
&=& C_0(x)\ I \nonumber \\
&& + \, C_1(x)\ m_q \nonumber \\
&& + \, C_3(x)\ \overline q q \nonumber \\
&& + \, C_{4.1}(x)\ G_{\mu \nu}^a G^{a \mu \nu}
   + \, C_{4.2}(x)\ m_q\, \overline q q \nonumber \\
&& + \, C_{5.1}(x)\ \overline q \, \sigma_{\mu \nu} {\lambda^a \over
2} \, q \, G^{a \mu \nu} 
   + \, C_{5.2}(x)\ m_q\, G_{\mu \nu}^a G^{a \mu \nu} \nonumber \\
&& + \, C_{6.1}(x)\ \overline q \Gamma q\ \overline q \Gamma q 
   + \, C_{6.2}(x)\ m_q\, \overline q\, \sigma_{\mu \nu} {\lambda^a
                  \over 2}\, q \, G^{a \mu \nu}
   + \, C_{6.3}(x)\ f_{abc} \, G_{\mu \nu}^a G_{\nu \lambda}^b
G_{\lambda \mu}^c \nonumber \\ 
&& + \, C_{7.1}(x)\ \overline q q\ G_{\mu \nu}^a G^{a \mu \nu} 
   + \, C_{7.2}(x)\ m_q\, \overline q \Gamma q\ \overline q \Gamma q 
   + \, C_{7.3}(x)\ m_q\, f_{abc} G_{\mu \nu}^a G_{\nu \lambda}^b
G_{\lambda \mu}^c \nonumber \\ 
&& + \, C_{8.1}(x)\ \overline q \Gamma q\ \overline q \Gamma
   \sigma_{\mu \nu} {\lambda^a \over 2} q G^{a \mu \nu}
   + \, C_{8.2}(x)\ m_q\, \overline q q\ G_{\mu \nu}^a G^{a \mu \nu}
\nonumber \\
&& + \, C_{8.3}(x)\ G_{\mu \nu}^a G^{a \mu \nu} G_{\rho \lambda}^b
G^{b \rho \lambda} \, ,
\end{eqnarray}%
\label{OPE}%
\end{mathletters}%
where $\mu$ is the normalization point at which the coefficient
functions and the operators are defined.  Here we have explicitly
included all operators up to dimension eight, to leading order in the
quark mass $m_q$, having the quantum numbers of the vacuum.  The first
digit of the subscript of the Wilson coefficients $C(x)$ gives the
energy dimension of the operator.  $\Gamma$ may take any of the 16
independent Dirac-$\gamma$ matrices.

   Upon Fourier transforming to momentum space, (\ref{OPE}) becomes an
expansion in $1/Q^2$, valid for large momentum transfers.  Thus one
may use perturbation theory to calculate the Wilson coefficients. 

   The contribution of the gluon operator term 4.1 in the expansion
for the proton is suppressed due a factor of $(2 \pi)^2$ which appears
in the denominator of the Wilson coefficient arising from integration
over the quark-loop momentum which introduces the factor $d^4 q / (2
\pi)^4 $.  Terms involving more than two gluon field strength tensors
$G_{\mu \nu}^a$ and products of $m_q$ and $G_{\mu \nu}^a G^{a \mu
\nu}$ are estimated to be small and are typically neglected.

The standard treatment of the OPE proceeds via inserting the explicit
forms of the interpolating fields into the two-point function of
(\ref{twopt}) and contracting out pairs of time-ordered quark-field
operators which are the fully interacting quark propagators of QCD.
Wick's theorem for the time-ordered product of two fermion fields
provides 
\begin{eqnarray}
\lefteqn{\langle \Omega \bigm | T \left \{ q^a (x), 
\overline q^b (0) \right \} \bigm | \Omega \rangle } \hspace{24pt}
\nonumber \\
&=& \langle 0 \bigm | T \left \{ q^a (x), \overline q^b (0) \right \} \bigm 
| 0 \rangle
+ \langle \Omega \bigm | : q^a (x), \overline q^b (0) : \bigm | \Omega
\rangle \, .
\end{eqnarray}
Taylor expansion of the normal ordered piece leads to
\begin{eqnarray}
: q^a_\alpha (x), \overline q^b_\beta (0) :\ &=&
          \ : q^a_\alpha (0), \overline q^b_\beta (0) :
\ +\ x^\mu : \partial_\mu q^a_\alpha (0), \overline q^b_\beta (0) :\
\nonumber \\
&& +\ {1 \over 2!} x^\mu x^\nu : \partial_\mu \partial_\nu
   q^a_\alpha (0), \overline q^b_\beta (0) :\ 
+ \cdots . 
\label{taylor}
\end{eqnarray}
The well known trick is to select the coordinate gauge $x^\mu \, A_\mu
= 0$ such that 
\begin{equation}
A_\mu(x) = - {1 \over 2} G_{\mu \nu}(0) \, x^\nu 
           - {1 \over 3} \left ( 
             \partial_\lambda G_{\mu \nu}(0)
             \right ) x^\lambda x^\nu + \cdots \, ,
\label{Aexp}
\end{equation}
and the partial derivatives of (\ref{taylor}) may be replaced by
covariant derivatives
\begin{equation}
\partial_\mu \to \nabla_\mu = \partial_\mu + i\, g\, A_\mu
\, .
\end{equation}
Generally, one works with (\ref{Aexp}) to leading order.  The
following relations allow an easy determination of the covariant
derivatives
\begin{mathletters}
\begin{equation}
\gamma^\mu \nabla_\mu  q_\alpha^a = - i m_q q_\alpha^a  \, ,
\end{equation}
\begin{equation}
\left [ \nabla^\mu, \nabla^\nu \right ] = i g_c G^{\mu \nu} \, ,
\end{equation}
\begin{equation}
\nabla^2 q_\alpha^a = - {1 \over 2} g_c \sigma \cdot G q_\alpha^a \, .
\end{equation}%
\end{mathletters}%
In summary, the quark correlators employed in calculating the OPE
under the usual assumption of vacuum saturation of the intermediate
states of composite operators include
\begin{eqnarray}
\lefteqn{\langle \Omega \bigm | T \left \{ q^a (x), 
\overline q^b (0) \right \} \bigm | \Omega \rangle = } \hspace{36pt}
\nonumber \\
         && {i \over 2 \pi^2 x^4} \, \gamma \cdot x \, \delta^{ab} 
             -{ m_q \over 2^2 \pi^2 x^2} \delta^{ab} 
             -{1 \over 2^2 \, 3} \langle \overline q q \rangle 
              \delta^{ab} \nonumber \\
         && +{i \over 2^4 \, 3} m_q \langle \overline q q \rangle 
           \, \gamma \cdot x \, \delta^{ab} 
          +{x^2 \over 2^6 \, 3} \langle \overline q g_c \sigma 
           \cdot G q \rangle \delta^{ab} \nonumber \\
         && -{i x^2 \over 2^7 \, 3^2} m_q
\langle \overline q g_c \sigma \cdot G q \rangle \, \gamma \cdot x \,
\delta^{ab}  
          -{ x^4 \over 2^{10} \, 3^3} \langle \overline q q \rangle 
\langle g_c^2 G^2 \rangle \delta^{ab}  \nonumber \\
         && + {i \over 2^5 \pi^2 x^2}
\bigl ( g_c G_{\alpha \beta}^n \bigr )
\bigl (  \gamma \cdot x \, \sigma^{\alpha \beta} + \sigma^{\alpha \beta} 
\, \gamma \cdot x  \bigr )
{\lambda^n_{ab} \over 2} \nonumber \\
&& + {1 \over 2^5 \pi^2} m_q \left [
\ln \left ( {-x^2 \Lambda^2 \over 4} \right ) + 2 \gamma_{EM} \right ]
\bigl ( g_c G_{\alpha \beta}^n \bigr )
\sigma^{\alpha \beta}
{\lambda^n_{ab} \over 2} \, .
\label{quarkcorr} \\
&& \nonumber \\
\lefteqn{\langle \Omega \bigm | T \left \{ q^a (x) \,
g_c  G_{\alpha \beta}^n \,
\overline q^b (0) \right \} \bigm | \Omega \rangle = } \hspace{36pt}
\nonumber \\
&& - { 1 \over 2^6 \, 3} \langle \overline q g_c \sigma \cdot G q \rangle
\sigma_{\alpha \beta}
{\lambda^n_{ab} \over 2}  \nonumber \\
&& + {i \over 2^8 \, 3} m_q \langle \overline q g_c \sigma \cdot
G q \rangle 
\bigl (  \gamma \cdot x \, \sigma_{\alpha \beta} + \sigma_{\alpha \beta} 
\, \gamma \cdot x  \bigr )
{\lambda^n_{ab} \over 2} \nonumber \\
&& + {x^2 \over 2^{10} \, 3^2} \langle g_c^2 G^2 \rangle 
\langle \overline q q \rangle
\sigma_{\alpha \beta}
{\lambda^n_{ab} \over 2} \, . 
\label{quarkgluoncorr}
\end{eqnarray}

   Finally the momentum space correlator is Borel transformed.  Here
one encounters the factorial suppression of higher-dimension operators
as indicated in
\begin{equation}
\widehat B \left [ {1 \over \left ( p^2 \right )^k } \right ] =
(-1)^k \, {1 \over (k-1)! } \, {1 \over (M^2)^{k-1} } \, .
\end{equation}
It is worth noting that the factorial suppression does not really set
in until one reaches the term $1/M^6$.  For the $\rho$ meson this term
corresponds to an operator of dimension eight, and for the nucleon the
dimensions are twelve and thirteen for the two nucleon sum rules
obtained from spin-1/2 interpolating fields.  Since the OPE for the
$\rho$-meson is traditionally truncated at dimension eight
\cite{shifman79}, the correlator should be more reliable than for the
nucleon sum rules which are truncated at dimension nine.

\subsection{Power Corrections from the Large-Order Behavior of
Perturbation Theory}
\label{renormalons}

   It is well known that the standard treatment of the OPE encounters
difficulties at large orders in the perturbative expansion of the
Wilson coefficients.  Perturbative expansions are divergent at large
orders and are asymptotic at best \cite{dyson52}.  In the standard
treatment, these divergences limit the accuracy to which the
perturbative coefficients may be determined.

   Renormalons \cite{hooft77}, a particular set of perturbative
graphs, provide a simple illustration of the factorial growth in the
contributions of perturbative graphs at large orders
\cite{david84,mueller85,david86,zakharov92,bigi94}.  The factorial
growth in large-order contributions associated with soft infrared (IR)
virtual momenta is problematic, as the series is not Borel summable.
As such, the Wilson coefficients, by themselves, are ill defined in
the standard treatment.  The uncertainty resulting from the
restriction to finite orders of the asymptotic expansion is commonly
referred to as the IR renormalon ambiguity.  Of course the use of
perturbative propagators to describe soft virtual momenta is
incorrect.  Instead, one must utilize fully dressed nonperturbative
propagators.  Hence, it is not surprising one encounters difficulties
in the IR regime, as one is simply encountering the Landau pole of the
perturbative coupling constant.

   A solution to the IR renormalon ambiguity problem was provided by
the ITEP group \cite{novikov85} through the introduction of the
normalization/separation scale $\mu$ of (\ref{OPEmu}).  The OPE
becomes a separation of scales, with virtual momenta lying above $\mu$
represented in the Wilson coefficients, and momenta below $\mu$ in the
vacuum expectation values (VEVs) of higher-dimension operators
beginning at dimension four.  The introduction of the scale $\mu$
eliminates the IR renormalon ambiguity of the perturbative
expansion.  However, in principle, both the Wilson coefficients and
the VEVs of operators contain perturbative as well as nonperturbative
physics. \footnote{For a particularly lucid discussion of why the
standard treatment of the OPE is phenomenologically successful, see
Ref. \protect\cite{david86}.}

   In principle, the OPE defined in (\ref{OPEmu}) is free of IR
renormalon ambiguities.  However, in practice it is difficult to
implement the formal procedure of truncating virtual momenta below
$\mu$ when calculating higher order $\alpha_s$ corrections to the
Wilson coefficients.  While these contributions are small for leading
order corrections, their inclusion is a source of error which can be
accommodated in the Monte-Carlo uncertainty analysis.

   In the ultraviolet (UV) virtual momenta regime, the renormalon
series is sign alternating and Borel summable.  Summation of the UV
series can give rise to a power correction proportional to
$\Lambda_{\rm QCD}^2 / Q^2$ which closely resembles the squared quark
mass contribution to the OPE \cite{zakharov92}.  For the vector
correlator, the UV renormalon gives rise to an effective mass of
roughly $(2 / \pi)^{1/2} \, \Lambda_{\rm QCD} / 3$, which is much
larger than the usual current quark mass the order of 5 MeV
\cite{zakharov92}.  Hence there is some concern that important power
corrections associated with UV renormalons has been overlooked in the
standard QCD-SR treatment.  Moreover, it is the power corrections that
provide the link to nonperturbative phenomena.  To the extent that the
power correction is scheme dependent and not reliably known, we will
refer to these quark-mass like power corrections as UV renormalon
uncertainties.

   Additional evidence for $1/q^2$-type power corrections independent
of renormalons has been obtained through a consideration of
constraints imposed by asymptotic freedom and analyticity on the
large-order behavior of perturbation theory \cite{brown92a,brown92b}.
In this case the perturbative series is not Borel summable.  Given the
absence of a local gauge-invariant field-operator product of dimension
two, some might argue that the coefficient of the singularity in the
Borel plane governing the large-order behavior of perturbative series
must have a zero.  However, perturbative estimates of the coefficient
do not provide evidence of such a zero \cite{brown92b} and the issue
is unsettled.

   Hence the issue of possible dimension-two power corrections is of
current interest.  The techniques presented in this manuscript might
be used to provide some phenomenological insight into the debate.
However, such considerations would take us too far afield and will be
deferred to a subsequent analysis \cite{leinweber96b}.  On the other
hand, it is interesting to look for discrepancies which might be
rooted in such power corrections, and this is done in the following
analysis.

\subsection{Ioffe Formula}

   To summarize the points of this section, we present the traditional
QCD Sum Rules for the nucleon \cite{ioffe83} obtained from the
consideration of the interpolating field $\chi_{\rm SR}/2$ of
(\ref{chiSR}) to operator dimension 8.
\begin{mathletters}
\begin{eqnarray}
{\rm At}\ \gamma \cdot p\, : \qquad
&& {1 \over 8} \, {1 \over \left ( 2 \pi \right )^4 } \, M^6 
+ {1 \over 32} \, {1 \over \left ( 2 \pi \right )^2 } \, 
\langle {\alpha_s \over \pi} \, G^2 \rangle \, M^2 
+ {1 \over 6} \, \langle \overline q q \rangle^2  \nonumber \\
&& 
+ {1 \over 24} \, \langle \overline q \, g \sigma \cdot G \, q \rangle
\, \langle \overline q q \rangle {1 \over M^2} \nonumber \\
&& \qquad 
= \lambda_N^2 \, e^{-M_N^2 / M^2} + \cdots \, . 
\label{ioffe1a} \\
{\rm At}\ 1\, : \qquad
&& - {1 \over 4} \, {1 \over \left ( 2 \pi \right )^2 } 
     \langle \overline q q \rangle \, M^4
+ { 5 \over 288} \langle \overline q q \rangle \,
  \langle {\alpha_s \over \pi} \, G^2 \rangle \nonumber \\
&& \qquad
 = \lambda_N^2 \, M_N \, e^{-M_N^2 / M^2} + \cdots \, .
\label{ioffe1b}%
\end{eqnarray}%
\label{tradit}%
\end{mathletters}%
The dots allow for excited state contributions.\footnote{The Wilson
coefficient of the dimension 7 operator in (\protect\ref{ioffe1b}) has
had an elusive history.  Early calculations
\protect\cite{chung84,belyaev83,ioffe85} appear to have calculated a
subset of the diagrams contributing to the dimension 7 operator
\protect\cite{leinweber90}.  However a more persistent error lies in a
factor of 3 correction in the term $[ x^2 / (2^{10} \, 3^2)] \left <
\overline q q \right > \left < g_c^2 G^2 \right > \sigma_{\alpha
\beta} \lambda_{ab}^n/2$ of (\protect\ref{quarkgluoncorr}) from that
used in previous analyses.}

   The Ioffe Formula for the nucleon mass is obtained from
(\ref{tradit}) by keeping the leading terms of the OPEs and assuming
dominance of the ground state on the right-hand side of the equations,
\begin{mathletters}
\begin{eqnarray}
{1 \over 8} \, {1 \over \left ( 2 \pi \right )^4 } \, M^6 
 &\simeq& \lambda_N^2 \, e^{-M_N^2 / M^2} \, , \\
- {1 \over 4} \, {1 \over \left ( 2 \pi \right )^2 } \langle \overline
q q \rangle \, M^4 
&\simeq& \lambda_N^2 \, M_N \, e^{-M_N^2 / M^2} \, .
\end{eqnarray}%
\label{bigassumption}%
\end{mathletters}%
The Ioffe formula follows from the ratio of these two equations,
\begin{mathletters}
\begin{eqnarray}
M_N &\simeq& - \, { 2 \, \left ( 2 \pi \right )^2 \, 
                    \langle \overline q q \rangle
                    \over 
                    M^2} \, ,  \\
    &=& 0.90\ {\rm GeV\ for\ } M = 1\ {\rm GeV} \, . \nonumber
\end{eqnarray}
If one prefers to eliminate the Borel mass by selecting $M = M_N$,
\begin{eqnarray}
M_N &\simeq& \left ( - \, 2 \, \left ( 2 \pi \right )^2 \, 
                     \langle \overline q q \rangle
                    \right )^{1/3} \, , \\
    &=& 0.97\ {\rm GeV} \, . \nonumber
\end{eqnarray}%
\end{mathletters}%
Here the quark condensate is estimated from the partially conserved axial
current (PCAC) relation and pion decay to give
\begin{eqnarray}
\langle \overline q q \rangle &=& - \, {m_\pi^2 \, f_\pi^2 \over
                                      m_u + m_d} \, , \label{GOR} \\
                              &=& (0.225\ {\rm GeV})^3 \, , \nonumber
\end{eqnarray}
which corresponds to $m_\pi = .140$ GeV, $f_\pi = 0.093$ GeV and the
average light quark mass is 7 MeV.  The qualitative features of chiral
symmetry breaking coupled with the quantitative success of these
relations has lead to enormous faith being placed into the validity of
the Ioffe formula.  Most practitioners expect this relation to be
valid, at least in a qualitative manner of proportionality.

\subsection{Excited State Contributions}

   The operator product expansion provides knowledge of the two-point
function (\ref{twopt}) in the near perturbative regime.  As such the
correlator represents contributions from both the ground state hadron
and excited states.  To maintain some predictive ability in the QCD-SR
approach, a ``continuum model'' for excited state contributions to the
correlator is introduced.

   Guided by the principle of duality, the leading terms of the OPE
surviving in the limit $M \to \infty$ are used to model the form of
the spectral density accounting for excited state contributions.
Consider, as an example, the identity operator for the nucleon
interpolator $\chi_{\rm SR}$ of (\ref{chiSR}).  Explicitly including
the excited state term of (\ref{Gp}) in equation (\ref{ioffe1a}) leads
to 
\begin{equation}
{1 \over 2} \, {1 \over ( 2 \, \pi)^4 } \, M^6 =
\int_0^\infty \xi_{\gamma \cdot p}(s) \, e^{-s/M^2} \, ds \, ,
\label{laplacetrans}
\end{equation}
for large Borel mass.  The Laplace transform readily provides
\begin{equation}
\xi_{\gamma \cdot p}(s) = {1 \over 2} \, {1 \over ( 2 \, \pi)^4 } \,
                          {1 \over 2!} \, s^2 \, .
\end{equation}
The continuum model contribution is obtained by introducing a
threshold, $s_0$, at which strength in the excited states becomes
significant.  Such a model is motivated well by the experimental cross
section for $e^+ \, e^- \to $ hadrons.  In addition, this approach
accommodates the finite widths and the multi-particle continuum of QCD.
Hence the excited state contributions associated with the identity
operator are
\begin{equation}
\int_{s_0 = w^2}^\infty \xi_{\gamma \cdot p}(s) \, e^{-s/M^2} \, ds 
= {1 \over 2} \, {1 \over ( 2 \, \pi)^4 } \, M^6 e^{-w^2/M^2} \,
\left ( {w^4 \over 2 M^4} + {w^2 \over M^2} + 1 \right ) \, .
\end{equation}
Continuum model contributions associated with $M^4$ and $M^2$ terms
are calculated in a similar manner.  Usually the continuum
contributions are placed on the OPE side of the sum rules.  Hence,
terms surviving in the limit $M \to \infty$ have the following factors
associated with them:
\begin{mathletters}
\begin{eqnarray}
M^6 \, : && 
\left [ 1 - e^{-w^2/M^2} \, \left ( {w^4 \over 2 M^4} + {w^2 \over M^2}
+ 1 \right ) \right ] \, , \\
M^4 \, : &&
\left [ 1 - e^{-w^2/M^2} \, \left ( {w^2 \over M^2} + 1 \right ) \right ]
\, , \\
M^2 \, : && 
\left [ 1 - e^{-w^2/M^2} \right ] \, .
\end{eqnarray}%
\label{continuum}%
\end{mathletters}%
The contributions of this model relative to the ground state, whose
properties one is really trying to determine, are not small.  They are
typically 10 to 50\% \cite{leinweber90}.  The validity of this
somewhat crude model is relied upon to cleanly remove the excited
state contaminations.

   In the lattice QCD investigation of the continuum model formulated
in Euclidean space \cite{leinweber95a}, the short time regime of
point-to-point lattice correlation functions is described well by the
QCD-Sum-Rule-inspired continuum model.  There, the Laplace transform
of the spectral density appears to be sufficient to render any
structure in the spectral density insignificant in the short Euclidean
time regime of point-to-point correlators.  Similar conclusions are
expected to hold for the Borel transformed sum rules under
investigation here.

   It should be noted that the spectral density of the physical world
is different than that encountered on the lattice.  For example, in
the nucleon channel one expects strength in the correlator above the
ground state to start at the $\pi \, p$ threshold \cite{lee95},
whereas in the lattice calculations such contributions are altered by
the heavier quark mass and the quenched approximation.  While the
lattice results are encouraging, one may find that in some cases the
continuum model is not sufficiently detailed when the continuum model
contributions are large.  An example of this will be presented in the
discussion surrounding the $\rho$-meson sum rules, where discrepancies
may be due to an over simplified continuum model.

\section{Uncertainty Estimates}
\label{uncert}

   The origin of the uncertainty estimate for the OPE lies in the
uncertainties assigned to the QCD parameters appearing in the
truncated OPE.  In this section we assess the uncertainties on these
QCD parameters.

In assigning these uncertainties one would like to select values
conservatively enough such that the QCD-SR approach can test the
validity of our present understanding of QCD.  QCD-SR predictions that
fail to agree with experiment at the $2\sigma$ level should indicate a
possibly interesting discrepancy.  At the same time one would also
like to select uncertainties in accord with the generally accepted
values in the literature.  In addition, the error assessments need to
reflect uncertainties associated with the factorization of higher
dimensional operators and small but never-the-less neglected
$\alpha_s$ corrections.  These considerations have been balanced in
the following.  While some may argue that some values are known better
than that indicated here, others will no doubt argue that the errors
are underestimated.

   In any event, one will learn for the first time how the
uncertainties in the QCD parameters are mapped into uncertainties in
the phenomenological fit parameters.  In the lattice QCD investigation
of Ref.\ \cite{leinweber95a} it was noted that interplay between the
pole position and continuum threshold lead to rather large
uncertainties in the fit parameters.  It will be interesting to see
how reasonable estimates of the QCD uncertainties are revealed in the
fit parameters.

   For the quark condensate, $\langle \overline q q \rangle$ we note
that there are two values commonly found in QCD-SR analyses.  PCAC and
pion decay considerations coupled with quark masses of $m_u + m_d =
11$ MeV lead to a value of $ - ( 0.250 )^3$ GeV${}^3$.  QCD-SR
considerations of $g_A$ and octet baryon magnetic moments
\cite{chiu85,chiu86,chiu87} prefer a smaller magnitude at $- ( 0.225
)^3$ GeV${}^3$.  Expecting the value to lie some where in this regime,
we take the average of these values with an uncertainty of half the
difference at an 80\% confidence level.  Introducing the standard
QCD-SR notation we define
\begin{equation}
a = - (2 \pi)^2 \langle \overline q q \rangle 
  = 0.52 \pm .05\ {\rm GeV}^3 \, .
\label{a}
\end{equation}
This value and uncertainty is also in accord with $a = 0.47 \pm 0.05\
{\rm GeV}^3$ obtained from more recent quark mass estimates
\cite{leutwyler96}.

   Early estimates of the gluon condensate $\langle \alpha_s/\pi \,
G^2 \rangle$ from charmonium sum rules \cite{shifman79} place the
value at $0.012 \pm 0.005$ GeV${}^4$, which is generally referred to
as the ``standard value''
\begin{equation}
b = (2 \pi)^2 \langle {\alpha_s \over \pi} \, G^2 \rangle
  = 0.47 \pm 0.20\ {\rm GeV}^4 \, .
\label{bStd}
\end{equation}
However, a number of more recent investigations place the gluon
condensate at much larger values
\cite{narison95,bertlmann88,marrow87}.  For example, a finite-energy
sum rule (FESR) analysis of the $\rho$-meson channel places the gluon
condensate at 2 to 5 times the standard value \cite{bertlmann88}.  In
this analysis the spectral density is taken from a fit of the
isospin-one channel of $e^+ \, e^-$ scattering data.  First results
from the ALEPH and CLEO II experiments \cite{narison95,ALEPH93} also
suggest a larger value at $\langle \alpha_s/\pi \, G^2 \rangle = 0.025
\pm 0.010$ GeV${}^4$.  A recent Laplace based sum rule analyses
\cite{gimenez91} suggests a value of $\langle \alpha_s/\pi \, G^2
\rangle \simeq 0.015 \pm 0.005$ GeV${}^4$, more in accord with the
standard value.  However, we are somewhat skeptical of the refinements
of this result as the analysis is based on an approach which is
criticized in the following.\footnote{For example, while the language
used to describe the sum rules in Ref. \protect\cite{gimenez91} is
somewhat different from that used here, the nature of the sum rules
are the same.  The concerns surrounding OPE convergence, continuum
models and the use of derivative sum rules discussed in this paper
still apply.}
The most reliable sum rules of Sec. \ref{nucleon13} also prefer a
larger gluon condensate in order to maintain consistency between the
sum rules.  There it is found that the nucleon mass may be reproduced
with a value of $\langle \alpha_s/\pi \, G^2 \rangle \simeq 0.03$
GeV${}^4$ which agrees with experimental estimate.  Since the gluon
condensate plays a critical role in the nucleon sum rules as part of a
factorized dimension seven operator we adopt an uncertainty of 50\%
and take
\begin{equation}
b = (2 \pi)^2 \langle {\alpha_s \over \pi} \, G^2 \rangle
  = 1.2 \pm 0.6\ {\rm GeV}^4 \, .
\label{b}
\end{equation}

   The mixed condensate is parameterized as $\langle \overline q \,
g_c \, \sigma \cdot G \, q \rangle = - m_0^2 \langle \overline q q
\rangle$.  Early analyses of baryons \cite{belyaev83} as well as more
recent analyses of heavy-light quark systems \cite{narison88} place
$m_0^2 = 0.80$ GeV${}^2$, while an analysis of $N3/2^-$ QCD-SR sum
rules demand $m_0^2$ to be smaller \cite{leinweber90} at $m_0^2 =
0.65$ GeV${}^2$.  Hence we take the average and half the difference as
the uncertainty.
\begin{equation}
m_0^2 = 0.72 \pm 0.08\ {\rm GeV}^2 \, .
\label{m_0^2}
\end{equation}
Since the parameter $m_0^2$ is selected via Monte Carlo, the value of
the mixed condensate is obtained by multiplying $m_0^2$ by the central
value of the quark condensate, as opposed to a randomly selected value
of the quark condensate.  This is in contrast to factorized operators
where both condensates appearing in the operator are selected via
Monte Carlo.

\begin{figure}[t]
\begin{center}
\epsfxsize=\hsize
\leavevmode
\epsfbox{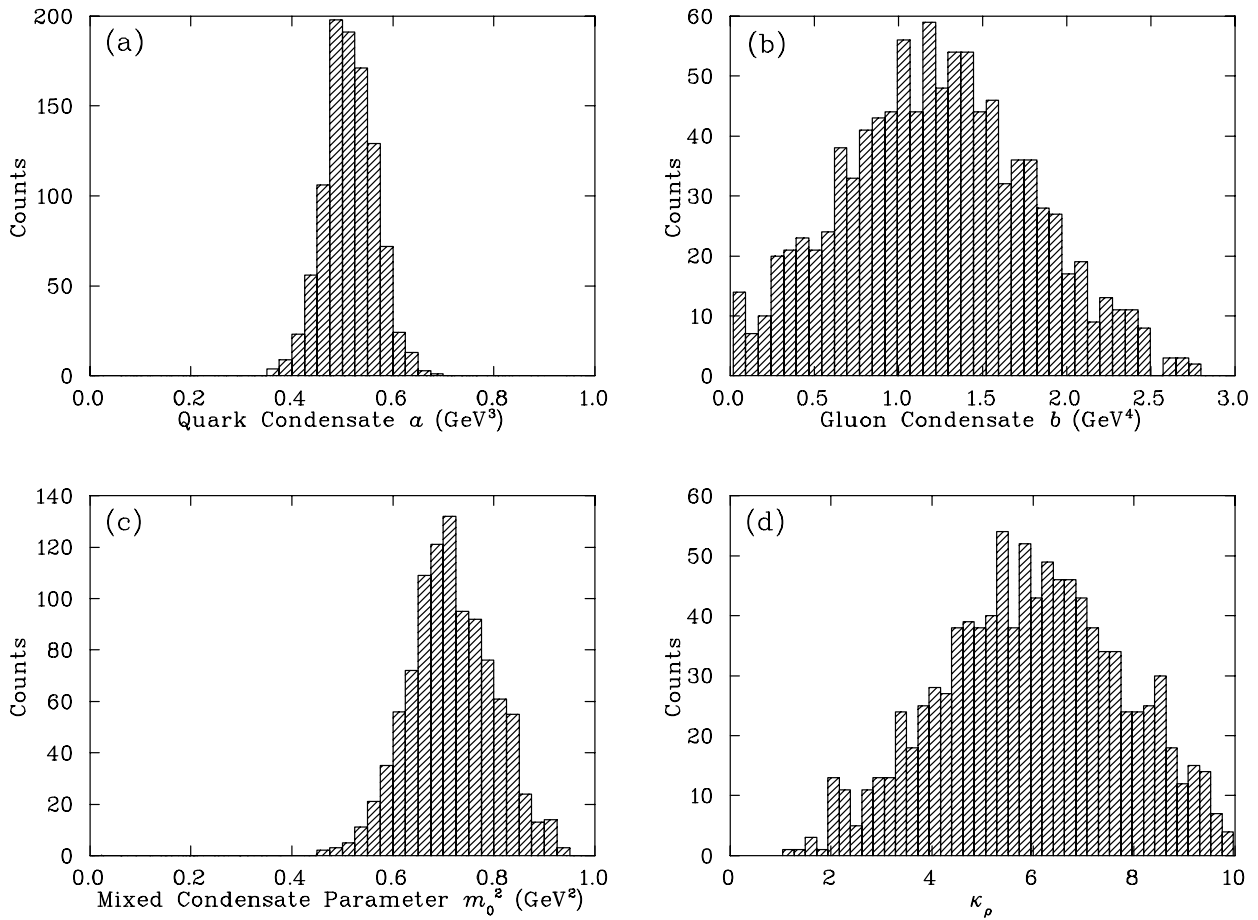}
\end{center}
\caption{ Histograms for QCD parameter distributions drawn from a
sample of 1000 QCD parameter sets.  Included are (a) the quark
condensate parameter of (\protect\ref{a}), (b) the gluon condensate
parameter of (\protect\ref{b}), (c) the mixed condensate parameter
$m_0^2$ and (d) the vacuum saturation violation parameter
$\kappa_\rho$.}
\label{distrib}
\end{figure}

   Relatively little is known about the magnitude of the
dimension-six, four-quark operators.  Early arguments placed the
values of these condensates within 10\% of the factorized values
\cite{novikov84}.  However other analyses claimed significant
violation of factorization
\cite{chung84,narison95,bertlmann88,marrow87,gimenez91,launer84,gimenez89}
for these operators in both nucleon and vector meson sum rules.
Parameterizing the four quark-operators as $\kappa \, \langle
\overline q q \rangle^2$, estimates of factorization violation place
$\kappa = 2$ or more.  For the nucleon we consider
\begin{equation}
\kappa_N = 2 \pm 1 \, ,\quad {\rm and} \quad 1.0 \le \kappa_N \le 4.0 \, .
\label{kappa_N}
\end{equation}
If one adopts the standard value for the gluon condensate, then
$\kappa_\rho \sim \kappa_N$ is sufficient to reproduce the
$\rho$-meson mass.  However, selection of the more recent and reliable
estimates of $b \sim 1.2\ {\rm GeV}^4$ demands $\kappa_\rho \sim 6$ to
reproduce the $\rho$-meson mass.  This value is in accord with the
observations of Refs.\ \cite{narison95,bertlmann88}, however the
uncertainty on this parameter is large.  Hence we adopt\footnote{With
$\kappa_\rho \sim 6$ as opposed to 2, even a small suppression of the
chirally even four-quark condensate at finite density can give rise to
substantial suppression of the $\rho$-meson mass and width.  Further
investigation of these issues is warranted.}
\begin{equation}
\kappa_\rho = 6 \pm 2 \, ,\quad {\rm and} 
\quad 1.0 \le \kappa_\rho \le 10.0 \, .
\label{kappa_rho}
\end{equation}
Figure \ref{distrib} displays histograms drawn from 1000 QCD parameter
sets for four of the QCD parameters discussed to this point.

   Variation of the QCD scale parameter $\Lambda_{\rm QCD}^{(3)}$ has
little effect on the results.  However, we restrict the values to the
conventional range adopted in QCD-SRs
\begin{equation}
\Lambda_{\rm QCD}^{(3)} = 0.15 \pm 0.04\ {\rm GeV}, \quad {\rm and}\quad
0.10\ {\rm GeV} \le \Lambda_{\rm QCD}^{(3)} \le 0.20\ {\rm GeV} \, .
\end{equation}
Variation of $\Lambda_{\rm QCD}^{(3)}$ above 200 MeV is only sensible
if the separation scale $\mu$ of the OPE, taken to be 500 MeV, is
increased to maintain plausible convergence of the power corrections
of the OPE.  A change in $\mu$ redefines the condensate values.  A
systematic determination of the new condensate values is beyond the
scope of this investigation and we choose to simply restrict the value
as above.

   It is perhaps interesting to note that if $\Lambda_{\rm QCD}^{(3)}$
is very large, say near the upper limit of the current world average
\cite{rpp94}, the OPE separation scale would necessarily become large.
Consequently, the Borel mass would be restricted to larger values, and
access to ground state properties may not be possible.  The correlator
may be dominated by strength from excited states, similar to the
present situation for the pseudoscalar pion correlator.  Fortunately,
the lower bound on the current world average for $\Lambda_{\rm
QCD}^{(3)}$ bodes well for present QCD sum rule analyses.  However, if
the current world average for $\Lambda_{\rm QCD}^{(3)}$ persists, a
comprehensive reformulation of the approach should be pursued.

   The $\rho$-meson sum rule requires knowledge of $\alpha_s/\pi$ at
the scale of 1 GeV${}^2$.  Since this scale is rather low for two-loop
perturbation theory, we adopt a 10\% uncertainty
\begin{equation}
{\alpha_s(1\ {\rm GeV}^2) \over \pi} = 0.117 \pm 0.012\, .
\end{equation}

   Finally, the $\rho$-meson sum rule requires knowledge of the
average light quark mass $m_q$.  Since it appears multiplied by the
quark condensate, we will use the relation of (\ref{GOR}) and equate
\begin{equation}
m_q \, \langle \overline q q \rangle = - \, {1 \over 2} \, 
                                       m_\pi^2 \, f_\pi^2 \, .
\end{equation}

   Monte-Carlo techniques might also be used to probe OPE truncation
errors by randomly selecting the size of the next term to follow the
truncated series.  The problem lies in determining a sensible and
realistic estimate without actually calculating the term.  Instead, we
will emphasize the importance of OPE convergence and select the Borel
analysis regime such that truncated terms are very unlikely to be
important.

\section{QCD-SR ANALYSIS}
\label{analysis}

\subsection{Monte-Carlo Uncertainty Analysis}

   Gaussian distributions are generated using established algorithms
for the conversion of uniform distributions on the interval (0,1)
\cite{rpp94}.  A particularly fast algorithm used in this
investigation is as follows.  If $u_1$ and $u_2$ are uniform on (0,1),
then construct $v_1 = 2 \, u_1 - 1$ and $v_2 = 2 \, u_2 - 1$.
Calculate $r^2 = v_1^2 + v_2^2$.  If $r^2 > 1$ start over.  Otherwise,
calculate
\begin{equation}
z_1 = v_1  \left ( {-2 \, \ln r^2 \over r^2} \right )^{1/2} \, , \quad
{\rm and} \quad
z_2 = v_2  \left ( {-2 \, \ln r^2 \over r^2} \right )^{1/2} \, .
\end{equation}
$z_1$ and $z_2$ are independent and normally distributed with mean 0
and variance 1.

   A set of Gaussianly-distributed randomly-selected condensate values
is generated, from which an OPE in Borel space, $\Pi^{\rm OPE}(M)$, is
constructed.  By repeating this procedure, an uncertainty for the OPE
may be determined.  Selecting $n_B$ evenly distributed points in the
Borel parameter space $(M_j, j = 1, 2, \ldots n_B)$, the standard
deviation in the OPE at the $j$'th Borel mass is given by 
\begin{equation}
\sigma_{{\rm OPE}}^2(M_j) = {1 \over n_C - 1 }
  \sum_{i=1}^{n_C} \left [ \Pi^{\rm OPE}_i(M_j) - \overline{\Pi^{\rm
OPE}}(M_j) \right ]^2  \, ,
\label{sigmaOPE}
\end{equation}
where $i$ denotes the $i$'th set of $n_C$ QCD parameter sets and 
\begin{equation}
\overline{\Pi^{\rm OPE}}(M_j) = {1 \over n_C} \sum_{i=1}^{n_C}
\Pi^{\rm OPE}_i(M_j) \, . 
\label{OPEbar}
\end{equation}
In practice, $n_C = 500$ was used in estimating the OPE uncertainty.
However, $n_C = 100$ was sufficient to get a stable uncertainty
estimate.  

   The uncertainties in the OPE are not uniform throughout the Borel
regime.  They are larger at the lower nonperturbative end of the Borel
region where uncertainties in the higher-dimensional vacuum
condensates dominate.  Hence, it is crucial that the appropriate
weight is used in the calculation of $\chi^2$.  Having determined the
OPE uncertainty, a $\chi^2$ measure is easily constructed.  For the
OPE obtained from the $k$'th set of QCD parameters, the
$\chi^2$ per degree of freedom is
\begin{equation}
{\chi_k^2 \over N_{DF}} = {1 \over n_B - n_p} \sum_{j=1}^{n_B} 
  { \left [ \Pi^{\rm OPE}_k(M_j) - 
            \Pi^{\rm SR}(M_j; \lambda_k, m_k, w_k) \right ]^2 
    \over
    \sigma_{{\rm OPE}}^2(M_j) } \, ,
\label{chisq}
\end{equation}
where $n_p$ is the number of phenomenological search parameters, and
$\Pi^{\rm SR}(M_j; \lambda_k, m_k, w_k)$ denotes the phenomenological
Spectral Representation of the QCD-SR.  In practice, $n_B = 51$ points
were used along the Borel axis.  $\chi^2$ is minimized by adjusting
the phenomenological parameters including the pole residue
$\lambda_{\cal O}$, the hadron mass $m$, and the continuum threshold
$w$.

   Distributions for the phenomenological parameters are obtained by
minimizing $\chi^2$ for many QCD parameter sets.  Provided the
resulting distributions are Gaussian, an estimate of the uncertainty
in the phenomenological results such as the hadron mass is obtained
from
\begin{equation}
\sigma_{m}^2 = {1 \over N - 1 }
  \sum_{k=1}^N \left ( m_k - \overline{m} \right )^2 \, , 
\quad {\rm with} \quad
\overline{m} = {1 \over N} \sum_{k=1}^N m_k \, .
\label{sigmam}
\end{equation}
In the event that the resultant distribution is not Gaussian, we will
report the median and asymmetric standard deviations from the median.
We note that (\ref{sigmam}) is the standard deviation of the
distribution and not the standard error obtained by further dividing
by $N$, the number of QCD parameter sets in the sample.  The standard
deviation is roughly independent of the number of QCD parameter sets
used in the analysis, and directly reflects the input parameter
uncertainties.  In the following, we generally select $N=1000$.  While
100 QCD parameter configurations are sufficient for obtaining reliable
uncertainty estimates, we also wish to explore correlations among the
QCD parameters and the phenomenological fit parameters.  The extra
sets aid in resolving more subtle correlations.

   Some care must be taken in the interpretation of the
$\chi^2/N_{DF}$.  While we are free to choose any number of points $n_B$
along the Borel axis for fitting the two sides of the sum rule, it is
important to recognize that the actual number of degrees of freedom is
determined by the number of terms in the OPE.  For the $\chi^2/N_{DF}$
to have true statistical meaning, a correlated $\chi^2$ calculation is
required.
\begin{eqnarray}
\chi_k^2  =  \sum_{j=1}^{n_B}
\sum_{j'=1}^{n_B} 
&& \left [ \Pi^{\rm OPE}_k(M_j) - \Pi^{\rm SR}(M_j; \lambda_k, m_k, w_k)
\right ] \times
\nonumber \\
&& \quad C_{j\,j'}^{-1} 
\left [ \Pi^{\rm OPE}_k(M_{j'}) - \Pi^{\rm SR}(M_{j'}; \lambda_k, m_k, w_k) 
\right ] 
\, ,
\label{corrchisq}
\end{eqnarray}
where $C_{j\,j'}^{-1}$ is the inverse covariance matrix.  The
covariance matrix may be estimated by
\begin{equation}
C_{j\,j'} = {1 \over (n_C - 1) } \, \sum_{i=1}^{n_C} \,
  \left [ \Pi^{\rm OPE}_i(M_j) - \overline{\Pi^{\rm OPE}}(M_j) \right ]
  \left [ \Pi^{\rm OPE}_i(M_{j'}) - \overline{\Pi^{\rm OPE}}(M_{j'}) \right ]
\label{covar}
\end{equation}
Unfortunately, for most of the fits considered here the covariance
matrix is ill-conditioned.  Inverting the covariance matrix via
singular value decomposition leads to well known pathological problems
\cite{michael95} such as best fit lines which do not pass through the
data points etc.  While one could experiment with different approaches
to stepping around this problem, we prefer to use the robust method of
determining best fit parameters by minimizing the uncorrelated
$\chi^2$ of (\ref{chisq}).  Even if some correlation among the data is
suspected, it is still acceptable to use the uncorrelated $\chi^2$ fit
provided the errors on the parameters are estimated using a
Monte-Carlo based procedure like that discussed here rather than from
the dependence of $\chi^2$ on the parameters
\cite{michael95,michael94}.  

   When simultaneously fitting multiple sum rules which may have some
fit parameters in common, (\ref{chisq}) is simply modified to
\begin{equation}
{\chi^2 \over N_{DF}} = {1 \over n_{\rm SR} \, n_B - n_p} \,
                  \sum_{i=1}^{n_{\rm SR}} \, \sum_{j=1}^{n_B} 
  { \left [ \Pi^{\rm OPE}_i(M_{ij}) - 
            \Pi^{\rm SR}_i(M_{ij}; \lambda_i, m, w_i) \right ]^2 
    \over
    \sigma_{{\rm OPE}}^2{}_i(M_{ij}) } \, ,
\label{chisqmulti}
\end{equation}
where $i$ runs over the considered sum rules.  (In this example, $m$
is a common parameter to all $n_{\rm SR}$ sum rules.)  The subscript
$i$ in $M_{ij}$ indicates that each sum rule has a unique set of $n_B$
Borel masses $M_j$ spanning the region of interest in Borel space.
The subscript $k$ in (\ref{chisq}), denoting a particular set of QCD
parameters, has been suppressed.  In the following discussion it will
become apparent that it is advantageous to weight the contributions in
the sum over the various sum rules by the size of their valid Borel
regimes.  If $V_B^i$ is the size of the valid Borel regime for the
$i$'th sum rule, then (\ref{chisqmulti}) is modified to
\begin{equation}
{\chi^2 \over N_{DF}} = {1 \over n_{\rm SR} \, n_B - n_p} \, 
                { n_{\rm SR} \over \sum_i^{n_{\rm SR}} V_B^i } \,
                  \sum_{i=1}^{n_{\rm SR}} \, \sum_{j=1}^{n_B} \,
  { V_B^i \, \left [ \Pi^{\rm OPE}_i(M_{ij}) - 
            \Pi^{\rm SR}_i(M_{ij}; \lambda_i, m, w_i) \right ]^2 
    \over
    \sigma_{{\rm OPE}}^2{}_i(M_{ij}) } \, .
\label{chisqweight}
\end{equation}

\subsection{Search Algorithm}
\label{algorithm}

   The optimization method utilized in this investigation is an
updated version of a direction-set routine by Powell originally
published in Ref. \cite{powell64}.  This algorithm finds the minimum
value of a function by iterative variation of the function parameters
which need not be independent.  Modifications of the original routine
improve the selection of conjugate directions in the search and avoid
the possible generation of linearly dependent search directions
\cite{zangwill67}.

   Powell's method is among the best methods for finding minima in a
multidimensional parameter space when derivatives are not readily
available \cite{brent73,numrec86}.  Part of the reason for the robust
nature of Powell's routine is the cautious criterion for ultimate
convergence.  Once a minimum is found, the optimal vector, $\vec a$,
in the parameter space is displaced by ten times the requested
accuracy.  The search proceeds from this new point until a minimum is
found at vector $\vec b$.  The minimum on the line joining $\vec a$
and $\vec b$ is found at $\vec c$.  Convergence is reached if the
components of the vectors $(\vec a - \vec c)$ and $(\vec b - \vec c)$
are all less than 10\% of the required accuracy.  Otherwise the
direction $(\vec a - \vec c)$ is utilized in a new search.

   The displacement of the first minimum by ten times the requested
accuracy plays a crucial role in removing the sensitivity of the
ultimate minimum from the initial parameter estimates.  While there is
sufficient flexibility in the algorithm parameters to search in the
locality of the initial parameter estimates, the opposite criteria has
been utilized in this investigation.  One of the main goals of this
investigation is to eliminate the possibility of fine tuning the
initial parameter estimates which otherwise can bias the results.
Instead, we are interested in the predictions of the QCD sum rules.

\section{\pmb{$\rho$}-MESON SUM RULES}
\label{rho}

   The fundamental QCD-SR for the $\rho$ meson follows from the
consideration of (\ref{twopt}), (\ref{chirho}), (\ref{rhoV}) and
(\ref{continuum}).  We consider the correlator, $\Pi_S(M^2)$, where
the continuum model is subtracted from the phenomenological side and
placed on the OPE side.
\begin{eqnarray}
\Pi_S(M^2) = 
f_\rho^2 \, e^{-M_\rho^2/M^2} &=& 
c_0 \, M^2 \, \left [  1-e^{w_\rho^2/M^2} \right ]
+ c_1 + {c_2\over M^2} + {c_3\over 2!\, M^4} + \cdots \nonumber \\
&& + {c_m \over (m-1)! \, (M^2)^{m-1} } + \cdots \, .
\label{rhoSR}
\end{eqnarray}
The Wilson coefficients are given by \cite{shifman79}
\begin{eqnarray}
&& c_0 = {1\over 8\pi^2} \left (1 + {\alpha_s \over \pi} \right),
\qquad c_1=0, \nonumber \\
&& c_2 = m_q \, \langle\overline{q}q\rangle
 + {1\over 24}\, \langle{\alpha_s\over \pi} G^2\rangle \, ,
\nonumber \\
&& c_3 = -{112\over 81} \, \pi \,\alpha_s \, \kappa_\rho \,
          \langle\overline{q}q\rangle^2 \, .
\label{rho-wilson}
\end{eqnarray}

The phenomenology of the QCD-SR is described by the
vector-meson pole of interest plus the continuum model accounting for
the contributions of all excited states.  By working in a region where
the pole dominates the phenomenology, one can minimize
sensitivity to the model and have assurance that it is the spectral
parameters of the ground state of interest that are being determined
by matching the sum rules. In practice, these considerations
effectively set an upper limit in the Borel parameter space, beyond
which the model for excited states dominates the phenomenological
side.

At the same time, the truncated OPE must be sufficiently convergent as
to accurately describe the true OPE.  Since the OPE is an expansion in
the inverse squared Borel mass, this consideration sets a lower limit
in Borel parameter space, beyond which higher order terms not present
in the truncated OPE are significant and important. Monitoring OPE
convergence is absolutely crucial to recovering nonperturbative
phenomena in the sum-rule approach, as it is the lower end of the
Borel region where the nonperturbative information of the OPE is most
significant.  As we shall see, this information must also be accurate.

In short, one should not expect to extract information on the ground
state spectral properties unless the ground state dominates the
contributions on the phenomenological side and the OPE is sufficiently
convergent.  In this investigation, we will analyze each individual
sum rule with regard to the above criteria.  A sum rule with an upper
limit in Borel space lower than the lower limit is considered invalid.
As a measure of the relative reliability of various sum rules we
consider the size of the regime in Borel space where both sides of the
sum rules are valid.  This quantity is used as a weight in the
calculation of the $\chi^2/N_{DF}$ as indicated in
(\ref{chisqweight}).  In addition, the size of continuum contributions
throughout the Borel region can also serve as a measure of
reliability, with small continuum model contributions being more
reliable.

   The fiducial Borel region is chosen \cite{leinweber90} such that
the highest-dimensional operator(s) (HDO) contribute no more than $\sim
10\%$ to the QCD side while the continuum contribution is less than
$\sim 50\%$ of the total phenomenological side (i.e., the sum of the
pole and the continuum contribution).  The former sets a criterion for
the convergence of the OPE while the latter controls the continuum
contribution.  While the selection of $50\%$ is obvious for pole
dominance, the selection of $10\%$ is a reasonably conservative
criterion that has not failed in practice.%
\footnote{Reasonable alternatives to the $10\%$ and $50\%$ criteria
are automatically explored in the Monte-Carlo error analysis, as the
condensate values and the continuum threshold change in each sample.}
See Ref.\ \cite{leinweber95f} for further specific examples.

\begin{figure}[p]
\begin{center}
\epsfysize=11.7truecm
\leavevmode
\setbox\rotbox=\vbox{\epsfbox{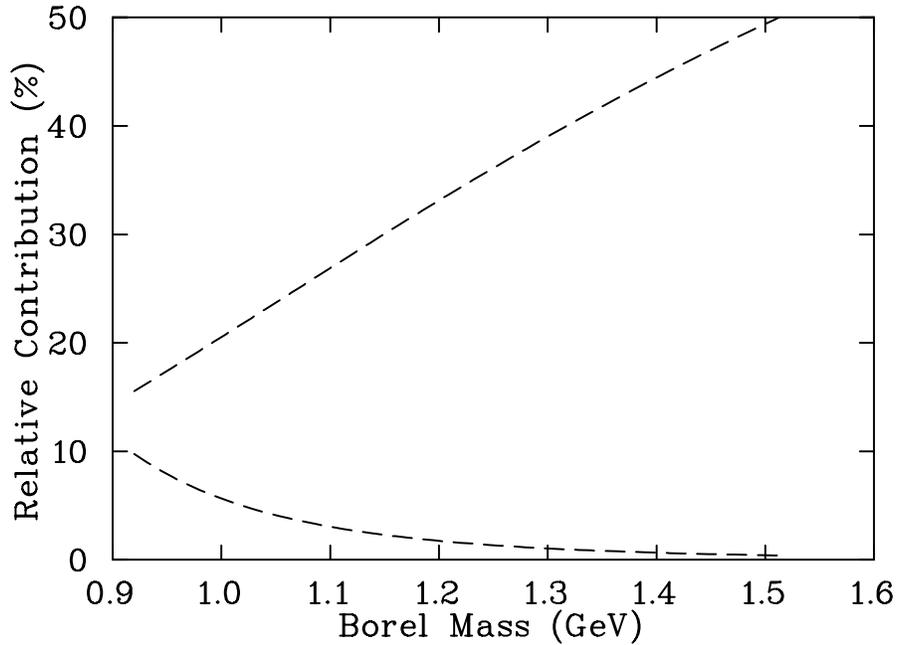}}\rotl\rotbox
\end{center}
\caption{The valid Borel window for the $\rho$-meson sum rule of
(\protect\ref{rhoSR}).  The relative HDO contributions limited to
$10\%$ of the OPE and the continuum model contributions limited to
50\% of the phenomenology are illustrated.  }
\label{SR1hdo}
\end{figure}

\begin{figure}[p]
\begin{center}
\epsfysize=11.7truecm
\leavevmode
\setbox\rotbox=\vbox{\epsfbox{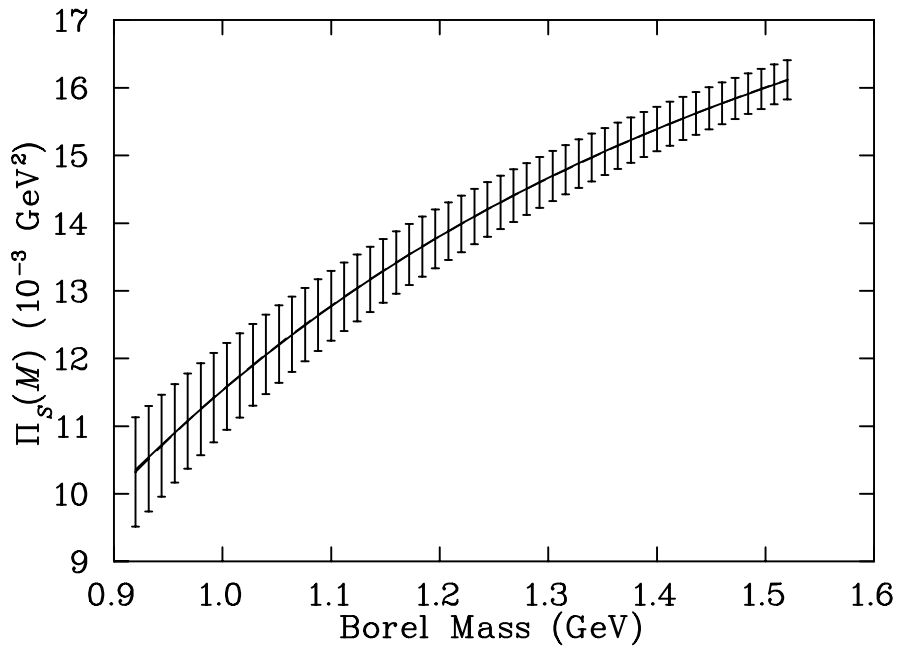}}\rotl\rotbox
\end{center}
\caption{A three parameter fit including $f_\rho$, $M_\rho$ and
$w_\rho$ of the $\rho$-meson sum rule of (\protect\ref{rhoSR}) for the
central values of the condensate distributions.  The error bars
illustrate the OPE uncertainty estimated at each of the 51 points
considered in the $\chi^2$ estimate.  Both the QCD$-$continuum
(dashed) and pole (solid) sides of (\protect\ref{rhoSR}) are plotted.}
\label{SR1fit}
\end{figure}

\begin{figure}[p]
\begin{center}
\epsfysize=11.7truecm
\leavevmode
\setbox\rotbox=\vbox{\epsfbox{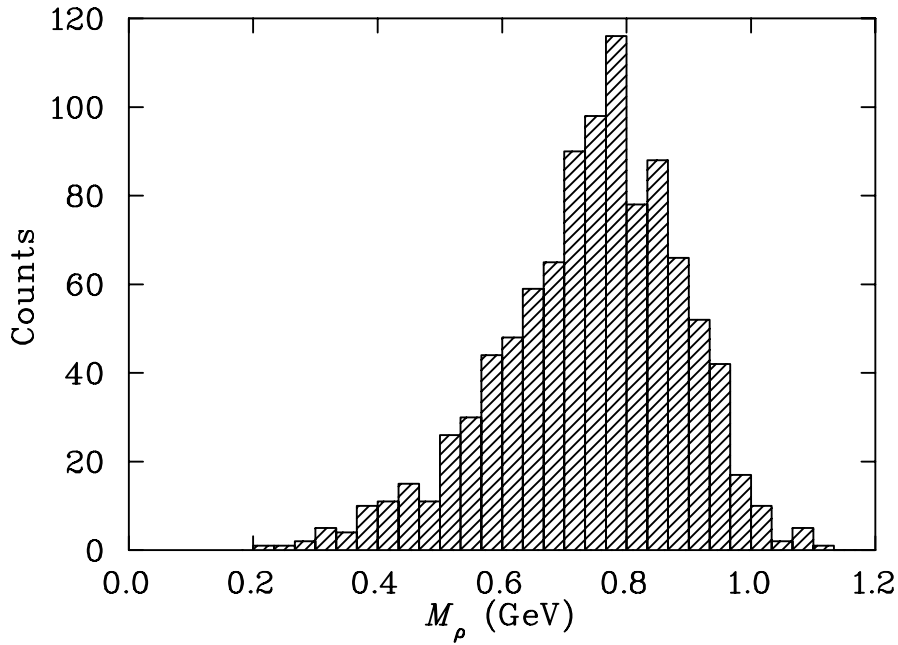}}\rotl\rotbox
\end{center}
\caption{Histogram of $\rho$-meson masses obtained from 1000 fits of
(\protect\ref{rhoSR}) for 1000 QCD parameter sets.  The distribution
corresponds to $M_\rho = 0.76 \, {}^{+0.12}_{-0.16}$ GeV.  }
\label{BinRhoMass}
\end{figure}

\begin{figure}[p]
\begin{center}
\epsfysize=11.7truecm
\leavevmode
\setbox\rotbox=\vbox{\epsfbox{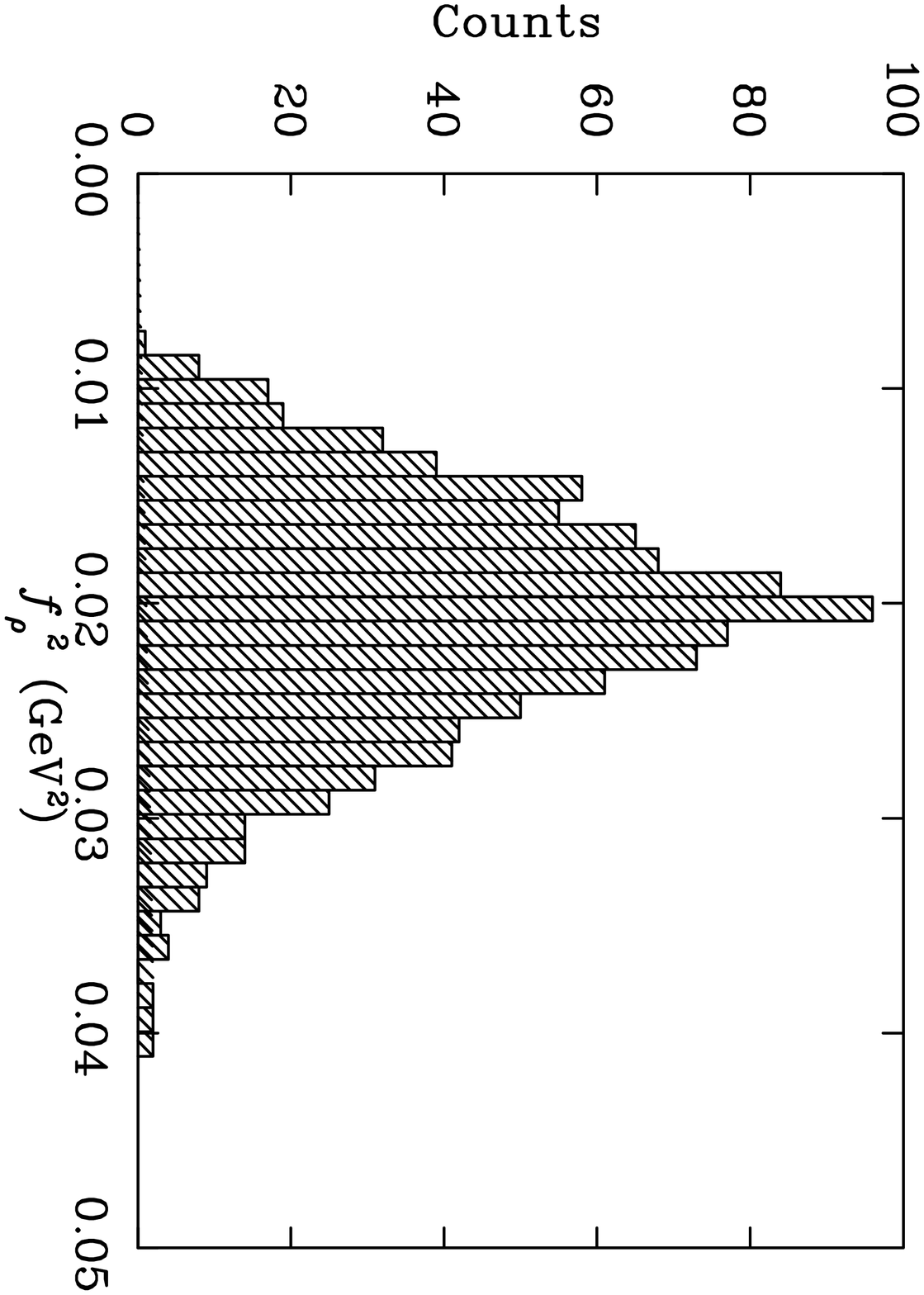}}\rotl\rotbox
\end{center}
\caption{Histogram of the squared $\rho$-meson decay constant obtained
from 1000 fits of (\protect\ref{rhoSR}) for 1000 QCD parameter sets.
The distribution corresponds to $f_\rho^2 = 0.021 \pm 0.006\ {\rm
GeV}^2$.  }
\label{BinRhof}
\end{figure}

   Figure \ref{SR1hdo} displays the valid Borel window for the
$\rho$-meson sum rule of (\ref{rhoSR}).  The HDO contributions limited
to $10\%$ of the OPE and the continuum model contributions limited to
50\% of the phenomenology are illustrated.  The continuum
contributions are the order of $15\%$ of the phenomenological
contributions at the lower end of the Borel regime.  Here, the pole of
interest truly dominates the QCD-SR.  Figure \ref{SR1fit} displays the
corresponding fit obtained by minimizing $\chi^2$ in a three parameter
search including $f_\rho$, $M_\rho$ and $w_\rho$.  Figure
\ref{BinRhoMass} displays the distribution of $\rho$-meson masses
obtained from the 1000 sum rule fits.  The distribution corresponds to
\begin{equation}
M_\rho = 0.76 \, {}^{+0.12}_{-0.16}\ {\rm GeV} \, .
\end{equation}
The uncertainty is somewhat larger than the commonly assumed 10\%.
Figure \ref{BinRhof} displays the distribution of the square of
$f_\rho$ which corresponds to
\begin{equation}
f_\rho = 0.144 \pm 0.021\ {\rm GeV} \, ,
\end{equation}
which agrees with the experimental value of $f_\rho = M_\rho / g_\rho
= 0.153$ GeV.  The continuum threshold is $w_\rho = 1.23 \pm 0.17$ GeV
which compares favorably with $1.31 \pm 0.04$ GeV (the $\rho'$ mass
less the half width), given the approximate nature of the continuum
model.

\subsection{Derivative Sum Rules and the Ratio Method}

In addition to the fundamental sum rule of (\ref{rhoSR}) one also
finds the use of sum rules obtained by differentiating
(\ref{rhoSR}) with respect to $1/M^2$.  For the $n$'th derivative
\begin{eqnarray}
f_\rho^2 \, ( M_\rho^2 )^n \, e^{-M_\rho^2/M^2} &=&
n! \, c_0 \, (M^2)^{n+1} \, E_n
+ (-1)^n \biggl [ c_{n+1} + {c_{n+2} \over M^2}
+ \cdots \nonumber \\
&& + {c_{m} \over (m-n-1)! \, (M^2)^{m-n-1} } + \cdots \biggr ] \ ,
\label{derivSR}
\end{eqnarray}
While these sum rules are obviously unnecessary for the extraction of
$\rho$-meson properties, the hope is that these additional sum rules
will provide new useful information that might be used to
further constrain the phenomenological properties.

It is possible to isolate the $\rho$-meson mass as a function of the
Borel mass, by taking a ratio of the first derivative sum rule with
the fundamental sum rule.  This approach has some aesthetic appeal, and
has widely become the method of choice for analyzing QCD-SRs.  The
continuum threshold is selected to make the ratio of the two sum rules
as flat as possible as a function of Borel mass.  Finally, the
$\rho$-meson mass is selected from the point at which the ratio is
most flat or stable.  Unfortunately, this method has some very
unpleasant shortcomings.

   The ratio method does not check the validity of each individual sum
rule.  It is possible to have individual sum rules that are not valid
while their ratio is flat as function of Borel mass.  In addition, the
ratio method cannot account for the fact that sum rules do not work
equally well.  The Borel region where a sum rule is valid can vary
from one sum rule to another.  Moreover, large uncertainties in one
sum rule can spoil the predictions of another sum rule.  Finally, the
continuum contributions to the sum rules are not monitored in the
ratio method. If the continuum contribution is dominant in a sum rule,
one should not expect to get reliable information about the lowest
resonance.

   The fact that sum rules do not work equally well deserves further
discussion.  In Ref.\ \cite{leinweber95f} a detailed comparison of
fundamental and derivative sum rules is carried out for the $\rho$
meson.  These arguments are not broadly known and are generally
applicable to any derivative Borel-improved QCD-SR.  Hence the
important observations are highlighted here.

   If one could carry out the OPE to arbitrary accuracy and use a
spectral density independent of models for excited state
contributions, the predictions based on (\ref{rhoSR}) and those based
on the derivative sum rules of (\ref{derivSR}) should be the same.  In
practical calculations, however, one has to truncate the OPE and use a
simple phenomenological ansatz for the spectral density. Thus it is
unrealistic to expect the sum rules to work equally well.  Compare the
$n$'th derivative sum rule of (\ref{derivSR}) with the direct sum rule
of (\ref{rhoSR}).
\begin{enumerate}
\item The perturbative contribution in the derivative sum rule has an
extra factor $n!$ relative to the corresponding term in
(\ref{rhoSR}), implying that the perturbative contribution is
more important in the derivative sum rules than in the fundamental sum
rule, and becomes increasingly important as $n$ increases.  Since the
perturbative term mainly contributes to the continuum of the spectral
density, maintaining dominance of the lowest resonance pole in the sum
rule will become increasingly difficult as $n$ increases.
\item In (\ref{rhoSR}), the term proportional to $c_{m}$ is suppressed
by a factor of $1/(m-1)!$, while it is only suppressed by $1/(m-n-1)!$
in the derivative sum rule ($m\,>\,n$). This implies that the
convergence of the OPE is much slower in the derivative sum rule than
in the fundamental sum rule.  Consequently, the poorly known high-order
power corrections are more important in the derivative sum rule than
in Eq.~(\ref{rhoSR}), and become more and more important as $n$
increases. If one would like to restrict the size of the last term of
the OPE to maintain some promise of OPE convergence, the size of the
Borel region in which the sum rules are believed to be valid is
restricted.
\item The power corrections proportional to $c_1, c_2, \cdots, c_n$ do
{\it not} contribute to the $n^{\rm th}$ derivative sum rule but do
contribute to (\ref{rhoSR}). If one truncates the OPE, part or
all of the nonperturbative information will be lost in the derivative
sum rules. It is also worth noting that the leading power corrections
are the most desirable terms to have.  They do not give rise to a term
in the continuum model and they are not the last term in the OPE,
whose relative contribution should be restricted to maintain OPE
convergence.
\end{enumerate}
In practice, the predictions based on the fundamental sum rule
of~(\ref{rhoSR}) are more reliable than those from the derivative
sum rules, which become less and less reliable as $n$ increases. This
can also be demonstrated by analyzing the sum rules numerically.

\begin{figure}[p]
\begin{center}
\epsfysize=11.7truecm
\leavevmode
\setbox\rotbox=\vbox{\epsfbox{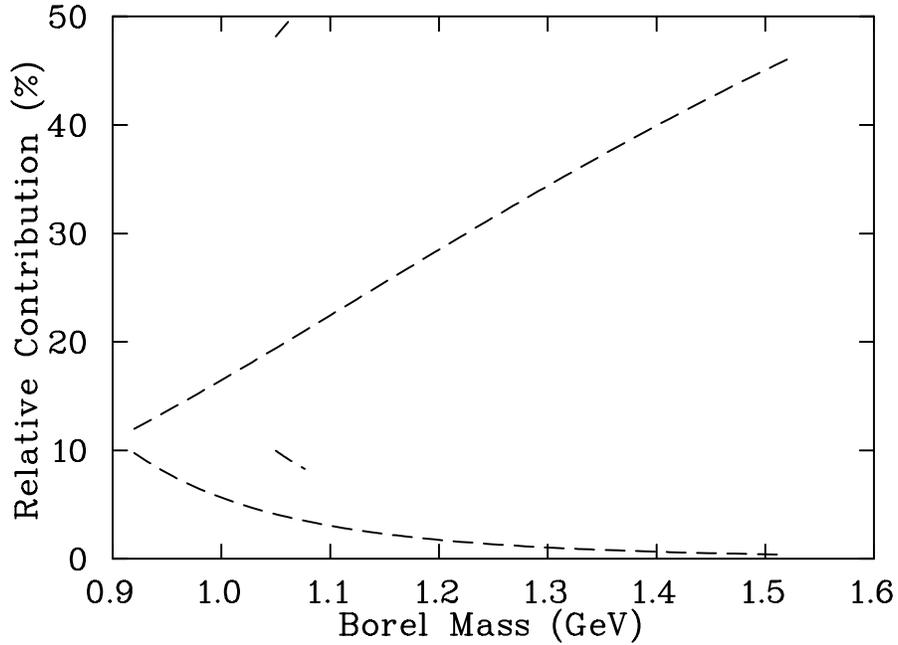}}\rotl\rotbox
\end{center}
\caption{The valid Borel windows for the fundamental $\rho$-meson sum
rule of (\protect\ref{rhoSR}) (dashed) and the first derivative sum
rule of (\protect\ref{derivSR}) (dot-dashed).  Both the relative HDO
contributions limited to $10\%$ and the continuum model contributions
limited to 50\% are illustrated.  }
\label{SR12hdo}
\end{figure}

   Figure \ref{SR12hdo} displays the valid Borel windows for a
simultaneous fit of the $\rho$-meson sum rules of (\ref{rhoSR}) and
the first derivative $(n=1)$ of (\ref{derivSR}).  The reliability of
the fundamental sum rule over the derivative sum rule is obvious.  The
valid Borel regime for the derivative sum rule is nearly nonexistent.

   All three points of the discussion surrounding derivative sum rules
are displayed in this figure.  Item 1 indicates the continuum model
contributions will be much larger in the derivative sum rule for a
given Borel mass.  Indeed the continuum model contribution never drops
below 45\% in the valid regime.  The relative contribution of the last
term in the truncated OPE is larger in the derivative sum rule for a
given Borel mass, in agreement with the expectations of items 2 and 3.

\begin{figure}[p]
\begin{center}
\epsfysize=11.7truecm
\leavevmode
\setbox\rotbox=\vbox{\epsfbox{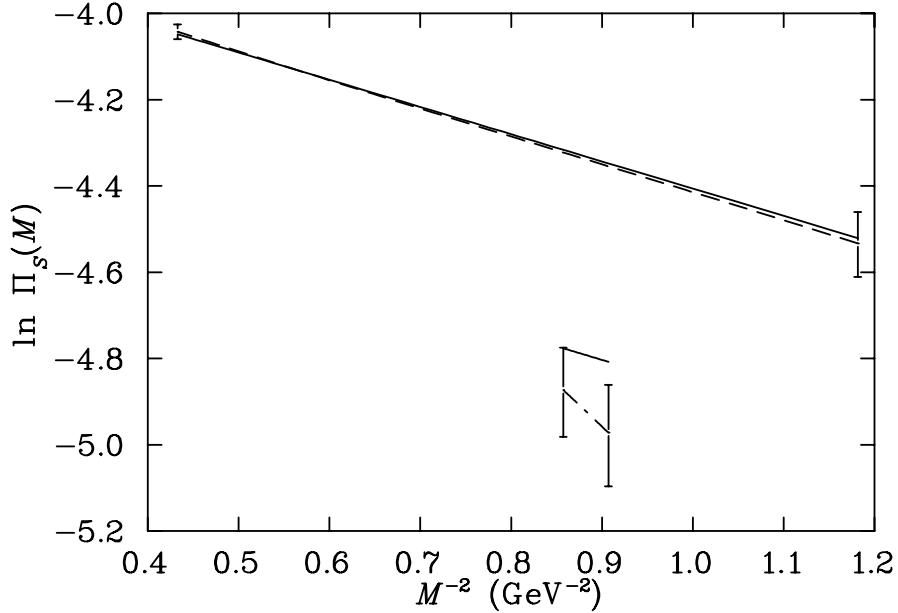}}\rotl\rotbox
\end{center}
\caption{A three parameter fit including $f_\rho$, $M_\rho$ and
$w_\rho$ of the fundamental $\rho$-meson sum rule of
(\protect\ref{rhoSR}) (dashed) and the first derivative sum rule of
(\protect\ref{derivSR}) (dot-dashed) for the central values of the
condensate distributions.  Solid lines indicate the ground state
contributions to these sum rules.  In this and the following plots
displaying sum rule fits, uncertainties are displayed only at the ends
of the Borel region for clarity.}
\label{SR12fit}
\end{figure}

   Figure \ref{SR12fit} displays the corresponding fit obtained by
minimizing the weighted $\chi^2/N_{DF}$ of (\ref{chisqweight}) in a
three parameter search including $f_\rho$, $M_\rho$ and $w_\rho$.
This time the logarithms of the left- and right-hand sides of
equations (\ref{rhoSR}) and (\ref{derivSR}) are plotted as a function
of the inverse squared Borel mass.  This method has the advantage of
revealing the possible breakdown of the OPE signified by deviations of
the dashed curves from linearity.  Uncertainties are displayed only at
the ends of the Borel region for clarity.

   While the central values of the best fit parameters $M_\rho = 0.77
\pm 0.07$ GeV, $f_\rho^2 = 0.022 \pm 0.002$ GeV${}^2$, and $w_\rho =
1.31 \pm 0.05$ GeV, remain essentially unchanged, the reduction of the
uncertainties comes from rather unreliable information as discussed
above.  These arguments are further supported by the failure of the
two sides of (\ref{derivSR}) for $n=1$ to agree as illustrated in
figure \ref{SR12fit}.  To further probe the reasons for this failure,
the analysis of these sum rules is repeated with a spectral
representation that accounts for the finite width of the $\rho$ meson.

   To account for the finite width of the $\rho$-meson, the $\delta$
function  in (\ref{rhoV}) is replaced by a normalized Breit-Wigner form 
\begin{equation}
\frac{1}{\pi}\, {M_\rho \, \Gamma_\rho \over
(s - M_\rho^2 + \Gamma_\rho^2 / 4)^2  + M_\rho^2 \,
\Gamma_\rho^2} \, ,
\label{breitwigner}
\end{equation}
and the integral over the $\rho$-meson ground state
\begin{equation}
I_\rho(M^2) = {M_\rho \, \Gamma_\rho \over \pi} \,
\int_{4m_{\pi}^2}^{\infty} 
{ds\, e^{-s/M^2} \over
(s - M_\rho^2 + \Gamma_\rho^2 / 4)^2  + M_\rho^2 \,
\Gamma_\rho^2} \, ,
\label{i-finite-widths}
\end{equation}
is handled numerically \cite{leinweber95e}.  Using the experimental
width of 151.5 MeV, a three parameter fit of the fundamental and
derivative sum rules with the Breit-Wigner form leads to essentially
the same results reported in figure \ref{SR12fit}.  The mass,
coupling, and continuum threshold all agree within uncertainties, and
have similar uncertainties.  Hence, these spectral properties may be
reliably estimated without precise information on the $\rho$ width.
However, $\rho-\omega$ mixing provides an interesting example where
finite mesonic widths can play a central role in arriving at the
correct physics \cite{leinweber95e,leinweber96a}.

   The origin of the discrepancy for the derivative sum rule is more
likely to lie in short-comings of continuum model.  For this sum rule
the continuum model contributions never drop below 45\%, which may be
too large for the simple model typically used in QCD-SR calculations.
Alternatively, poorly known higher order terms, not present in the
truncated OPE may become important in the derivative sum rule.  In the
derivative sum rule, the next term in the series is
suppressed by merely $2!\,$.

   The importance of carefully monitoring the validity of QCD-SRs is
nicely displayed in the recent debate over the behavior of
vector-meson masses in nuclear matter.  There an analysis involving
invalid sum rules leads to the conclusion that vector-meson masses
increase in finite density \cite{koike95}, in contradiction to a
previous analysis \cite{hatsuda92} also based on the ratio method.
The debate is resolved in Ref.\ \cite{leinweber95f} where valid sum
rules indicate that vector-meson masses decrease for increasing
nuclear matter density.

   Having firmly established the analysis procedure in the well
understood $\rho$-meson channel, we are ready to apply the same
approach to the nucleon sum rules.  We will find that many results
contradict the conventional wisdom.

\section{NUCLEON SUM RULES:  SPIN-1/2 INTERPOLATORS}
\label{nucleon11}

   Here we focus on the Borel improved QCD Sum Rules for the
generalized interpolator of (\ref{chiO}).  The sum rules proportional
to the correlator $\Pi_S(M)$ where continuum model contributions are
subtracted from the phenomenological side and placed on the OPE side
are
\begin{mathletters}
\begin{eqnarray}
(\gamma \cdot p):\quad
&&{5 + 2 \beta + 5 \beta^2 \over 64} \, M^6 \, L^{-4/9} 
\left [ 1 - e^{-w_1^2/M^2}
\left ( {w_1^4 \over 2 M^4} + {w_1^2 \over M^2} + 1 \right ) \right ]
\nonumber \\
&&\quad +{5 + 2 \beta + 5 \beta^2 \over 256} \, b \, M^2 \, L^{-4/9} 
\left [ 1 - e^{-w_1^2/M^2} \right ] 
\nonumber \\
&&\quad +{7 - 2 \beta - 5 \beta^2 \over 24} \, a^2 \, L^{4/9} 
        -{13 - 2 \beta - 11 \beta^2 \over 96} \, 
         { m_0^2 \, a^2 \over M^2} \, L^{-2/27} 
\nonumber \\
&&\qquad = \widetilde \lambda_{\cal O}^2 \, e^{-M_N^2/M^2} \, ,
\label{nucl11atP} 
\end{eqnarray}
and
\begin{eqnarray}
(1):\quad
&&{7 - 2 \beta  - 5 \beta^2 \over 16} \, a \, M^4 
\left [ 1 - e^{-w_2^2/M^2}
\left ( {w_2^2 \over M^2} + 1 \right ) \right ] \nonumber \\
&&\quad -{3 (1 - \beta^2) \over 16} \, m_0^2 \, a \, M^2 \, L^{-14/27} 
\left [ 1 - e^{-w_2^2/M^2} \right ] \nonumber \\
&&\quad +{19 + 10 \beta - 29 \beta^2 \over 2^7 \, 3^2} \, a \, b  
\nonumber \\
&&\qquad = \widetilde \lambda_{\cal O}^2 \, M_N \, e^{-M_N^2/M^2} \, ,
\label{nucl11at1}%
\end{eqnarray}%
\label{nucl11}%
\end{mathletters}%
where the Dirac-$\gamma$ structures are indicated on the left, and
$\widetilde\lambda_{\cal O} = (2 \pi)^2 \lambda_{\cal O}$.  The
factors of $L = \left [ \log \left ( M/\Lambda_{\rm QCD} \right ) /
\log \left ( \mu/\Lambda_{\rm QCD} \right ) \right ]$ account for the
anomalous scaling of the operators and interpolators in the leading
logarithmic approximation \cite{yang93}.

\subsection{Traditional QCD-SR Analysis}

\begin{figure}[p]
\begin{center}
\epsfysize=11.7truecm
\leavevmode
\setbox\rotbox=\vbox{\epsfbox{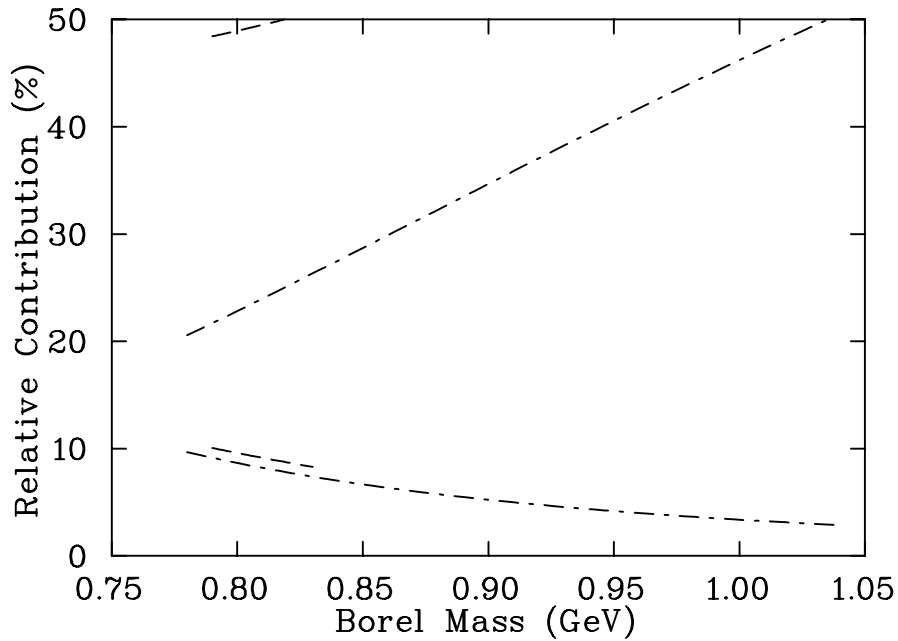}}\rotl\rotbox
\end{center}
\caption{The valid Borel windows for the nucleon sum rules of
(\protect\ref{nucl11atP}) (dashed) and (\protect\ref{nucl11at1})
(dot-dashed).  Both the relative HDO contributions limited to $10\%$
of the OPE and the continuum model contributions limited to 50\% of
the phenomenology are illustrated.  }
\label{SR34K2hdo}
\end{figure}

   Let us begin with the traditional or standard nucleon sum rule
analysis where $\alpha_s$ corrections are ignored.  The more common
interpolator of (\ref{chiSR}) is utilized by setting $\beta = -1$ in
(\ref{nucl11}).  In addition, the traditional condensate parameter set
with $b = 0.47 \pm 0.20\ {\rm GeV}^4$ is selected in order to evaluate
the validity of previously published results.  For a preliminary
analysis of the validities of the sum rules, the nucleon mass is fixed
to 1 GeV.  Figure \ref{SR34K2hdo} illustrates the valid Borel regimes
for these two sum rules found by implementing the procedure discussed
for the $\rho$-meson.

   A comparison of figure \ref{SR34K2hdo} with the analogous
figure \ref{SR12hdo} for the $\rho$-meson suggests the nucleon sum
rules will not be as reliable.  In the nucleon sum rules, the
continuum model contributes 20\% or more to the phenomenological side
of the sum rules.  In particular, the continuum model contribution in
the sum rule of (\ref{nucl11atP}) exceeds that for the derivative
$\rho$-meson sum rule which has already been shown to be incompatible
with the fundamental sum rule.  It should not be surprising to find a
similar disaster for the nucleon case.

   As in the $\rho$-meson case, there is an analogous ratio method for
obtaining the nucleon mass that has also become popular.  Here the
nucleon mass is isolated as a function of the Borel mass by taking the
ratio of equations (\ref{nucl11at1})/(\ref{nucl11atP}).  The mass is
usually extracted from a regime in which the ratio is flat.  Again OPE
convergence and ground-state dominance criteria are not checked in
this approach, and too often the mass is extracted from a regime in
which the continuum model dominates.

\begin{figure}[p]
\begin{center}
\epsfysize=11.7truecm
\leavevmode
\setbox\rotbox=\vbox{\epsfbox{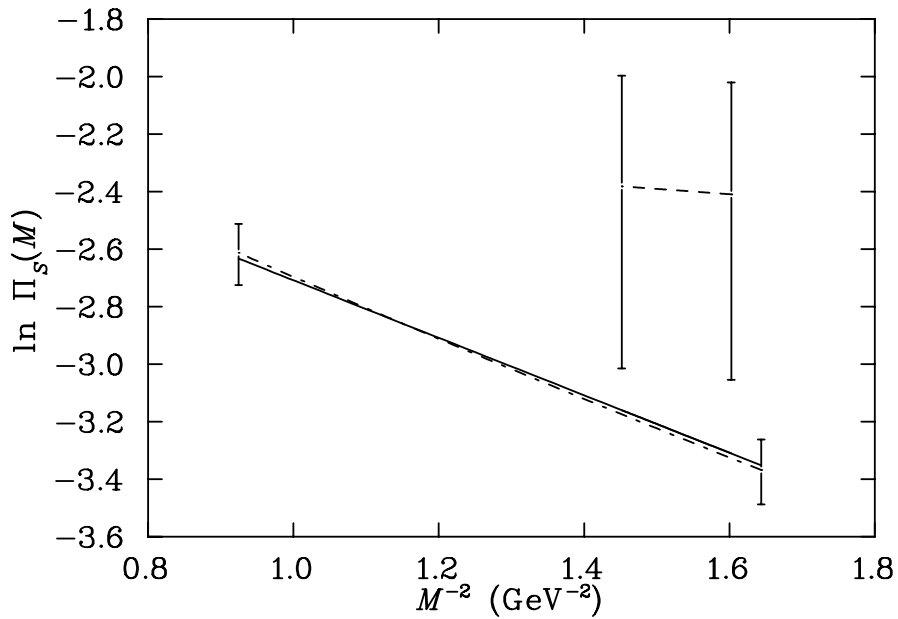}}\rotl\rotbox
\end{center}
\caption{ Three parameter fit of (\protect\ref{nucl11atP}) (dashed)
and (\protect\ref{nucl11at1}) (dot-dashed) to the ground state
contribution (solid) for the residue $\lambda_{\cal O}$, and the two
continuum model thresholds of $w_1$ and $w_2$.  The small Borel regime
and the relatively huge OPE uncertainties of the sum rule of
(\protect\ref{nucl11atP}) indicate that this sum rule plays a
negligible role in determining the spectral properties of the
nucleon. }
\label{SR34K2fit}
\end{figure}

   Figure \ref{SR34K2fit} illustrates the three parameter fit of the
residue $\lambda_{\cal O}$, and the two continuum model thresholds of
$w_1$ and $w_2$.  The small Borel regime and the relatively huge OPE
uncertainties of the sum rule of (\ref{nucl11atP}) indicate that this
sum rule plays a negligible role in determining the spectral
properties of the nucleon.  This is a perfect example of a case where
large uncertainties in the sum rule of (\ref{nucl11atP}) would spoil
the predictive ability of (\ref{nucl11at1}) if the ratio method was
used.  The origin of the large uncertainties lies in the poorly known
value of the four-quark condensate.  The resultant fit parameters for
this sum rule are also unphysical as the continuum threshold lies at
zero below the nucleon pole.  In other words, the present form of the
sum rule at $(\gamma \cdot p)$ is unable to provide any information on
the ground state that is consistent with the more reliable sum rule of
(\ref{nucl11at1}) at the structure 1.

   Unfortunately, the community has incorrectly identified
(\ref{nucl11atP}) as the more reliable sum rule.  This belief is based
on the simple fact that the leading term of the OPE (the identity
operator) has a well known vacuum expectation value.  However, the
$\alpha_s$ corrections for the identity operator \cite{hatsuda95} are
uncomfortably large.  The first order corrections are independent of
$\beta$ and are
\begin{equation}
{\alpha_s \over \pi} \left ( {53 \over 12} + \gamma_E \right ) \, ,
\end{equation}
where $\gamma_E = \psi(1) \simeq 0.58$.  This correction is $\sim
50$\% and suggests perturbative corrections are completely out of
control for this sum rule.  Since the magnitude of the
identity operator contribution is closely tied to the size of
continuum model contributions, it is not surprising to find the valid
Borel regime for this sum rule disappears when $\alpha_s$
corrections are added.  

   Another factor in the traditional preference of (\ref{nucl11atP})
over (\ref{nucl11at1}) is that the second term of (\ref{nucl11at1})
vanishes for Ioffe's choice of nucleon interpolator $(\beta =
-1)$.  This is really nothing more than a good reason to exclude the
interpolator $\chi_{\rm SR}$ of (\ref{chiSR}) as an optimal
interpolator.

   When one considers the monstrous leading-order $\alpha_s$
corrections, the possibility of a significant dimension-two power
correction arising from a summation of the perturbative series
\cite{brown92b,zakharov92}, and the large uncertainties associated
with the value of the four-quark condensate, one cannot help but reach
the conclusion that the QCD side of the sum rule of (\ref{nucl11atP})
is really unknown.

\subsection{Nucleon Sum Rule at \pmb{$\gamma \cdot p$}}

   Since the sum rule of (\ref{nucl11atP}) has traditionally been the
nucleon sum rule of choice, it is interesting to consider the
predictions of this sum rule alone.  Applying the same techniques as
used for the $\rho$-meson to (\ref{nucl11atP}), a three parameter fit
of $\lambda_{\cal O}$, $w_1$ and $M_N$ yields
\begin{equation}
\widetilde\lambda_{\cal O}^2 = 0.15 \pm 0.05\ {\rm GeV}^6 \, , \quad
w_1 = 0.74 \pm 0.19\ {\rm GeV} \, , \quad
M_N = 0.44 \pm 0.05\ {\rm GeV} \, .
\label{failure}
\end{equation}
While setting $\kappa_N = 1$ slightly increases the nucleon mass to
0.47 GeV, adding leading order $\alpha_s$ corrections has a negligible
effect on the fit parameters.  Consideration of the more recent
condensate parameter set discussed in Sec.\ \ref{uncert} leads to
similar results
\begin{equation}
\widetilde\lambda_{\cal O}^2 = 0.14 \pm 0.10\ {\rm GeV}^6 \, , \quad
w_1 = 0.71 \pm 0.5\ {\rm GeV} \, , \quad
M_N = 0.43 \pm 0.09\ {\rm GeV} \, .
\label{SR3mod}
\end{equation}
This may be the first QCD sum rule to predict spectral properties that
significantly contradict experiment.\footnote{Note that this is unlike
the case of the pseudoscalar pion current \protect\cite{novikov81}
where a check of OPE convergence and ground state dominance indicates
the absence of a valid Borel regime.} As such, this may be the most
interesting of all QCD sum rules, and we shall return to this sum rule
after examining the selection of the most optimal interpolator for
nucleon sum rule analyses.

   The predictions of (\ref{nucl11atP}) summarized in (\ref{failure})
are extremely different from that found in the literature.  Hence it
is important to understand how these new results come about.
A critical examination of the nucleon ground state results reported in
Ref.\ \cite{leinweber90} by the present author sheds considerable
light on the importance of the rigorous QCD-SR analysis presented
here.  

   Ref. \cite{leinweber90} was the first analysis to give careful
regard to the issues of OPE convergence and ground-state dominance of
the phenomenological side.  Moreover, all phenomenological parameters
were included as fit parameters to be determined by the sum rules.
However, there are three important differences between the analysis
presented here and that of Ref.\ \cite{leinweber90}.
\begin{enumerate}
\item The standard logarithmic measure \cite{belyaev83} of sum rule
agreement was used to fit the sum rules.  This approach fails to apply
an appropriate weight to sum rule discrepancies, unlike the $\chi^2$
measure of (\ref{chisq}).
\item The algorithm used to fit the sum rules did not employ the
cautious convergence criteria discussed in Section \ref{algorithm}.
Implementation of the new criteria aids in avoiding insignificant
local minima.
\item Only the standard condensate parameter set was considered in
determining the nucleon mass.  Here 1000 sets are considered.
\end{enumerate}
The optimal parameter set found in Ref.\ \cite{leinweber90}
corresponds to a local minimum which is sensitive to the measure of
sum rule agreement, the initial parameter estimates and the condensate
parameter set.  

   Other authors have managed to keep the nucleon mass closer to the
experimental result by extracting the result from a regime in Borel
space that provides the correct answer or fixing the continuum
threshold to values suggested by the particle data tables.  As we
shall see in Section \ref{correl}, fixing the continuum threshold to
phenomenologically motivated values largely determines the mass.
Hence you {\it can} get anything you want provided you are clever
enough to fix the threshold to the {\it right} value.  And since the
excited states are modeled there is tremendous freedom in the
threshold parameter space.  Of course, such input may not be in
agreement with the content of the OPE.  In any event, such results
must not be regarded as predictions of QCD sum rules.

\subsection{Optimal Mixing of Spin-1/2 Interpolators}
\label{OptMixSect}

   Having established the nucleon sum rule of (\ref{nucl11at1}) at the
structure 1 to be the potentially more reliable sum rule, it is
important to determine the optimal mixing of spin-1/2 interpolators
for this sum rule.

   It has long been established that there are two independent
interpolating fields with no derivatives having the quantum numbers of
spin 1/2 and isospin 1/2.  Both are expected to excite the ground
state nucleon from the vacuum.  As such, there is a history of
argument over the optimal nucleon interpolating field to be used in
analyses
\cite{leinweber90,chung84,ioffe81,leinweber95b,chung82,ioffe83,%
furnstahl92,jin94,dosch89a}
and this issue is recognized to be of paramount importance
\cite{shifman92}.

   While some advocate an interpolating field for which the leading
terms of the OPE are stationary with respect to the interpolating
field mixing parameter \cite{chung84,dosch89a}, others argue that a
balance between OPE convergence and QCD continuum contributions must
be maintained \cite{leinweber90,ioffe83,furnstahl92,jin94}.  Ideally,
one would like to simply calculate with alternate interpolating fields
and confirm that the nucleon properties remain unchanged.  However,
previous implementations of the QCD-SR approach have hindered any firm
conclusions.  The rigorous analysis introduced here will shed
considerable light on these issues.

   Fortunately, the lattice approach to QCD is not plagued with the
same limitations and the debate over the optimal interpolating field
was resolved in Ref.\ \cite{leinweber95b}.  There is was found that,
to a good approximation, $\chi_2$ of (\ref{chiN2}) excites pure QCD
continuum.  Since $\chi_2$ has negligible overlap with the ground
state nucleon, it is tempting to simply conclude that the optimal
interpolating field is $\chi_1$.  While this is certainly the case for
lattice QCD investigations, it is not the case for QCD-SR analyses.
By including $\chi_2$ components in an interpolating field, one can
reduce the continuum contributions excited by $\chi_1$ and allow a
broader Borel analysis window in which better contact with the ground
state of interest is obtained.

   One of the most difficult things to monitor in the QCD-SR approach
is whether the OPE is sufficiently convergent for a particular value
of Borel mass.  The lattice results of Ref.\ \cite{leinweber95b}
indicate the $\chi_2 \overline \chi_2$ correlator has the fastest
converging OPE, as its overlap with the nucleon ground state is
negligible.  Similarly, the combination $\chi_1 \overline \chi_1$
produces an OPE with the slowest convergence, as this correlation
function is dominated by the ground state nucleon for small Borel
masses.

   Hence, errors made in truncating the OPE are dominated by errors in
the $\chi_1 \overline \chi_1$ component of the general correlator.
The relative error in the OPE truncation can be reduced by adding
$\chi_2$ components to the correlator.  However, the $\chi_2$
components in the OPE are simply subtracted off again by the continuum
model terms.  Hence the relevant error is the absolute error.  For
$|\beta|
\mathrel{\raise.3ex\hbox{$<$\kern-.75em\lower1ex\hbox{$\sim$}}} 1$,
this error is dominated by $\chi_1 \overline \chi_1$ components of the
correlator.  As a result, OPE truncation errors are approximately
independent of $\beta$.  This crucial point has been neglected in
previous arguments regarding the optimal nucleon interpolating field.

   Since the errors made in truncating the OPE are dominated by errors
in the $\chi_1 \overline \chi_1$ component of the general correlator
the criteria of a 10\% limit on the HDO contribution to the OPE should
be applied to sum rules when $\beta = 0$.  Once this lower limit is
established, it may be used for all $|\beta|
\mathrel{\raise.3ex\hbox{$<$\kern-.75em\lower1ex\hbox{$\sim$}}} 1$.
We will refer to this approach for determining the valid Borel regime
as the ``refined'' method in the following.

\begin{figure}[t]
\begin{center}
\epsfysize=11.7truecm
\leavevmode
\setbox\rotbox=\vbox{\epsfbox{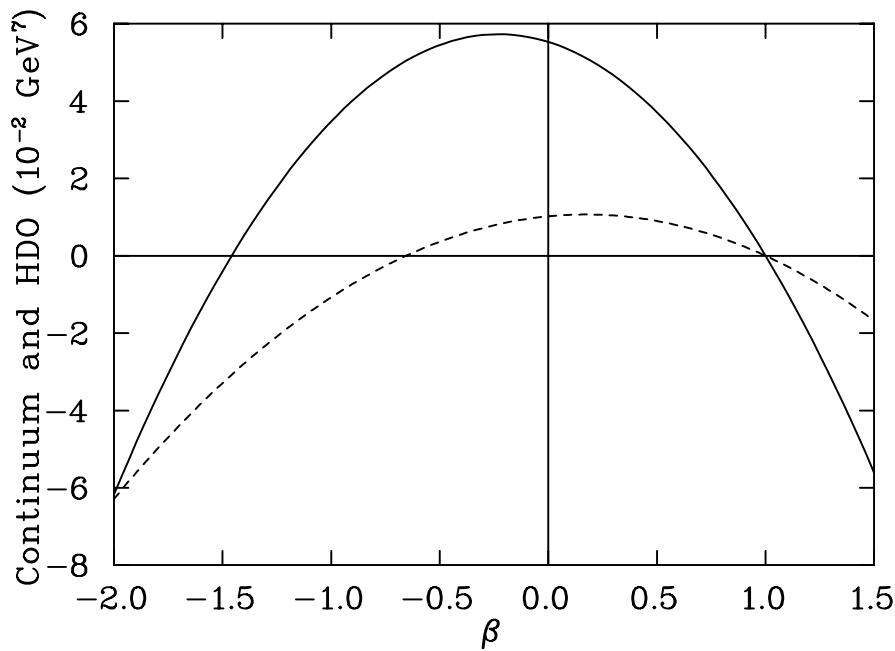}}\rotl\rotbox
\end{center}
\caption{ Continuum model (solid curve) and HDO (dashed curve)
contributions to the Borel improved QCD-SR of
(\protect\ref{nucl11at1}) plotted as a function of the interpolating
field mixing parameter $\beta$.}
\label{optcon}
\end{figure}

   The optimal nucleon interpolator must involve $\chi_1$, as this
interpolator is required to maintain overlap with the ground state.
The task is to determine the optimal mixing of $\chi_2$.  Since
$\chi_2$ has negligible overlap with the nucleon, the ground state
contribution to a sum rule is also independent of $\beta$.  Hence, the
size of the continuum model contribution is the predominant factor in
determining the optimal interpolator.  Figure \ref{optcon} illustrates
the contributions of the continuum model terms in (\ref{nucl11at1})
for $M=0.94$ GeV and $w_2 = 1.4$ GeV.  The following discussion is not
dependent on the precise values of these parameters.  The first point
to be made is that contributions from the continuum model are largest
for $\beta \sim -0.2$.  This selection of mixing is the worst possible
choice for extracting information on the ground state nucleon.

   Figure \ref{optcon} also indicates it is possible to have vanishing
continuum model contributions at $\beta \simeq -1.5$ or $\beta = 1$.
However, we are relying on the continuum model to account for strength
in the correlator that does not have its origin in the ground state.
Without a continuum model, one would need to include additional poles
on the right-hand-side of (\ref{nucl11at1}) to account for positive
and negative parity excitation strength.  For $\beta < -1.5$ or $\beta
> 1.0$ the correlator is negative indicating the sum rule is saturated
by a negative parity state.

   Thus the optimal interpolator is $\beta \sim -1.2$ or $\beta \sim
0.8$.  To discriminate between these two regimes, we turn to the
higher-dimension-operator contributions which do not contribute to the
continuum model.  It is these terms that provide crucial information
on whether the strength in the correlator lies in the ground state or
the excited states.  If these terms are absent, the optimal fit of the
correlator is obtained when $\widetilde\lambda_{\cal O} \to 0$ and
$w_2 \to 0$. In this case the continuum model and the OPE become
equivalent via the Laplace transform of (\ref{laplacetrans}) and the
fit is perfect.  Hence the higher-dimension-operator contributions
should be large in magnitude.  A change in sign from the leading terms
of the OPE will also assist in distinguishing ground state strength
from excited state strength as the change in the curvature of the
correlator will be more prominent.

   The last term of (\ref{nucl11at1}) is independent of the continuum
model, and its value is plotted as a function of beta in Figure
\ref{optcon}.  The contributions are larger for $\beta \sim -1.2$ than
for $\beta \sim+0.8$.  In addition, the sign of the contribution is
opposite that of the continuum model contributions.  Hence the
preferred regime is $\beta \sim -1.2$.  These arguments are supported
by the numerical analysis described below.

   In summary, the lattice results indicate OPE convergence, and
ground state pole contributions are approximately independent of
$\beta$.  Considerations of the size of continuum model contributions
and the sign and magnitude of operators independent of the continuum
model leads to the preferred value of
\begin{equation}
\beta = -1.2 \pm 0.1 \, .
\end{equation}
A more precise determination of $\beta$ will depend on the details
of limits for continuum model and HDO contributions, condensate
values, and other parameters of the sum rules.

   At $\beta = -0.2$, where the leading term of the OPE is
stationary with respect to $\beta$ \cite{chung84,dosch89a}, the
continuum contributions are maximal.  The positive value and small
magnitude of the HDO contribution indicates the stability of the
leading terms of the OPE will not be realized as stability in the
ground state mass, coupling, nor in the continuum threshold.

   Another point of view argues that the optimal interpolating field
is one that eliminates direct instanton contributions from the
correlator \cite{novikov81}.  However, such arguments are based on the
assumption that direct instantons necessarily contribute to the
nucleon correlator in a significant manner.  Section \ref{instantons}
will demonstrate that direct instantons are not necessary to maintain
sum rule self consistency for a wide range of interpolating fields.
The $\beta$ dependence of the direct instanton contributions to
(\ref{nucl11at1}) is proportional to $13 \beta^2 + 10 \beta + 13 \ne
0$.  Minimization of instanton contributions would place $\beta =
-5/13$, very near $\beta = -0.6$ where the HDO term of
(\ref{nucl11at1}) vanishes and where continuum contributions are near
their peak.  Isolation of ground state contributions is nearly
impossible in this regime.  Therefore minimization of direct instanton
contributions does not lead to an optimal interpolating field.

   While it is important to establish the optimal mixing of
interpolating fields for QCD-SR analyses, one should not overlook the
fact that there is a range of values for $\beta$ where the sum rules
are expected to work.  Moreover, the ground state contribution to all
these sum rules is equivalent to the 1\% level \cite{leinweber95b}.
In other words, the right-hand side of (\ref{nucl11at1}) for a single
pole plus continuum model is independent of $\beta$.  After the first
sum rule is written down, additional sum rules may be introduced with
merely one new fit parameter (the continuum threshold) per sum rule.
Since direct instanton contributions to the sum rules are not
independent of $\beta$ \cite{forkel93}, one has an excellent
opportunity to see if direct instanton contributions really are
necessary to maintain sum-rule consistency.  This issue is addressed
in Section \ref{instantons} below.

\subsection{``Ioffe Formula''}

   Having discovered that $\chi_2$ has negligible overlap with the
nucleon ground state, it might be interesting to return to the ``Ioffe
Formula'' this time generalized for arbitrary $\beta$
\begin{equation}
M_N \simeq - \, \left [ 
             { 7 - 2\, \beta - 5\, \beta^2 \over
               5 + 2\, \beta + 5\, \beta^2 } \right ] \,
             { 4 \, \left ( 2 \pi \right )^2 \, 
             \langle \overline q q \rangle
             \over 
             M^2} \, .
\end{equation}
For $\beta = 0$, $M_N = 2.5$ GeV when $M = 1$ GeV, or $M_N = 1.4$ GeV
when $M = M_N$ GeV.  The explicit dependence on the choice of
interpolating field and the Borel mass limits the significance of the
``Ioffe formula'' to a qualitative role at best.

\subsection{Nucleon Sum Rule at 1 for \pmb{$\beta$}=\pmb{$-$}1.2}
\label{SubSecNucl11at1}

   Here we focus on the more reliable sum rule of (\ref{nucl11at1}) at
the optimal interpolator mixing of $\beta = -1.2$.  Fortunately, the
$\alpha_s$ corrections for the leading order operator of the OPE are
much smaller than for the identity operator at the order of 10\%.  The
$\beta$-dependent coefficient of the quark condensate in
(\ref{nucl11at1}) becomes \cite{hatsuda95}
\begin{equation}
7 \left ( 1 + {15 \over 14} {\alpha_s \over \pi} \right ) 
- 2 \, \beta \left ( 1 + {3 \over 2} {\alpha_s \over \pi} \right )
- 5 \, \beta^2 \left ( 1 + {7 \over 10} {\alpha_s \over \pi} \right )
\, .
\end{equation}
These corrections are small relative to uncertainties in the other
terms of the correlator and can be absorbed by a small shift in the
condensate itself.  Since our QCD parameter set is extracted from
analyses with neglected $\alpha_s$ corrections, we shall do the same
here. 

\begin{figure}[p]
\begin{center}
\epsfysize=11.7truecm
\leavevmode
\setbox\rotbox=\vbox{\epsfbox{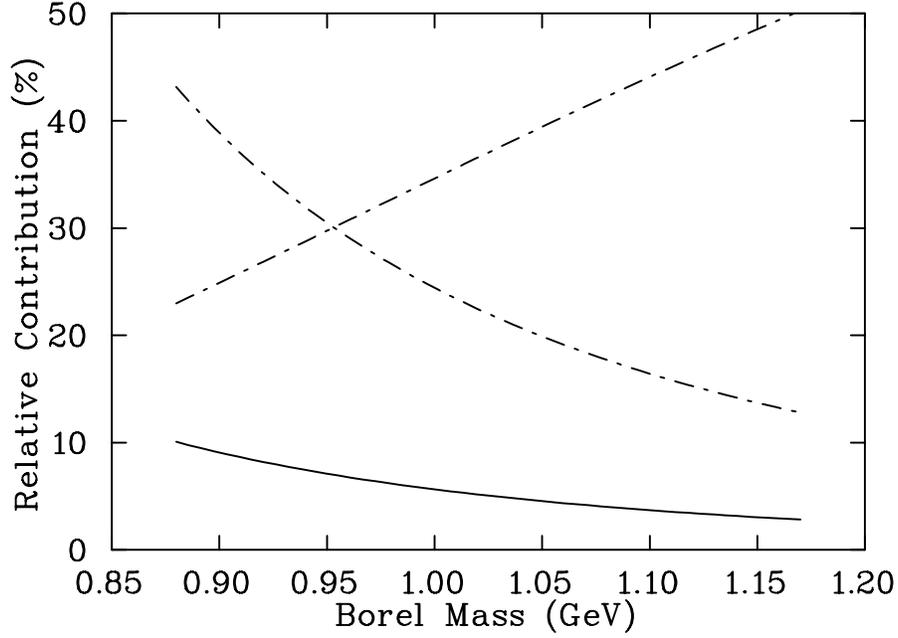}}\rotl\rotbox
\end{center}
\caption{The valid Borel regime for the sum rule of
(\protect\ref{nucl11at1}) at the optimal $\beta = -1.2$.  Relative HDO
and continuum model contributions are illustrated (dot-dash curves).
The HDO contributions are less than 10\% of the OPE contributions for
this fit when $\beta = 0$, as indicated by the solid curve.  }
\label{SR4hdo}
\end{figure}

   Armed with knowledge of the optimal interpolator, and a refined
understanding of OPE convergence tests, we now present the QCD-SR
predictions for the nucleon spectral properties from spin-1/2
interpolators.  Figure \ref{SR4hdo} displays the valid Borel regime
for this sum rule.  The HDO contributions are less than 10\% of the
OPE contributions for this fit when $\beta = 0$, as indicated by the
solid curve.  As described above, contributions from correlator
components known to be more convergent increase the HDO operator
contributions making resolution of the pole from the continuum more
probable.

\begin{figure}[p]
\begin{center}
\epsfysize=11.7truecm
\leavevmode
\setbox\rotbox=\vbox{\epsfbox{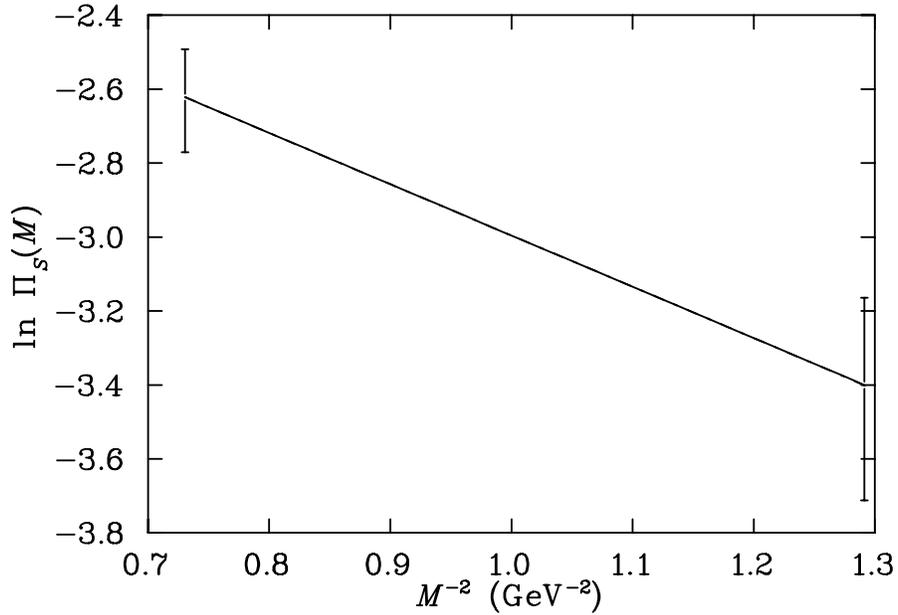}}\rotl\rotbox
\end{center}
\caption{ Three parameter fit of (\protect\ref{nucl11atP}) for the
residue $\lambda_{\cal O}$, the continuum model threshold $w_2$, and
the nucleon mass.  The near perfect linearity of the OPE$-$continuum
side (hidden dot-dash line) of $\Pi_S(M)$ supports the arguments
surrounding the refined determination of the valid Borel regime .}
\label{SR4fit}
\end{figure}

\begin{figure}[p]
\begin{center}
\epsfysize=11.7truecm
\leavevmode
\setbox\rotbox=\vbox{\epsfbox{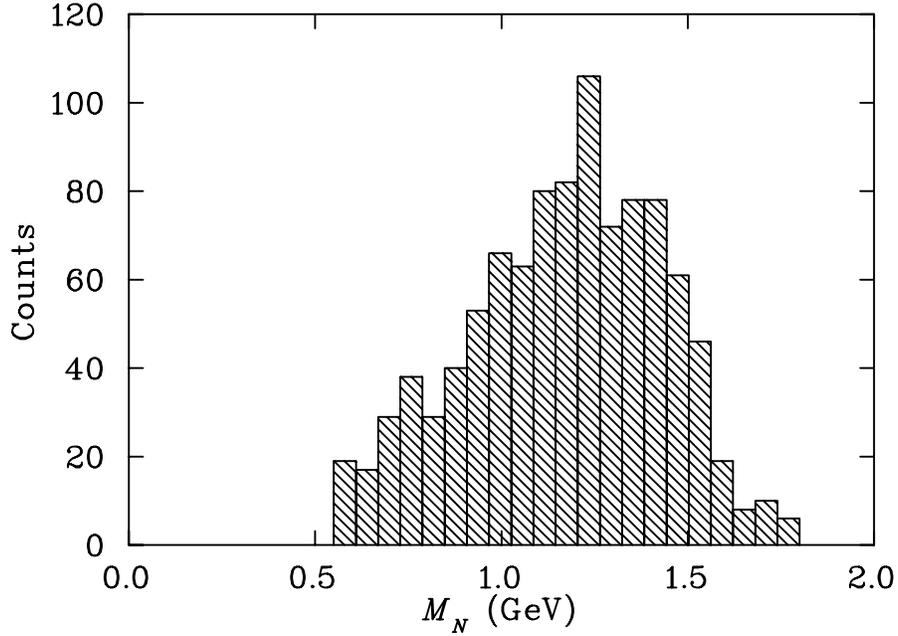}}\rotl\rotbox
\end{center}
\caption{Histogram for the nucleon mass obtained from fits of
(\protect\ref{nucl11at1}) at the optimal $\beta = -1.2$ for 1000 QCD
parameter sets.  }
\label{SR4binM}
\end{figure}

\begin{figure}[p]
\begin{center}
\epsfysize=11.7truecm
\leavevmode
\setbox\rotbox=\vbox{\epsfbox{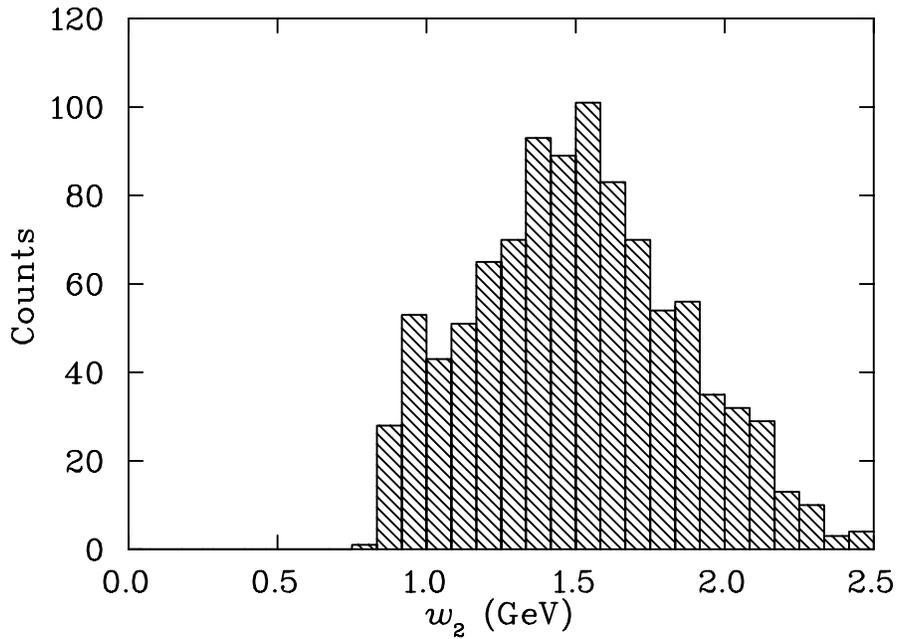}}\rotl\rotbox
\end{center}
\caption{Histogram for the continuum threshold obtained from fits of
(\protect\ref{nucl11at1}) at the optimal $\beta = -1.2$ for 1000 QCD
parameter sets.  }
\label{SR4binW}
\end{figure}

\begin{figure}[t]
\begin{center}
\epsfysize=11.7truecm
\leavevmode
\setbox\rotbox=\vbox{\epsfbox{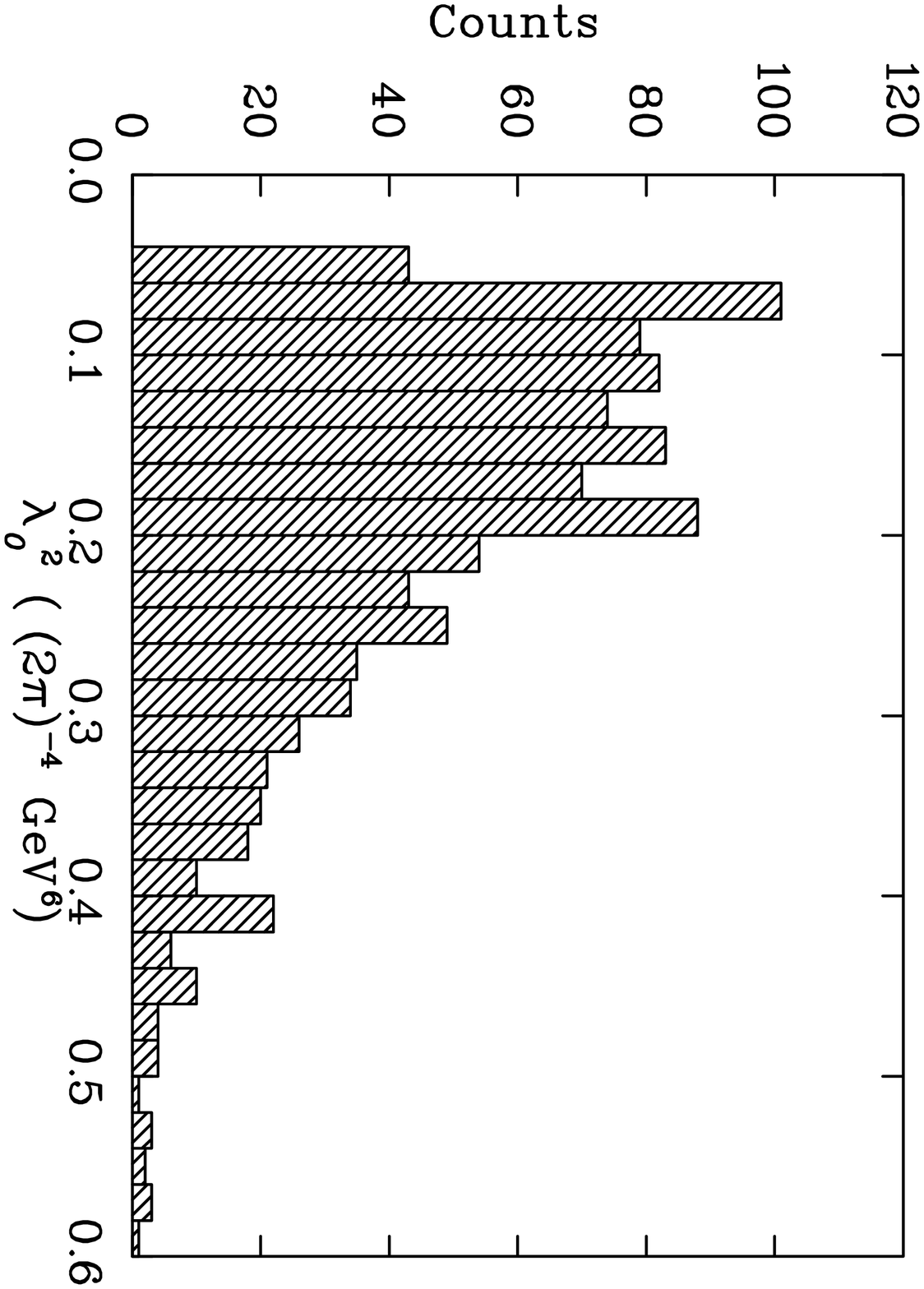}}\rotl\rotbox
\end{center}
\caption{Histogram for the residue of the nucleon pole obtained from
fits of (\protect\ref{nucl11at1}) at the optimal $\beta = -1.2$ for
1000 QCD parameter sets.  }
\label{SR4binLambda}
\end{figure}

   Figure \ref{SR4fit} displays the fit for this sum rule.  The near
perfect linearity of the OPE$-$continuum side of $\Pi_S(M)$ supports
the arguments presented above.  Despite 40\% contributions of the HDO
term to this sum rule, the OPE$-$continuum side displays the
anticipated behavior of reasonable OPE convergence.  The optimal fit
parameters obtained from consideration of 1000 QCD parameter sets are
\begin{equation}
\widetilde\lambda_{\cal O}^2 = 0.20 \pm 0.12\ {\rm GeV}^6 \, , \quad
w_2 = 1.53 \pm 0.41\ {\rm GeV} \, , \quad
M_N = 1.17 \pm 0.26\ {\rm GeV} \, .
\end{equation}
The uncertainty of the nucleon mass is more than double the
traditional expectation at 260 MeV.  Figures \ref{SR4binM},
\ref{SR4binW}, and \ref{SR4binLambda} display histograms for the
nucleon mass, threshold and residue respectively.  These histograms
will aid in the discussion of the following section.

   Table \ref{CompareSR} compares the prediction for $\lambda_{\cal
O}$ with other estimates of $\lambda_{\rm SR} = 2\, \lambda_{\cal O}$.
This residue corresponds to Ioffe's interpolator of (\ref{chiSR}) and
is expected to be independent of $|\beta|
\mathrel{\raise.3ex\hbox{$<$\kern-.75em\lower1ex\hbox{$\sim$}}} 1$ to
the order of 1\% \cite{leinweber95b}.

\begin{table}[b]
\caption{Comparison of predictions for $\lambda_{\rm SR} = 2\,
         \lambda_{\cal O}$ for
         various approaches to QCD.}
\label{CompareSR}
\setdec 0.0(0)
\begin{tabular}{llc}
Approach             &Reference                
  &$\lambda_{\rm SR}$    \\
                     &                         
  &($\times 10^{-2}$ GeV${}^3$) \\
\tableline
QCD Sum Rule         &This work
  &\dec 2.2$\pm 0.7$   \\
Lattice (mean-field improved) 
                     &Leinweber \cite{leinweber95b}
  &\dec 2.7$\pm 0.5$  \\
Lattice (conventional renormalization) 
                     &Gavela {\it et al.} \cite{gavela89}  
  &\dec 2.4     \\
Lattice (coordinate space)  
                     &Chu {\it et al.} \cite{chu93b}       
  &\dec 2.2$\pm 0.4$  \\
Instanton Liquid     &Schafer {\it et al.} \cite{schafer94a}
  &\dec 3.2$\pm 0.1$  \\
Baryon wave functions $(x^2 \to 0)$
                     &Brodsky {\it et al.} \cite{brodsky84}
  &\dec 12.     \\
Quark Model          &Thomas and McKellar \cite{thomas83}  
  &\dec  8.     \\
Bethe-Salpeter amplitude 
                     &Tomozawa  \cite{tomozawa81} 
                                                           
  &\dec 2.5     \\
Quark Model          &Milosevic {\it et al.} \cite{milosevic82} 
                                                           
  &\dec  2.     \\
MIT Bag Model        &Donoghue and Golowich \cite{donoghue82}  
                                                           
  &\dec 1.27    \\
\end{tabular}
\end{table}

   In the past the community has emphasized the importance of the
independence of the nucleon mass from the Borel mass parameter.  It is
common to find a plot of the nucleon mass extracted from the ratio
method discussed above as a function of the Borel mass.  While
independence of the spectral parameters from the Borel mass is
certainly desirable, the ratio method is potentially problematic in
practice.  We have already seen that the reliabilities and validities
of two sum rules are usually different and could even result in
separate valid regimes.  This feature cannot be revealed in the ratio
of the two sum rules.  In addition, one can always achieve flatness in
the ratio method in the large Borel mass region, where both sides of
the sum rules are dominated by the continuum model.  However, one
learns little about the lowest pole in this case.

   Contact with the plateau criterion may be made by rewriting the sum
rule of interest solving for the spectral parameter in question.  For
example, one can express the nucleon mass in the sum rule
of (\ref{nucl11at1}) as,
\begin{equation}
M_N = M \left[  \ln\, \left({\widetilde\lambda_{\cal O}^2 \over
\Pi_S(M)} 
\right)\right]^{1 / 2}\ ,
\label{MN-Borel}
\end{equation}
where $\Pi_S(M)$ denotes the left-hand side of (\ref{nucl11at1}).
Indeed a plot of $M_N$ versus the Borel parameter $M$ is very flat
within the valid region.  This information is already apparent in
figure \ref{SR4fit} where the near perfect linearity of the
OPE$-$continuum side of the sum rule indicates independence of the
fitted mass from the Borel parameter.  For a similar analysis of the
$\rho$-meson in finite density nuclear matter see \cite{leinweber95f}. 

\begin{figure}[t]
\begin{center}
\epsfysize=11.7truecm
\leavevmode
\setbox\rotbox=\vbox{\epsfbox{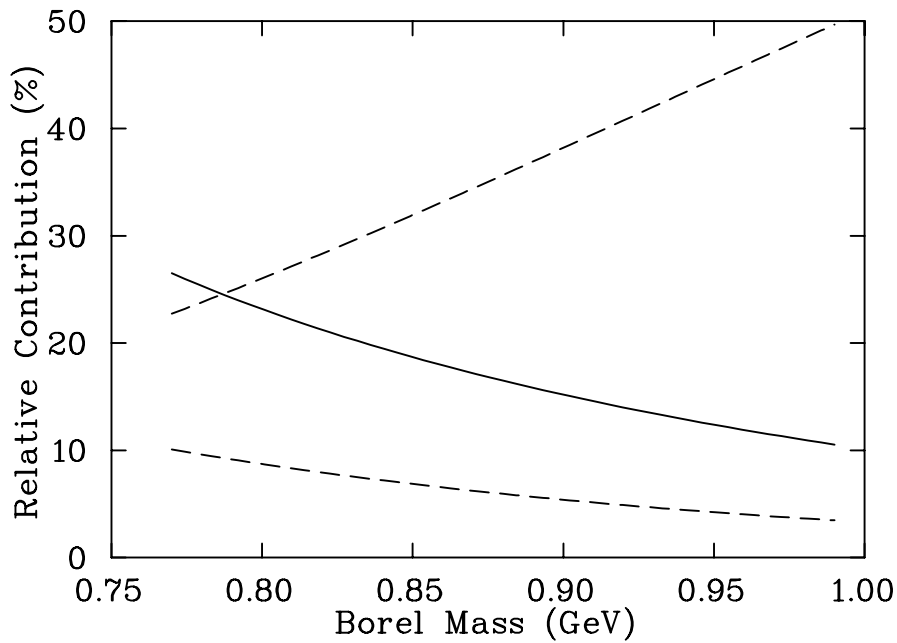}}\rotl\rotbox
\end{center}
\caption{ The naive Borel regime for (\protect\ref{nucl11atP}) at
$\beta = -1$.  The solid curve illustrates the HDO contributions to
the slowest converging component of the correlator $(\beta = 0)$.  The
failure of this curve to drop below 10\% before the continuum model
contributions exceed 50\% indicates the absence of a valid Borel
regime.}
\label{SR3K2alphasRef}
\end{figure}

   Returning to the sum rule at the structure $\gamma \cdot p$ of
(\ref{nucl11atP}), one finds the valid Borel regime disappears when
the refined OPE convergence criteria is applied.  The dashed curves of
figure \ref{SR3K2alphasRef} display the naive Borel regime for
(\ref{nucl11atP}) at $\beta = -1$ used in the fit of (\ref{SR3mod}).
The solid curve illustrates the HDO contributions to the slowest
converging component of the correlator $(\beta = 0)$.  This curve
fails to drop below 10\% before the continuum model contributions
exceed 50\%.  The HDO of the faster converging components reduces the
HDO contribution of the slowest converging component to give the
illusion of reasonable OPE convergence at $\beta = -1$.  This is most
likely the explanation for the failure of the sum rule of
(\ref{nucl11atP}); the absence of a valid Borel regime.

\subsection{Nucleon Sum Rule at 1 for \pmb{$\beta$}= +0.8}
\label{SecNucl11at1.8}

   Having considered the most favorable interpolator for the nucleon,
let us now examine the QCD-SR predictions for a less than favorable
interpolator.  At $\beta= +0.8$, figure \ref{optcon} indicates the
continuum model contributions can be reasonably small.  Hence it
should be possible to extract ground state information that is
reliable.  However, the HDO term is of the same sign and small
relative to the leading order contributions.  The difficulty here will
be resolving the ground-state pole from the approximate continuum
model.

   The lower limit of the valid Borel regime is determined by the OPE
contributions which are independent of the phenomenological fit
parameters and independent of $\beta$.  Hence this limit is known to
be 0.88 GeV.  The inherent instability of this sum rule makes it
difficult to determine the upper limit.  Figure \ref{optcon} suggests
that continuum contributions should be similar to that for $\beta =
-1.2$ and therefore we take an upper limit of 1.17 GeV, as in Figure
\ref{SR4hdo}.

\begin{figure}[p]
\begin{center}
\epsfysize=11.7truecm
\leavevmode
\setbox\rotbox=\vbox{\epsfbox{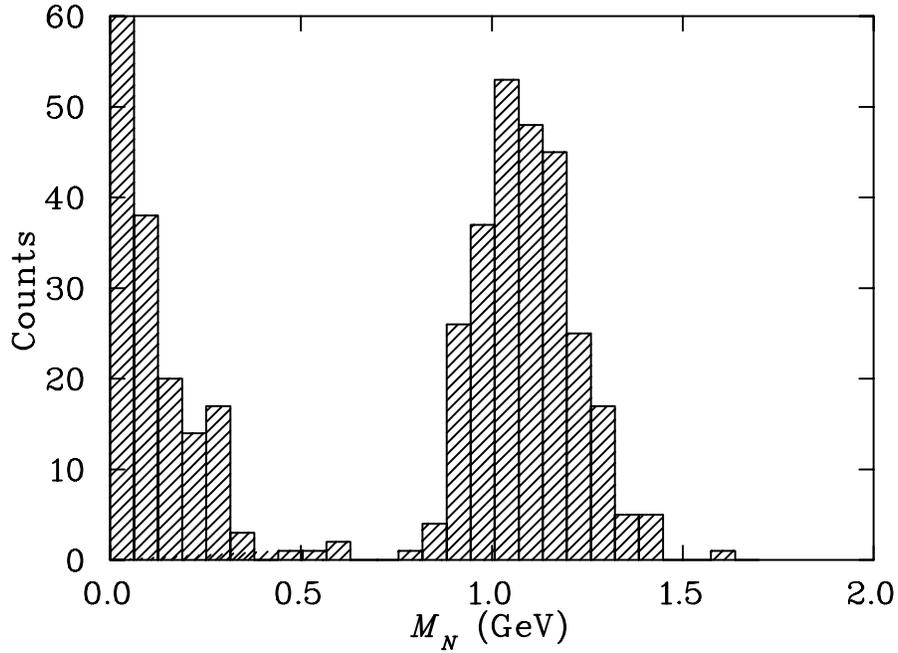}}\rotl\rotbox
\end{center}
\caption{Histogram for the nucleon mass obtained from fits of
(\protect\ref{nucl11at1}) at $\beta = +0.8$ for 1000 QCD parameter
sets.  When the pole is resolved from the continuum, the histogram
compares favorably with that of figure \protect\ref{SR4binM}. }
\label{SR4at0.8binM}
\end{figure}

\begin{figure}[p]
\begin{center}
\epsfysize=11.7truecm
\leavevmode
\setbox\rotbox=\vbox{\epsfbox{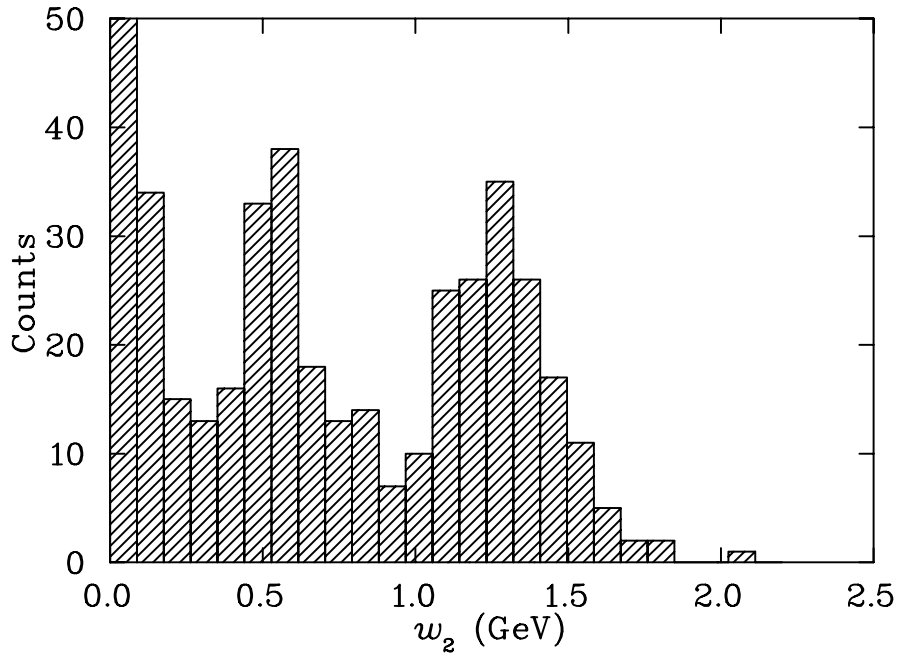}}\rotl\rotbox
\end{center}
\caption{Histogram for the continuum threshold obtained from fits of
(\protect\ref{nucl11at1}) at $\beta = +0.8$ for 1000 QCD parameter
sets.  The three types of sum rule fits are clearly resolved. }
\label{SR4at0.8binW}
\end{figure}

\begin{figure}[t]
\begin{center}
\epsfysize=11.7truecm
\leavevmode
\setbox\rotbox=\vbox{\epsfbox{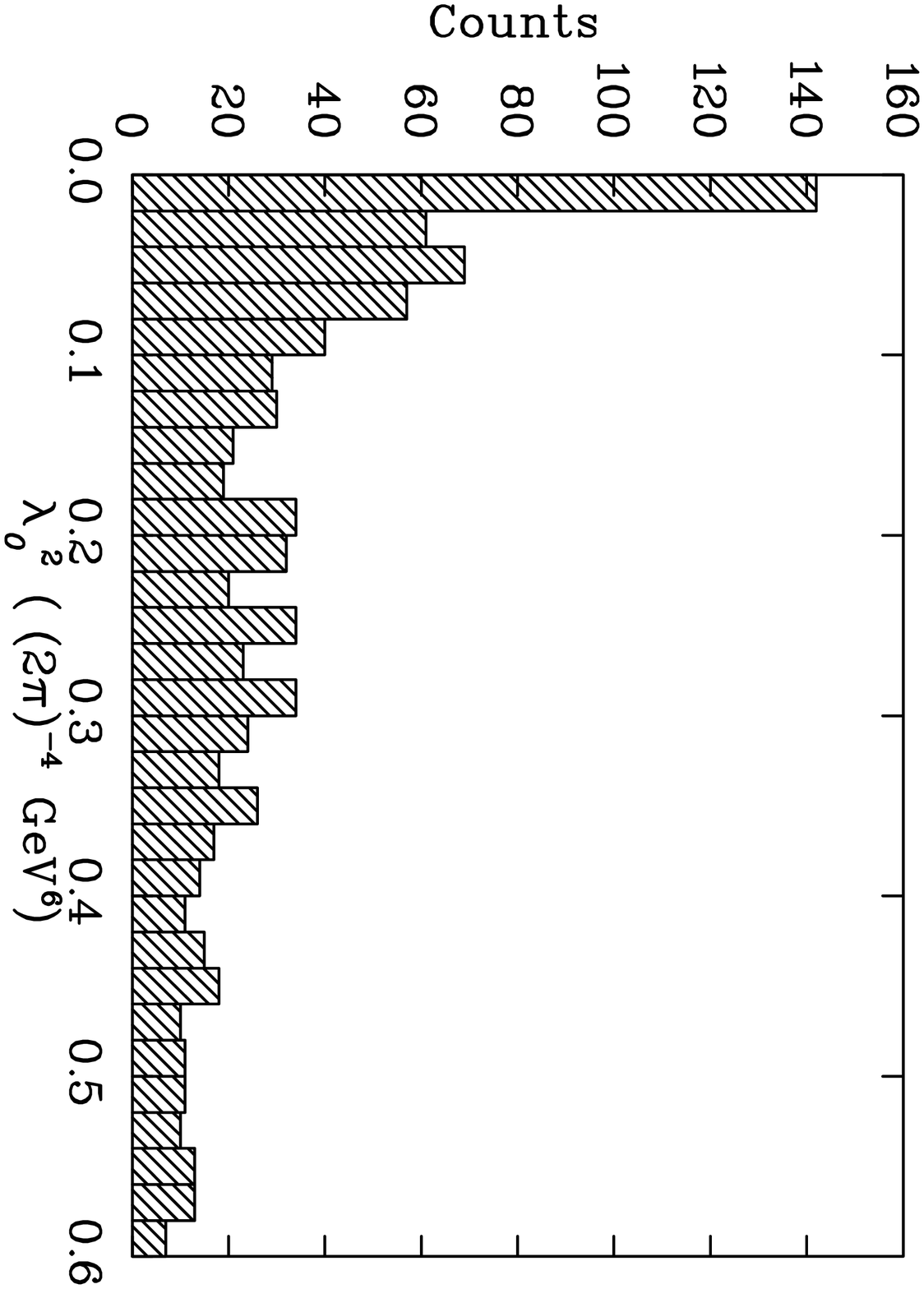}}\rotl\rotbox
\end{center}
\caption{Histogram for the residue of the nucleon pole obtained from
fits of (\protect\ref{nucl11at1}) at $\beta = +0.8$ for 1000 QCD
parameter sets.  Additional counts at small $\lambda_{\cal O}^2$
allows for additional strength from the continuum model.}
\label{SR4at0.8binLambda}
\end{figure}

   Figures \ref{SR4at0.8binM}, \ref{SR4at0.8binW}, and
\ref{SR4at0.8binLambda} display histograms for the nucleon mass,
threshold and residue obtained from (\ref{nucl11at1}) for $\beta =
+0.8$.  Figure \ref{SR4at0.8binM} resolves two regimes for the nucleon
mass and figure \ref{SR4at0.8binW} suggests the presence of at least
three regimes.

   In section \ref{correl} the nature of these three regimes are
resolved via a correlation analysis.  As mentioned earlier there are
two ways to fit a sum rule.  If the terms which do not contribute to
the continuum model are insufficient to resolve the pole, a good fit
may be obtained by letting the continuum threshold and the residue of
the pole go to zero.  One then has an approximate Laplace transform
relating the two sides of the sum rule.  These fits are represented by
the left-most shoulder of the histograms.

   An intermediate regime arises when the sum rules are capable of
resolving a threshold at which strength in the spectral density
begins.  However, a gap between ground and excited states is not
resolved.  The peak at 0.5 GeV in the continuum threshold is
representative of these fits.  For these fits the nucleon mass is
scattered about typical values but with with a small residue.  The
smaller residue compensates for strength from the continuum model.

   Finally, the third regime represented by the broad right-most peak
in the continuum threshold histogram corresponds to more standard fits
where a gap between the ground and first excited state has been
resolved.

   It is encouraging to see the similarity in the nucleon mass
distributions of figures \ref{SR4at0.8binM} and \ref{SR4binM} when the
sum rules resolve a pole plus continuum fit.  This agreement comes
about without the need for direct instanton contributions.

\section{NUCLEON SUM RULES:  SPIN-1/2 AND 3/2 INTERPOLATORS}
\label{nucleon13}

\subsection{Sum Rule Derivation}

   The spin-3/2 interpolating field for the nucleon is
\cite{chung82,belyaev83} 
\begin{equation}
\chi_{3/2}^\mu(x) =  \epsilon^{abc}
\biggl [
\left ( {u^a}^{\rm T}(x)\, C \sigma_{\rho \lambda}\, d^b (x) \right )
\, \sigma^{\rho \lambda} \gamma^\mu\, u^c (x) 
- \left ( {u^a}^{\rm T}(x)\, C \sigma_{\rho \lambda}\, u^b (x) \right )
\, \sigma^{\rho \lambda} \gamma^\mu \, d^c (x)
\biggr ] \, .
\label{chi3/2}
\end{equation}
Nucleon to vacuum matrix elements of $\chi_{3/2}^\mu$ are defined as
\begin{mathletters}
\begin{eqnarray}
\langle 0 | \chi_{3/2}^\mu | 1/2^+ \rangle &=&
\left ( \alpha \, p^\mu + \lambda_{3/2}\, \gamma^\mu \right ) \gamma_5
\, u(p) \, , \\
\langle 0 | \chi_{3/2}^\mu | 1/2^- \rangle &=&
\left ( \alpha \, p^\mu + \lambda_{3/2}\, \gamma^\mu \right ) 
u(p) \, .
\end{eqnarray}%
\label{3/2matrix}%
\end{mathletters}%
The appearance of $\gamma_5$ on the right-hand side matches the parity
of $\chi_{3/2}^\mu$.  The property $\gamma_\mu \, \chi_{3/2}^\mu = 0$
provides the relation
\begin{equation}
\alpha = { 4 \, \lambda_{3/2} \over \pm M_N} \, .
\end{equation}
In an analogous manner we work with the generalized spin-1/2
interpolator 
\begin{equation}
\chi_{\cal O}^\mu = \gamma^\mu \gamma_5 \, \chi_{\cal O} \, ,
\end{equation}
which has the matrix elements of (\ref{3/2matrix}) with $\alpha = 0$
and $\lambda_{3/2} \to \lambda_{\cal O}$.  Overlap of the spin-3/2
interpolating field with the generalized spin-1/2 interpolator of
(\ref{chiO}) gives rise to four Dirac-$\gamma$ structures
\begin{equation}
\langle 0 | \chi_{\cal O}^\mu | 1/2^\pm \rangle \langle 1/2^\pm |
\overline\chi_{3/2}^\nu | 0 \rangle =
\lambda_{\cal O} \lambda_{3/2} \left \{ - \gamma^\mu \gamma^\nu \left
( \gamma \cdot p \pm M_N \right ) \pm {4 \over M_N} \gamma^\mu p^\nu
\gamma \cdot p - 2 \gamma^\mu p^\nu \right \} \, ,
\end{equation}
from which three independent sum rules are obtained.  
\begin{mathletters}
\begin{eqnarray}
\left ( \gamma_\mu \gamma_\nu \, \gamma \cdot p \right ) : \quad
&&\kappa_N \, a^2 L^{20/27} 
- { 9 - \beta \over 24} \, { m_0^2 a^2 \over M^2} L^{2/9} \nonumber \\
&&\qquad = \widetilde\lambda_{\cal O} \widetilde\lambda_{3/2} \,
e^{-M^2_N/M^2} \, , 
\label{nucl13atZ} \\
&&  \nonumber \\
\left ( \gamma_\mu \gamma_\nu \right ) : \quad
&&{1 \over 2} a\, M^4\, L^{8/27} \left [ 1 - e^{-w_3^2/M^2}
\left ( {w_3^2 \over M^2} + 1 \right ) \right ]
+ { 1 + 3\, \beta \over 96}\, a \, b \, L^{8/27} \nonumber \\
&&\qquad =  \widetilde\lambda_{\cal O} \widetilde\lambda_{3/2} \,
M_N\, e^{-M^2_N/M^2} \, , 
\label{nucl13atGG} \\
&& \nonumber \\
\left ( \gamma_\mu p_\nu \, \gamma \cdot p \right ) : \quad
&&{1 \over 2} a\, M^2 \, L^{8/27} \left [ 1 - e^{-w_3^2/M^2} \right ]
- {3 - \beta \over 16}\, m_0^2 a\, L^{-2/9} \nonumber \\
&& \quad - {1 + 3\, \beta \over 96}\, {a\, b \over M^2} L^{8/27}
\nonumber \\ 
&&\qquad = {\widetilde\lambda_{\cal O} \widetilde\lambda_{3/2} \over
M_N} e^{-M^2_N/M^2} \, . 
\label{nucl13atGqq}%
\end{eqnarray}%
\label{nucl13}%
\end{mathletters}%
The corresponding Dirac-$\gamma$ structure from which the sum rule is
obtained is indicated on the left.  The sum rule at the structure
$\gamma_\mu p_\nu$ is identical to that for $\gamma_\mu \gamma_\nu
\, \gamma \cdot p$.

The leading term of the sum rule at the structure $\gamma_\mu
\gamma_\nu \, \gamma \cdot p$ is the four-quark condensate.  As such
the OPE of this sum rule is poorly known.  In addition, there are no
terms from which a continuum model may be constructed.  For an
analysis and discussion of this sum rule under the assumption of
factorization see \cite{leinweber90}.

The sum rule at the structure $\gamma_\mu q_\nu \, \gamma \cdot p$ is
a very favorable sum rule.  The dimensionality of the structure at
which the sum rule is extracted allows the HDO terms of the sum rule
to approach the factorial suppression regime of the Borel transform
thus improving the truncated OPE convergence.  Moreover, there are two
terms which are not used in the continuum model.  Hence there should
be better resolution of the pole from the continuum.

The mixed condensate is absent in (\ref{nucl13atGG}) for all $\beta$.
With only two pieces of information on the OPE side of the sum rule,
this sum rule is of little use on it's own.  Fortunately, both
(\ref{nucl13atGG}) and (\ref{nucl13atGqq}) are extracted from Dirac
$\gamma$-structures in which even- and odd-parity excitations
contribute with opposite signs.  Hence it is reasonable to set the
continuum model thresholds equal for these two sum rules and this
provides greater stability in the fit parameters.  

Since the terms of (\ref{nucl13}) used in the continuum model are
independent of the interpolator mixing parameter $\beta$, the obvious
selection for the optimal mixing is $\beta = 0$.  Of course, if the
interpolator $\chi_2$ of (\ref{chiN2}) doesn't contribute to the
continuum model, nor to the ground state, it will be important to
demonstrate that different values of $\beta$ lead to similar results.
The expectation is that the continuum threshold will be dependent on
$\beta$ while the mass and residue remain unchanged for $|\beta|
\mathrel{\raise.3ex\hbox{$<$\kern-.75em\lower1ex\hbox{$\sim$}}} 1$.
This dependence is examined in Section \ref{instantons} examining the
necessity of direct instanton contributions.  

   Ioffe's selection of $\beta = -1$ as the optimal spin-1/2
interpolator is reasonably close to the determination of $\beta =
-1.2$ in Section \ref{OptMixSect} for spin-1/2 to spin-1/2
interpolators.  However, the optimal mixing is specific to each
individual sum rule.  Here we have found an optimal value quite
different from Ioffe's preference and, as we shall see, Ioffe's
selection of $\beta \sim -1$ fails to provide a sum rule with a valid
regime.

\subsection{Sum Rule Analysis}

\begin{figure}[p]
\begin{center}
\epsfysize=11.7truecm
\leavevmode
\setbox\rotbox=\vbox{\epsfbox{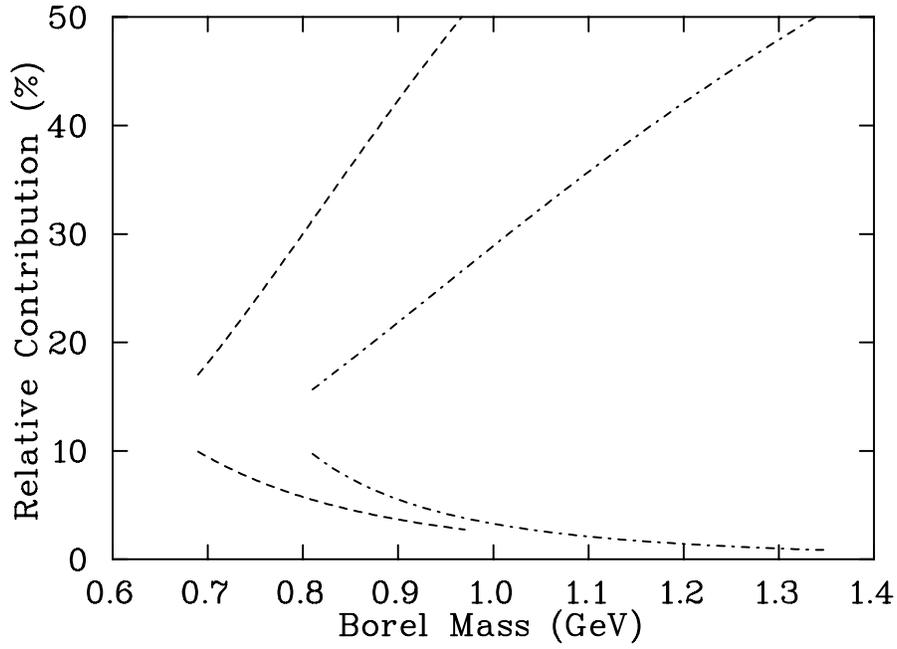}}\rotl\rotbox
\end{center}
\caption{The valid Borel windows for the nucleon sum rules of
(\protect\ref{nucl13atGG}) (dashed) and (\protect\ref{nucl13atGqq})
(short dash-dot) at the optimal $\beta = 0$.  The relative HDO
contributions limited to $10\%$ and the continuum model contributions
limited to 50\% are illustrated. }
\label{SR467hdo}
\end{figure}

   We begin by considering a simultaneous analysis of
(\ref{nucl11at1}) at $\beta = -1.2$, and (\ref{nucl13atGG}) and
(\ref{nucl13atGqq}) at $\beta = 0$.  A valid Borel regime could not be
found for (\ref{nucl11at1}) when analyzed in conjunction with the two
additional sum rules of (\ref{nucl13atGG}) and (\ref{nucl13atGqq}).
Hence we will analyze (\ref{nucl13atGG}) and (\ref{nucl13atGqq})
alone.  Figure \ref{SR467hdo} displays the valid Borel regimes for
these two sum rules.

\begin{figure}[p]
\begin{center}
\epsfysize=11.7truecm
\leavevmode
\setbox\rotbox=\vbox{\epsfbox{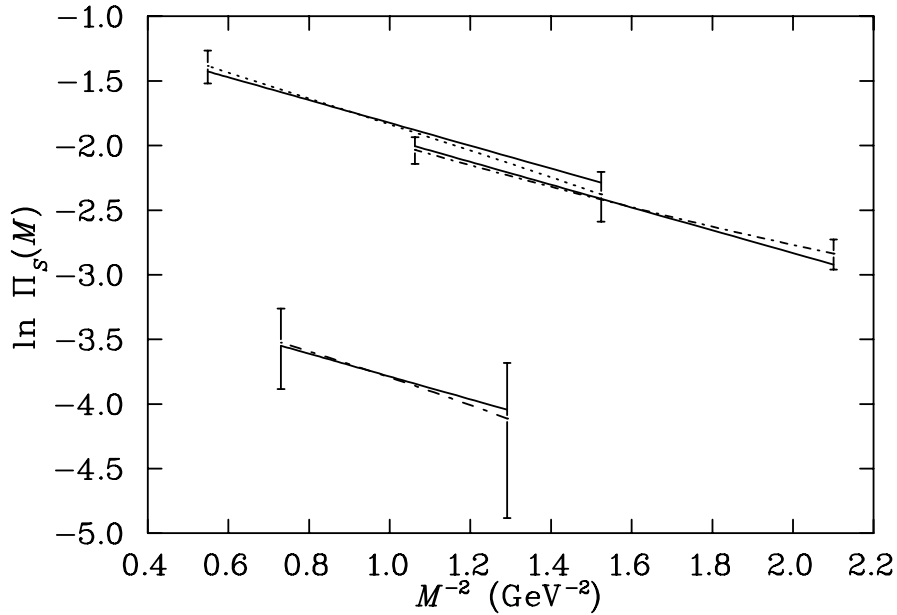}}\rotl\rotbox
\end{center}
\caption{A three parameter fit including $\lambda_{3/2}$, $M_N$ and
$w_3$ of the nucleon sum rules of (\protect\ref{nucl13atGG}) (dashed)
and (\protect\ref{nucl13atGqq}) (short dash-dot) at $\beta = 0$.  The
sum rule of (\protect\ref{nucl11at1}) at $\beta = -1.2$ (long
dash-dot) is also illustrated to demonstrate the consistency of all
three sum rules.  The solid lines illustrate the ground state
contributions. }
\label{SR467fit}
\end{figure}

\begin{figure}[t]
\begin{center}
\epsfysize=11.7truecm
\leavevmode
\setbox\rotbox=\vbox{\epsfbox{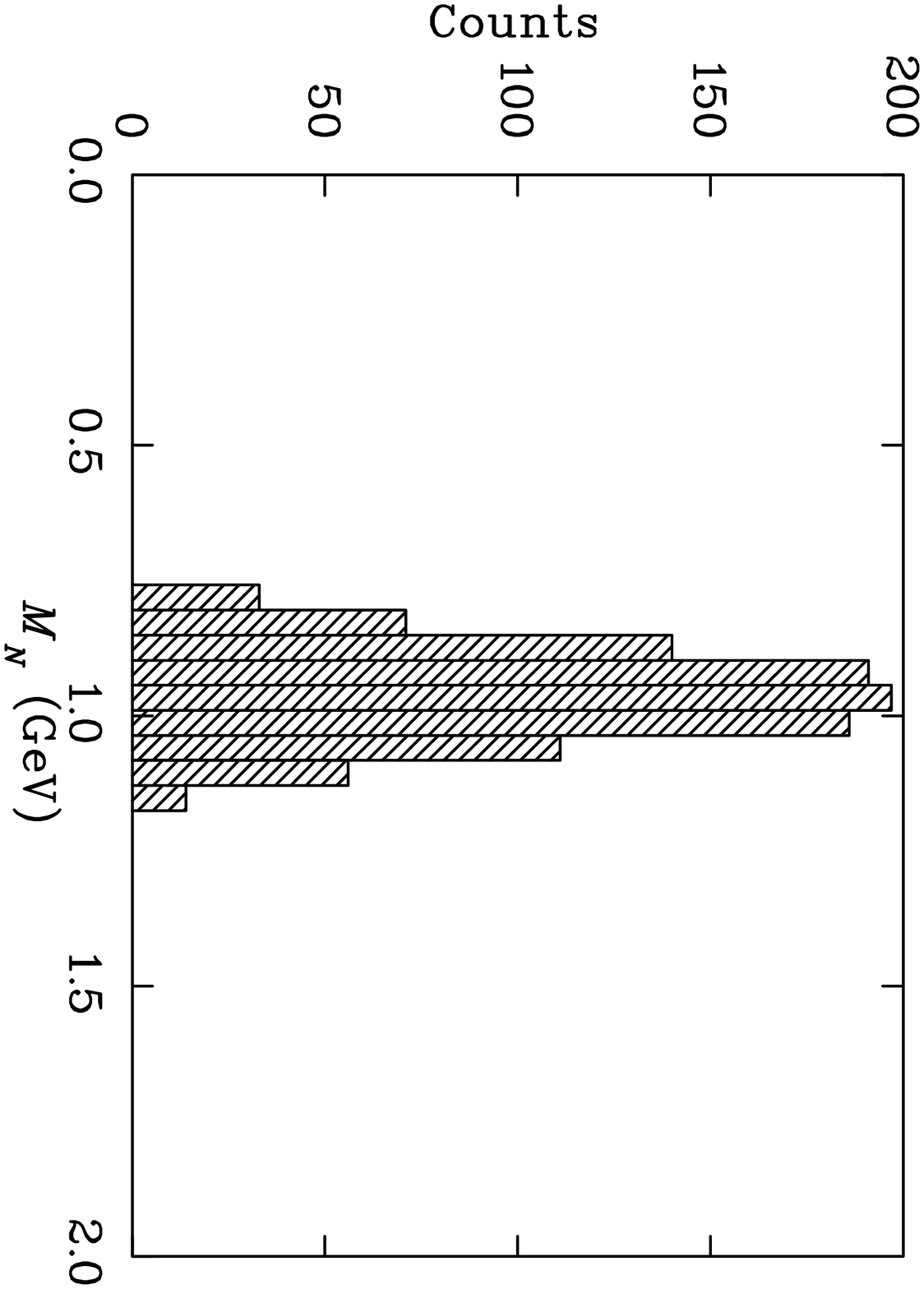}}\rotl\rotbox
\end{center}
\caption{Histogram for the nucleon mass obtained from fits of
(\protect\ref{nucl13atGG}) and (\protect\ref{nucl13atGqq}) at the
optimal $\beta = 0$ for 1000 QCD parameter sets. }
\label{SR67binM}
\end{figure}

   The corresponding three parameter fit of the residue
$\widetilde\lambda_{\cal O} \widetilde\lambda_{3/2}$, threshold,
$w_3$, and nucleon mass is indicated in figure \ref{SR467fit}.  The
sum rule of (\ref{nucl11at1}) is also displayed for the Borel regime
considered in Sec.\ \ref{SubSecNucl11at1}.  The linearity of the
OPE$-$continuum sides of the subtracted sum rules over a relatively
broad Borel regime indicates the OPE is under reasonable control.
However, since the continuum model dominates the phenomenology of
(\ref{nucl11at1}), we do not consider it further. The resultant fit
parameters obtained from the consideration of 1000 QCD parameters sets
are
\begin{mathletters}
\begin{equation}
M_N = 0.96 \pm 0.08\ {\rm GeV} \, ,
\end{equation}
\begin{equation}
\widetilde\lambda_{\cal O} \, \widetilde\lambda_{3/2} 
                             = 0.41 \pm 0.14\ {\rm GeV}^6 \, , \quad
w_3 = 1.3 \pm 0.2\ {\rm GeV} \, . 
\end{equation}%
\end{mathletters}%
The uncertainty of 100 MeV for the nucleon mass is a significant
improvement over the traditional sum rule of (\ref{nucl11at1}) alone
as illustrated in figure \ref{SR67binM}.  The value of the threshold
lies roughly at the value suggested by the particle data tables of
approximately 1.27 GeV; the Roper mass of 1.440 GeV less the half
width of 0.175 GeV.  The value for $\lambda_{3/2}$ is a prediction
that awaits vindication from alternative approaches.

The presence of a valid Borel regime common to both sum rules of
(\ref{nucl13atGG}) and (\ref{nucl13atGqq}) provides an opportunity to
apply the Monte-Carlo uncertainty analysis to the ratio method.  The
ratio of equations (\ref{nucl13atGG})/(\ref{nucl13atGqq}) provides a
ground state contribution equal to $M_N^2$ and independent of the
residues $\lambda_{\cal O} \lambda_{3/2}$.  The optimal fit parameters
are
\begin{equation}
M_N = 1.05 \pm 0.08\ {\rm GeV} \, , \quad
w_3 = 1.7 \pm 0.5\ {\rm GeV} \, ,
\end{equation}
and agree with the previous results.  The difference in the central
values is due to the different Borel regimes used in the fit.  Despite
taking a ratio of sum rules, the uncertainty in the nucleon mass
remains nearly unchanged.  The fit obtained from the full covariance
matrix $\chi^2$ of (\ref{corrchisq}) provides similar results and
yields $\chi^2/N_{DF} = 1.34$.  This acceptable value suggests the
distributions selected for the QCD parameters are in reasonable accord
with the approximations inherent in the QCD-SR approach.  Future
studies might use the $\chi^2/N_{DF}$ to place restrictions on QCD
parameter sets selected in the Monte-Carlo approach.

Table \ref{ChiSquare} summarizes an investigation of $\chi^2/N_{\rm
DF}$ and uncertainties obtained from the analysis of
(\ref{nucl13atGG}) and (\ref{nucl13atGqq}) at $\beta = 0$ for various
relative errors of condensate input parameters.  Results are based on
samples of 200 condensate parameter sets.  The ratio results are
obtained from the ratio of these sum rules analyzed in the Borel
regime common to both sum rules.  A comparison of the uncertainties
for the nucleon mass obtained from the ratio method and the standard
analysis advocated here indicates there is very little if anything to
be gained in the ratio approach to sum rule analysis.

\begin{table}[t]
\caption{An investigation of $\chi^2/N_{\rm DF}$ and uncertainties
obtained from the analysis of (\protect\ref{nucl13atGG}) and
(\protect\ref{nucl13atGqq}) at $\beta = 0$ for various relative errors
of condensate input parameters.  Results are based on samples of 200
condensate parameter sets.  Ratio results are obtained from the ratio
of these sum rules analyzed in the Borel regime common to both sum
rules.  The last row summarizes the results for the condensate
parameter set estimated in Sec. \protect\ref{uncert}. }
\label{ChiSquare}
\setdec 0.00(00)
\begin{tabular}{ccccccc}
           &\multicolumn{3}{c}{Sum Rule Ratio} 
           &\multicolumn{3}{c}{Standard Analysis} \\
Condensate &$\chi^2/N_{\rm DF}$ &$M_N$ &$w_3$ &$M_N$ &$w_3$ 
           &$\widetilde\lambda_{\cal O} \widetilde\lambda_{3/2}$ \\
Uncertainty&                    &(GeV) &(GeV) &(GeV) &(GeV) &(GeV${}^6$) \\
\tableline
20\%       &0.11  &1.04(12) &1.78(91) &0.96(11) &1.29(26) &0.40(15) \\
15\%       &0.26  &1.04(10) &1.73(75) &0.96(10) &1.29(22) &0.40(12) \\
13\%       &0.69  &1.04(9)  &1.72(68) &0.95(9)  &1.28(19) &0.40(10) \\
12\%       &1.13  &1.04(8)  &1.70(65) &0.95(8)  &1.27(17) &0.39(9)  \\
10\%       &2.22  &1.04(6)  &1.67(47) &0.95(7)  &1.26(14) &0.39(8)  \\
 5\%       &18.74 &1.02(3)  &1.47(6)  &0.95(3)  &1.26(7)  &0.38(4)  \\
\tableline
Estimated  &1.34  &1.05(8)  &1.72(52) &0.96(8)  &1.28(20) &0.41(14) \\
\end{tabular}
\end{table}

Table \ref{ChiSquare} indicates a $\chi^2/N_{\rm DF} \simeq 1$ is
obtained when the QCD input parameters are specified with 12\%
uncertainties.  Hence, the best one can determine phenomenological
parameters with the current implementation of QCD-SRs is 10\% accuracy
for the nucleon mass and 25\% accuracy for the residue.  This large
uncertainty for the residue does not bode well for the analysis of
three-point functions where the quantity of interest appears
multiplied by the reside.  The two-point function is used to normalize
the three-point function, and it appears that this normalization is
not well determined.  Reduction of this uncertainty to the 10\% level
requires 5\% uncertainties on the QCD input parameters.  However, the
$\chi^2/N_{\rm DF} = 18.7$ indicates this level of accuracy is beyond
the current state of the art.  Of course, this is not surprising since
approximations at the level of 10\% have been made in deriving the sum
rules.

\section{CORRELATIONS}
\label{correl}

\subsection{\pmb{$\chi^2$} Measures of Association}

   To quantitatively examine the correlations among the QCD and
phenomenological parameters, we utilize a contingency table analysis
of two distributions \cite{numrec86} to determine the significance,
and strength of correlations.  For the correlations found to be
significant, the linear correlation coefficient is calculated.

   Determinations of the significance, probability and strength of
correlations can be based on $\chi^2$ measures of association.
Null-hypothesis probabilities, Cramer's V, and the Contingency
Coefficient $C$ are a few of the better known measures of association
which may be obtained from the contingency table analysis described
below.

   A contingency table analysis of association between two variables
proceeds by binning the two-dimensional distribution into an $I \times
J$ grid \cite{numrec86}.  If $N_{ij}$ is the number of points in bin
$i,j$, then the row and column totals are
\begin{equation}
N_{i \cdot} = \sum_j^J N_{ij} \, , \qquad N_{\cdot j} = \sum_i^I N_{ij} \, ,
\end{equation}
and the total number of points in the distribution is
\begin{equation}
N = \sum_{i,j} N_{ij} = \sum_i^I N_{i \cdot} = \sum_j^J N_{\cdot j} 
\, . 
\end{equation}
The null hypothesis is the assumption of no association between the
two variables.  By measuring the fit of this assumption to the data we
can learn about the significance of the association.  Under the null
hypothesis, the fraction of points in bin $i,j$, of column $j$ is
independent of $j$.  That is,
\begin{equation}
{n_{ij} \over N_{\cdot j} } =  {N_{i \cdot} \over N} \, , \qquad
n_{ij} = {N_{i \cdot} \, N_{\cdot j} \over N} \, ,
\end{equation}
where $n_{ij}$ is the number of points in bin $i,j$ predicted by the
null hypothesis.  The $\chi^2$ is
\begin{equation}
\chi^2 = \sum_{i,j} { \left ( N_{ij} - n_{ij} \right )^2 \over
                      n_{ij} } \, ,
\end{equation}
and the number of  degrees of freedom are
\begin{equation}
\nu = I\, J - I - J + 1 \, ,
\end{equation}
where the row and column totals used in the hypothesis are subtracted
and the addition of one accounts for the equality of the sum of row
and column totals.

   Knowledge of $\chi^2/\nu$ allows an estimate of the probability for
the null hypothesis to be true.  A small probability indicates
significant association between the data.  We calculate the standard
null-hypothesis significance parameter
\begin{equation}
P_{\rm NH} = {1 \over \Gamma(\nu/2)} \,
  \int_{\chi^2/2}^\infty e^{-t} \, t^{\nu/2 - 1} \, dt \, ,
\end{equation}
where $\Gamma$ is the gamma function.

Cramer's $V$ is a normalized measure of the strength of association
independent of the size of the contingency table
\begin{equation}
V = \left ( {\chi^2 \over N\, \min(I-1,J-1) } \right )^{1/2} \, .
\end{equation}
This measure lies in the range
\begin{equation}
0 \le V \le 1 \, .
\end{equation}
$V=1$ indicates there is a unique square for each $i,j$ holding all
points in row $i$ and column $j$.  $V=0$ indicates there is no
association.

   The Contingency Coefficient $C$ is also a measure of the strength
of the correlation.  It is defined as
\begin{equation}
C = \sqrt{ { \chi^2 \over \chi^2 + N } } \, ,
\end{equation}
and is dependent on the size of the contingency table.  Since all
the contingency tables used in this analysis are the same size, this
poses no problem.  Its range is
\begin{equation}
0 \le C < 1 \, ,
\end{equation}
with larger values indicating stronger association.

   We also consider the significance of association based on the
symmetrical uncertainty coefficient $U$ determined via an examination
of the entropy of the distribution \cite{numrec86}.  In essence this
measure describes how much entropy is lost in one parameter given
knowledge of the other.  This measure is bounded by
\begin{equation}
0 \le U \le 1 \, ,
\end{equation}
and large values of $U$ indicate significant loss of entropy implying
significant dependency among the parameters.

   Once a distribution has been found to be significant, the linear
correlation coefficient is calculated
\begin{equation}
r = { \sum_i ( x_i - \overline x ) ( y_i - \overline y ) \over
\left [ \sum_i ( x_i - \overline x )^2 \right ]^{1/2}
\left [ \sum_i ( y_i - \overline y )^2 \right ]^{1/2} } \, .
\end{equation}
The linear correlation coefficient also measures the strength of the
association and lies in the range
\begin{equation}
-1 \le r \le 1 \, .
\end{equation}
When $r = +1$ the points lie on a perfect line with positive slope,
while $r = -1$ indicates the points lie on a perfect line with
negative slope.  $r = 0$ indicates there is no correlation.  While the
linear correlation coefficient is a poor measure of the significance
of correlation, it has the advantage of identifying the sign of
significant correlations.

   Hence the procedure is as follows.  The goodness of fit parameter
is utilized to identify associations that are significant.  This
parameter has the advantage of presenting a clean determination of
which associations are significant.  Cramer's $V$, the contingency
coefficient $C$, and the symmetrical uncertainty coefficient are used
to determine which QCD parameters play the dominant role in
determining hadronic spectral properties.  Finally, the
proportionality of the association is measured by the linear
correlation coefficient.  Since the uncertainties of the QCD parameter
distributions can affect the strengths of the associations, we utilize
1000 QCD parameter sets in which all QCD parameters have an equal
relative uncertainty of 15\%.  However, the illustrations of this
section show the actual correlations for the QCD parameters described
in Section \ref{uncert}.

\subsection{\pmb{$\rho$}-meson Associations}

   Table \ref{RhoCorr} identifies the QCD parameters having
significant correlations with the spectral properties of the
$\rho$-meson.  Here the quark condensate plays a dominant role in
determining all the spectral properties.  This is simply a reflection
of the fact that it appears squared in the last term of (\ref{rhoSR}).
The correlations for $\kappa_\rho$ show a linearized version of
essentially the same correlation.  The $\alpha_s$ correlations are
similar to that for $\kappa_\rho$ and reflect the fact that the
$\alpha_s$ corrections to the identity operator are small relative to
one.  An interesting aspect is the absence of significant gluon
condensate correlations for $f_\rho^2$ and $w_\rho$.

\begin{table}[t]
\caption{QCD Parameters having significant roles in determining
$\rho$-meson spectral properties.  Measures of association include
Cramer's $V$, the Contingency Coefficient $C$, the Symmetrical
Uncertainty Coefficient $U$, and the linear correlation coefficient
$r$. }
\label{RhoCorr}
\setdec 0.00
\begin{tabular}{lccccc}
Spectral Property  &Parameter &$V$ &$C$ &$U$ &$r$ \\    
\tableline
$\rho$-meson Mass $M_\rho$
  &$\langle \overline q q \rangle$ 
  &\dec 0.34 &\dec 0.82 &\dec 0.23 &\dec $+$0.78 \\
  &$\langle {\alpha_s \over \pi} \, G^2 \rangle$
  &\dec 0.20 &\dec 0.63 &\dec 0.07 &\dec $-$0.27 \\
  &$\kappa_\rho$
  &\dec 0.19 &\dec 0.62 &\dec 0.08 &\dec $+$0.37 \\
  &$\alpha_s/\pi$
  &\dec 0.18 &\dec 0.61 &\dec 0.07 &\dec $+$0.38 \\
\tableline
Threshold $w_\rho$ of (\protect\ref{rhoSR})
  &$\langle \overline q q \rangle$ 
  &\dec 0.36 &\dec 0.83 &\dec 0.24 &\dec $+$0.80 \\
  &$\alpha_s/\pi$
  &\dec 0.21 &\dec 0.66 &\dec 0.09 &\dec $+$0.42 \\
  &$\kappa_\rho$
  &\dec 0.19 &\dec 0.62 &\dec 0.08 &\dec $+$0.37 \\
\tableline
Pole Residue $f_\rho^2$ of (\protect\ref{rhoSR})
  &$\langle \overline q q \rangle$ 
  &\dec 0.34 &\dec 0.82 &\dec 0.23 &\dec $+$0.78 \\
  &$\alpha_s/\pi$
  &\dec 0.21 &\dec 0.66 &\dec 0.09 &\dec $+$0.46 \\
  &$\kappa_\rho$
  &\dec 0.19 &\dec 0.62 &\dec 0.08 &\dec $+$0.36 \\
\end{tabular}
\end{table}

\begin{figure}[p]
\begin{center}
\epsfysize=11.7truecm
\leavevmode
\setbox\rotbox=\vbox{\epsfbox{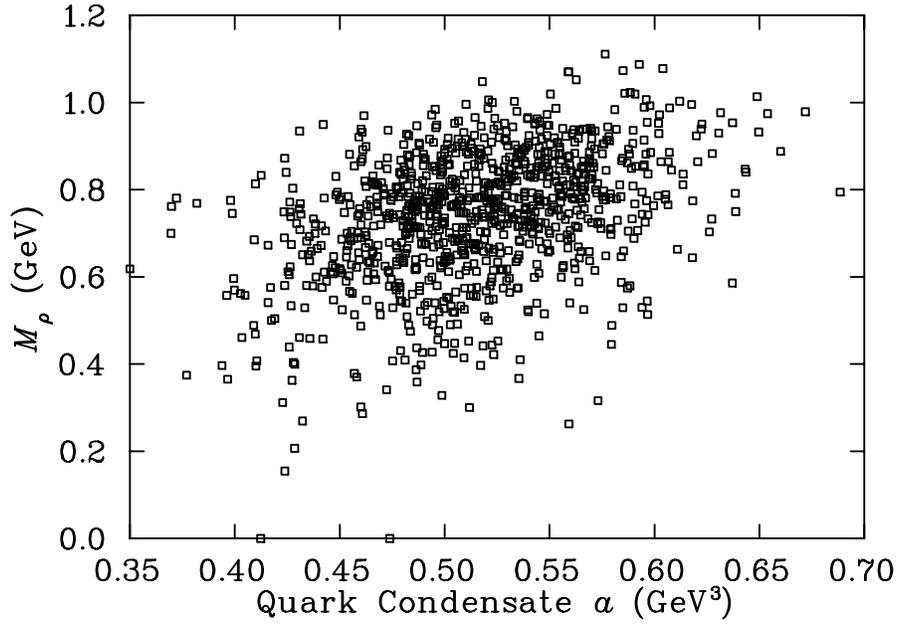}}\rotl\rotbox
\end{center}
\caption{ Scatter plot illustrating the association between the
$\rho$-meson mass and the quark condensate $a$ estimated in Section
\protect\ref{uncert}.  }
\label{rhoMa}
\end{figure}

\begin{figure}[p]
\begin{center}
\epsfysize=11.7truecm
\leavevmode
\setbox\rotbox=\vbox{\epsfbox{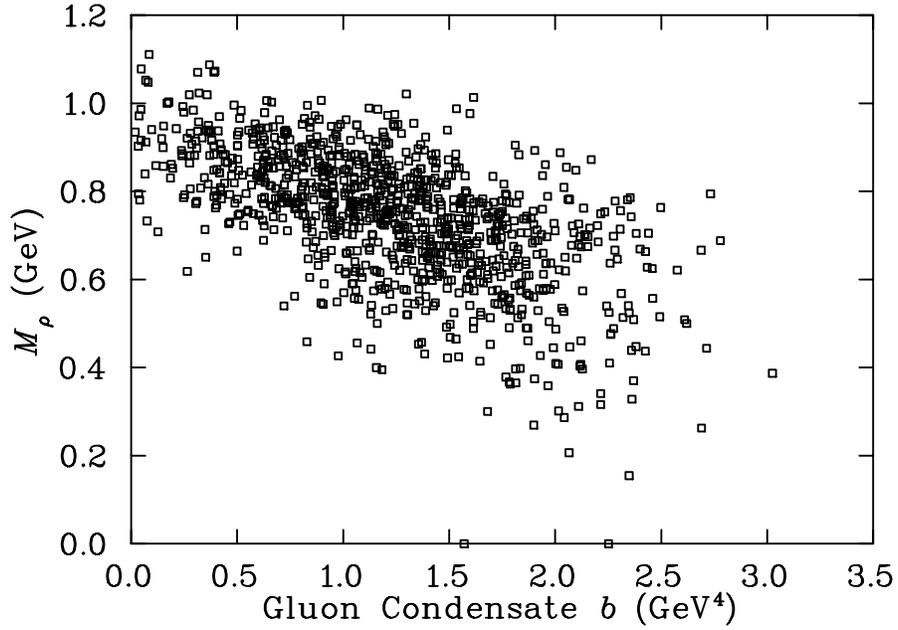}}\rotl\rotbox
\end{center}
\caption{ Scatter plot illustrating the association between the
$\rho$-meson mass and the gluon condensate $b$ estimated in Section
\protect\ref{uncert}.  }
\label{rhoMb}
\end{figure}

   The correlations between the $\rho$-meson mass and two of the more
interesting QCD parameters estimated in Section \ref{uncert} are
illustrated in figures \ref{rhoMa} and \ref{rhoMb}.  The relatively
large uncertainty in the gluon condensate estimate reveals an
anti-correlation with the mass.  Qualitatively similar plots hold for
the residue and continuum threshold.

\subsection{Nucleon Associations}

   Table \ref{nucl11at1Corr} identifies the QCD parameters having
significant correlations with the spectral properties of the nucleon
obtained from an analysis of (\ref{nucl11at1}) at $\beta = -1.2$.
Here the gluon condensate plays the dominant role in determining the
mass and continuum threshold.  The mixed condensate and finally the
quark condensate play lesser roles.  The predominant role of the quark
condensate is to determine the residue of the nucleon pole.
Fortunately, the quark condensate and nucleon mass are at least
positively correlated.

\begin{table}[b]
\caption{QCD Parameters having significantly correlated roles in
determining nucleon spectral properties via an analysis of
(\protect\ref{nucl11at1}) at $\beta = -1.2$.  Measures of association
include Cramer's $V$, the Contingency Coefficient $C$, the Symmetrical
Uncertainty Coefficient $U$, and the linear correlation coefficient
$r$.}
\label{nucl11at1Corr}
\setdec 0.00
\begin{tabular}{lccccc}
Spectral Property  &Parameter &$V$ &$C$ &$U$ &$r$ \\    
\tableline
Nucleon Mass $M_N$
  &$\langle {\alpha_s \over \pi} \, G^2 \rangle$
  &\dec 0.30 &\dec 0.77 &\dec 0.20 &\dec $+$0.76 \\
  &$\langle \overline q \, g \, \sigma \cdot G \, q \rangle$   
  &\dec 0.22 &\dec 0.66 &\dec 0.09 &\dec $-$0.47  \\
  &$\langle \overline q q \rangle$ 
  &\dec 0.23 &\dec 0.68 &\dec 0.09 &\dec $+$0.46 \\
\tableline
Threshold $w_2$ of (\protect\ref{nucl11at1})
  &$\langle {\alpha_s \over \pi} \, G^2 \rangle$
  &\dec 0.31 &\dec 0.77 &\dec 0.22 &\dec $+$0.79 \\
  &$\langle \overline q \, g \, \sigma \cdot G \, q \rangle$   
  &\dec 0.22 &\dec 0.64 &\dec 0.09 &\dec $-$0.44  \\
  &$\langle \overline q q \rangle$ 
  &\dec 0.22 &\dec 0.65 &\dec 0.09 &\dec $+$0.43 \\
\tableline
Pole Residue $\lambda_{\cal O}^2$ of (\protect\ref{nucl11at1})
  &$\langle \overline q q \rangle$ 
  &\dec 0.30 &\dec 0.78 &\dec 0.17 &\dec $+$0.69 \\
  &$\langle {\alpha_s \over \pi} \, G^2 \rangle$
  &\dec 0.24 &\dec 0.70 &\dec 0.15 &\dec $+$0.63 \\
  &$\langle \overline q \, g \, \sigma \cdot G \, q \rangle$   
  &\dec 0.16 &\dec 0.56 &\dec 0.07 &\dec $-$0.29  \\
\end{tabular}
\end{table}

\begin{figure}[p]
\begin{center}
\epsfysize=11.7truecm
\leavevmode
\setbox\rotbox=\vbox{\epsfbox{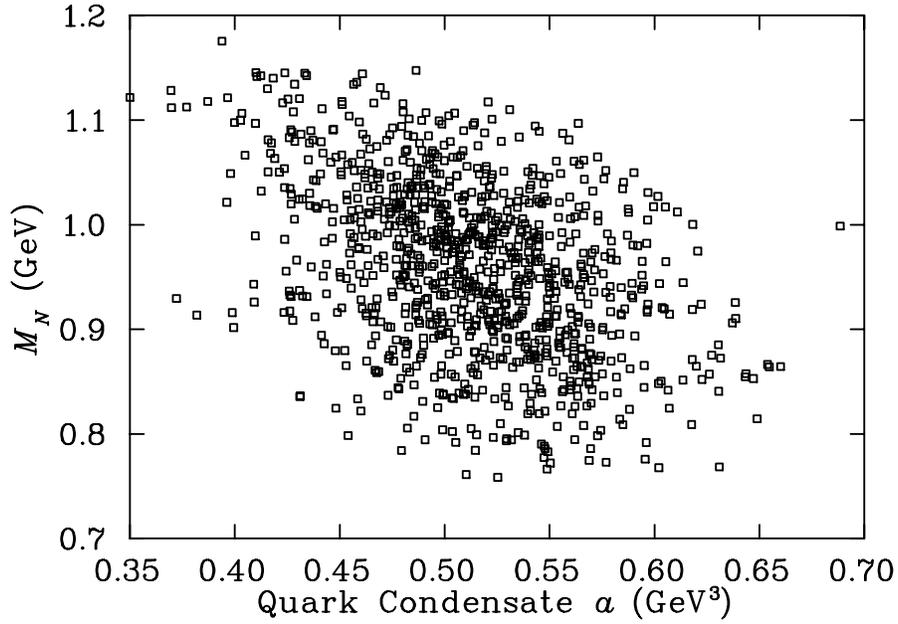}}\rotl\rotbox
\end{center}
\caption{The association between the nucleon mass and the quark
condensate $a$ estimated in Section \protect\ref{uncert}.  The nucleon
mass is extracted from the simultaneous analysis of sum rules
(\protect\ref{nucl13atGG}) and (\protect\ref{nucl13atGqq}) at $\beta =
0$.}
\label{nuclMa}
\end{figure}

\begin{figure}[p]
\begin{center}
\epsfysize=11.7truecm
\leavevmode
\setbox\rotbox=\vbox{\epsfbox{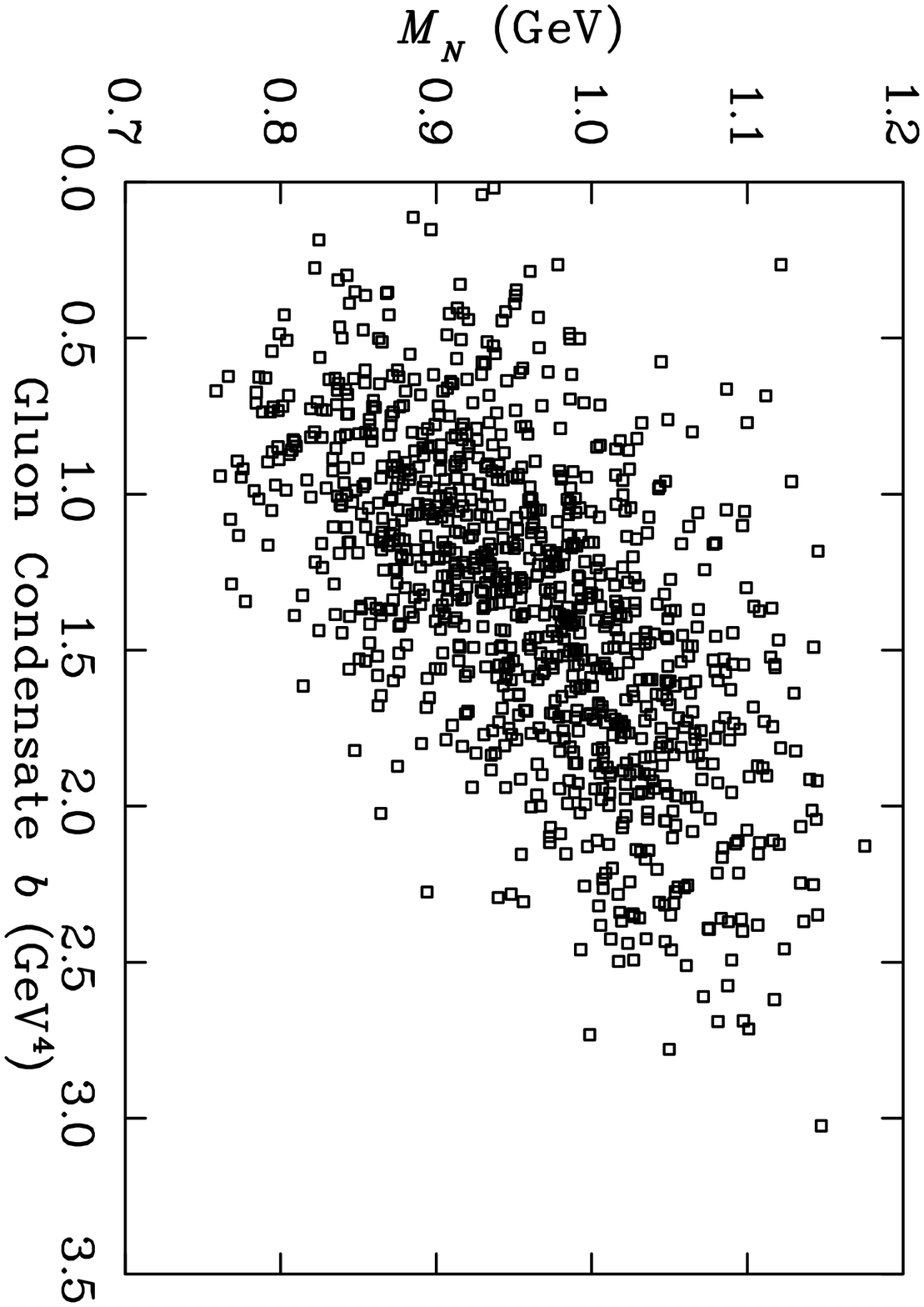}}\rotl\rotbox
\end{center}
\caption{The association between the nucleon mass and the gluon
condensate $b$ estimated in Section \protect\ref{uncert}.  The nucleon
mass is extracted from the simultaneous analysis of sum rules
(\protect\ref{nucl13atGG}) and (\protect\ref{nucl13atGqq}) at $\beta =
0$.}
\label{nuclMb}
\end{figure}

\begin{figure}[t]
\begin{center}
\epsfysize=11.7truecm
\leavevmode
\setbox\rotbox=\vbox{\epsfbox{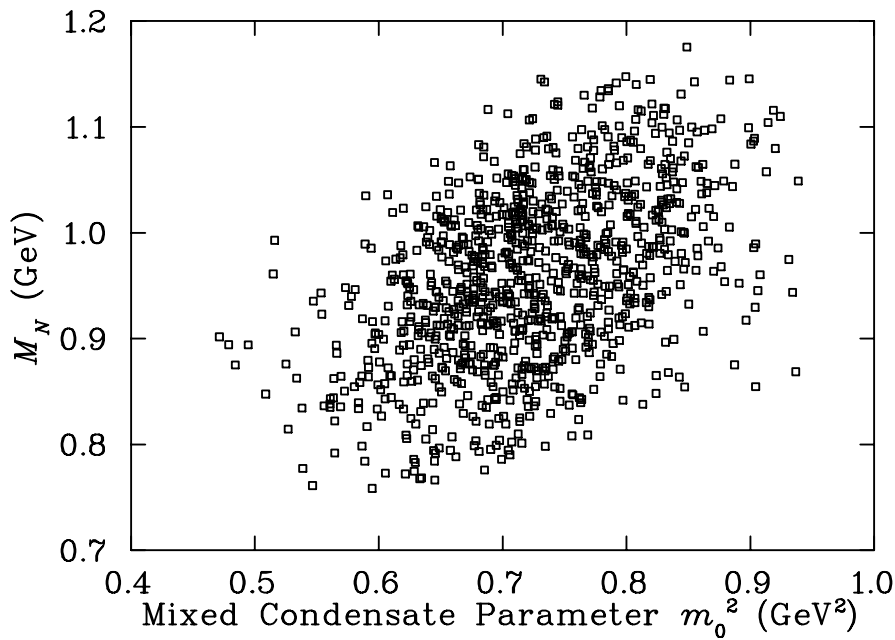}}\rotl\rotbox
\end{center}
\caption{The association between the nucleon mass and the mixed
condensate parameter $m_0^2$ estimated in Section
\protect\ref{uncert}.  The nucleon mass is extracted from the
simultaneous analysis of sum rules (\protect\ref{nucl13atGG}) and
(\protect\ref{nucl13atGqq}) at $\beta = 0$.}
\label{nuclMm0sq}
\end{figure}

   However, these correlations are dependent on the choice of
interpolating field.  Table \ref{NucleonCorr} identifies the QCD
parameters having significant correlations with the spectral
properties of the nucleon obtained from the analysis of
(\ref{nucl13atGG}) and (\ref{nucl13atGqq}) at $\beta = 0$.  This time
the quark condensate plays the dominant role in determining the
nucleon mass.  This role is also shared with the mixed condensate
$\langle \overline q \, g \, \sigma \cdot G \, q \rangle$.  Perhaps
the most interesting result is that the magnitude of the quark
condensate is anticorrelated with the nucleon mass, in complete
contradiction with the ``Ioffe formula''.  Figures \ref{nuclMa},
\ref{nuclMb} and \ref{nuclMm0sq} illustrate the correlations between
the nucleon mass and the quark, gluon, and mixed condensates
respectively.

\begin{table}[t]
\caption{QCD Parameters having significantly correlated roles in
determining nucleon spectral properties via a simultaneous analysis of
(\protect\ref{nucl13atGG}) and (\protect\ref{nucl13atGqq}) at $\beta =
0$.  Measures of association include Cramer's $V$, the Contingency
Coefficient $C$, the Symmetrical Uncertainty Coefficient $U$, and the
linear correlation coefficient $r$.}
\label{NucleonCorr}
\setdec 0.00
\begin{tabular}{lccccc}
Spectral Property  &Parameter &$V$ &$C$ &$U$ &$r$ \\    
\tableline
Nucleon Mass $M_N$
  &$\langle \overline q q \rangle$ 
  &\dec 0.31 &\dec 0.79 &\dec 0.18 &\dec $-$0.68 \\
  &$\langle \overline q \, g \, \sigma \cdot G \, q \rangle$   
  &\dec 0.26 &\dec 0.73 &\dec 0.17 &\dec $+$0.67  \\
  &$\langle {\alpha_s \over \pi} \, G^2 \rangle$
  &\dec 0.16 &\dec 0.54 &\dec 0.07 &\dec $+$0.21 \\
\tableline
Threshold $w_3$ of 
  &$\langle \overline q q \rangle$ 
  &\dec 0.32 &\dec 0.80 &\dec 0.17 &\dec $-$0.66 \\
(\protect\ref{nucl13atGG}) and (\protect\ref{nucl13atGqq})
  &$\langle \overline q \, g \, \sigma \cdot G \, q \rangle$   
  &\dec 0.26 &\dec 0.71 &\dec 0.15 &\dec $+$0.64  \\
  &$\langle {\alpha_s \over \pi} \, G^2 \rangle$
  &\dec 0.16 &\dec 0.55 &\dec 0.08 &\dec $+$0.52 \\
\tableline
Pole Residue $\lambda_{\cal O}\, \lambda_{3/2}$ 
  &$\langle \overline q \, g \, \sigma \cdot G \, q \rangle$   
  &\dec 0.34 &\dec 0.82 &\dec 0.25 &\dec $+$0.82  \\
of (\protect\ref{nucl13atGG}) and (\protect\ref{nucl13atGqq}) 
  &$\langle \overline q q \rangle$ 
  &\dec 0.17 &\dec 0.56 &\dec 0.08 &\dec $-$0.37 \\
  &$\langle {\alpha_s \over \pi} \, G^2 \rangle$
  &\dec 0.16 &\dec 0.56 &\dec 0.07 &\dec $+$0.33 \\
\end{tabular}
\end{table}

   What this indicates is that the intimate relationship between the
quark condensate and the nucleon mass suggested in the ``Ioffe
formula'' is invalid.  Considering the assumptions made in deriving
the ``Ioffe formula'' one might expect the quark condensate value to
be tied more tightly to the continuum model than the ground state
pole, and this is borne out in Table \ref{NucleonCorr}.  There the
quark condensate plays the dominant role in determining the continuum
model threshold.

   Of course, all the condensate values themselves are correlated by
the theory of QCD, and this correlation has not been included in this
analysis.  It may very well be that the nucleon mass is positively
correlated with the quark condensate when the higher dimension
condensates are also correlated with the quark condensate.  The
important point to recognize is that the correlation does not come
about through some direct connection such as the ``Ioffe formula''.

   Moreover, we are extracting nucleon properties in a Borel regime in
which these leading order condensates just happen to be the dominant
ones.  One could also extract the mass from much deeper in the
non-perturbative regime, perhaps by a lattice QCD calculation.  There,
it is much higher dimension operators which account for the same
nucleon mass \cite{leinweber95a}.  Therefore, it is really impossible
to identify a particular condensate for giving rise to ground state
hadron properties.

   The only true statement that may be made about the relationship
between condensates and spectral properties is what we have known all
along: The condensates reflecting chiral symmetry breaking play a
significant role in determining hadron properties.

\newpage

\subsection{Phenomenological Correlations}
\label{PhenomCorr}

\begin{figure}[t]
\begin{center}
\epsfysize=11.7truecm
\leavevmode
\setbox\rotbox=\vbox{\epsfbox{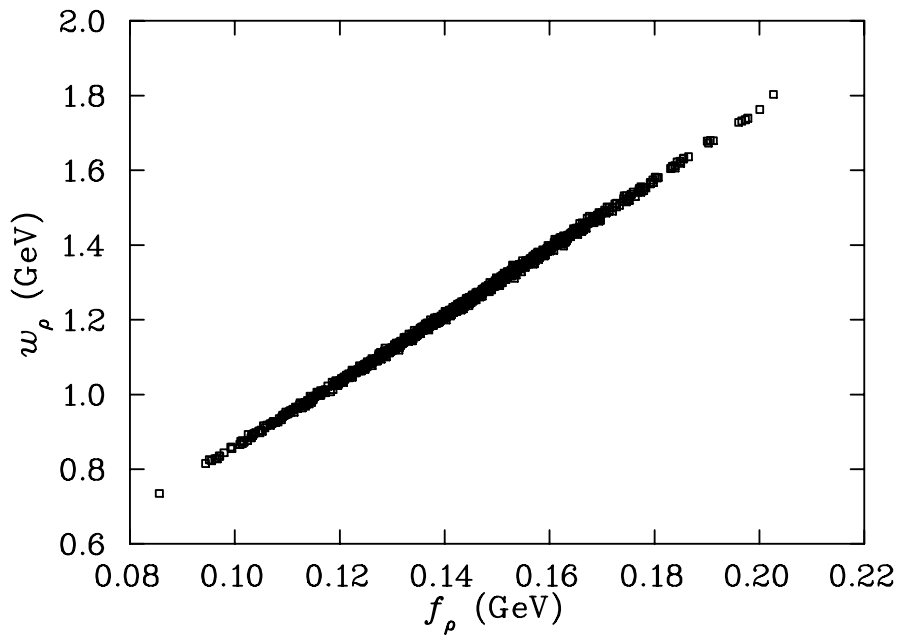}}\rotl\rotbox
\end{center}
\caption{ The association between the $\rho$-meson continuum threshold
and the decay constant obtained from 1000 fits of
(\protect\ref{rhoSR}).  }
\label{rhowlambda}
\end{figure}

\begin{figure}[p]
\begin{center}
\epsfysize=11.7truecm
\leavevmode
\setbox\rotbox=\vbox{\epsfbox{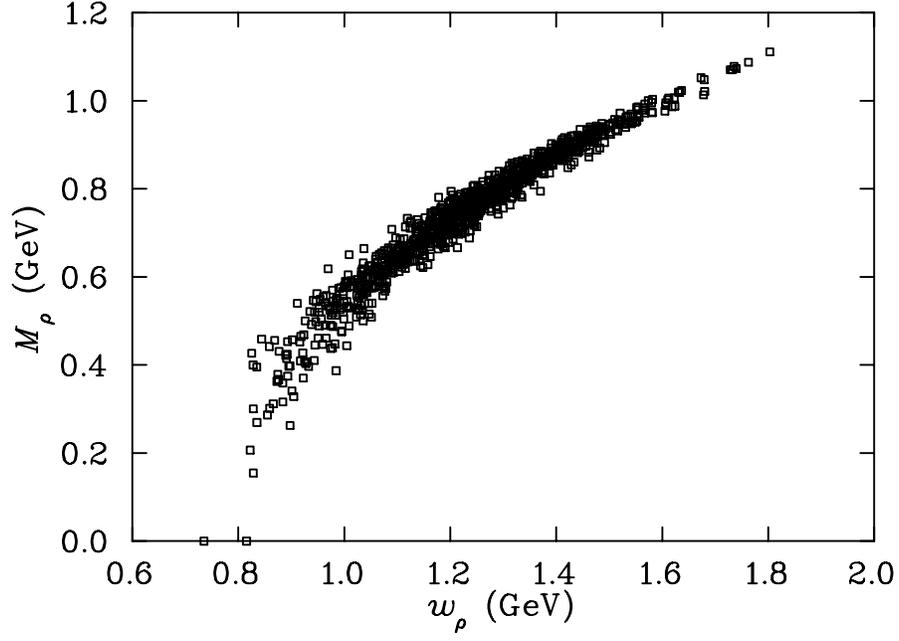}}\rotl\rotbox
\end{center}
\caption{ The association between the $\rho$-meson mass and the
continuum threshold obtained from 1000 fits of (\protect\ref{rhoSR}).
}
\label{rhoMw}
\end{figure}

\begin{figure}[p]
\begin{center}
\epsfysize=11.7truecm
\leavevmode
\setbox\rotbox=\vbox{\epsfbox{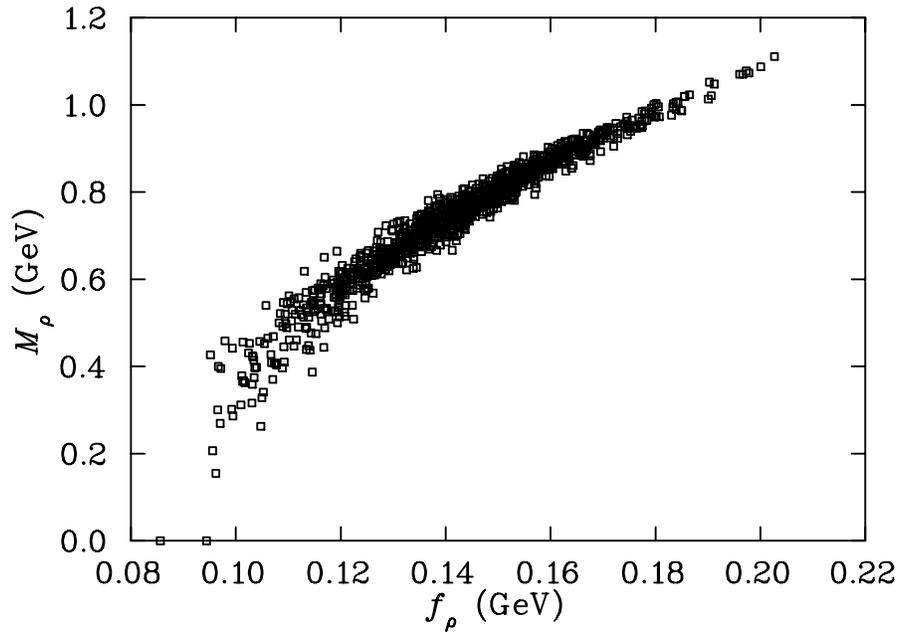}}\rotl\rotbox
\end{center}
\caption{ The association between the $\rho$-meson mass and the
decay constant obtained from 1000 fits of
(\protect\ref{rhoSR}).  }
\label{rhoMlambda}
\end{figure}

   Correlations among the phenomenological fit parameters themselves
are qualitatively different from the correlations with the QCD
parameters.  Figures \ref{rhowlambda}, \ref{rhoMw} and
\ref{rhoMlambda} display the close correspondence between all three of
the fit parameters, $w_\rho$, $f_\rho$, and $M_\rho$ obtained from
1000 fits of (\ref{rhoSR}).  This close association between the
$\rho$-meson spectral parameters suggests an underlying constraint on
the fit parameters.  Indeed, local duality necessarily leads to
correlations among the spectral parameters.  To examine the
correlations of local duality in an analytical but qualitative
fashion, consider the finite energy sum rules (FESR)
\cite{krasnikov83} obtained from
\begin{equation}
\int_0^{s_0} \, \phi(s) \, \rho_V(s) \, ds =
{1 \over \pi} \int_0^{s_0} \, \phi(s) \, {\rm Im} \, \Pi(s+i\epsilon)
\, ds 
\, .
\end{equation}
Here, $\rho_V(s)$ is given by (\ref{rhoVb}) and $\Pi(q^2)$ is the QCD
evaluation of (\ref{twopt}) at the structure $g_{\mu \nu} - p_\mu
p_\nu / p^2$.  The weighting function $\phi(s)$ for FESR is taken to
be $1,\ s,\ s^2,\ldots$.  These functions weight excited state
contributions in favor of ground state contributions and, as such,
FESR are inappropriate for a quantitative determination of ground
state properties.  However, the first moment and, to a lesser extent,
the second moment of the FESRs can provide some general insight into
the relationships among the spectral parameters.  Taking $\phi(s) = 1$
and $\phi(s) = s$ yields the following two relationships 
\begin{mathletters}
\begin{eqnarray}
8 \pi^2 \, f_\rho^2 - s_0  &=& 0 \, , 
\label{FESR1} \\
8 \pi^2 \, f_\rho^2 \, M_\rho^2 - {s_0^2 \over 2} &=&
8 \pi^2 \, m_q \, \langle\overline{q}q\rangle
 + {\pi^2 \over 3}\, \langle{\alpha_s\over \pi} G^2\rangle 
\simeq 0 \, .
\label{FESRs}%
\end{eqnarray}%
\label{FESRrho}%
\end{mathletters}%
In a qualitative sense, the contributions from the OPE on the
right-hand side of (\ref{FESRs}) are negligible.  In this case
(\ref{FESRrho}) implies
\begin{mathletters}
\begin{eqnarray}
w_\rho &=& 2\pi \, \sqrt{2} \, f_\rho \, , 
\label{FESRa} \\
M_\rho &=& {1 \over \sqrt{2}} \, w_\rho \, ,
\label{FESRb} \\
M_\rho &=& 2\pi \, f_\rho \, .
\label{FESRc}%
\end{eqnarray}%
\label{FESRrel}%
\end{mathletters}%
These relationships, capturing the features of local duality in a
qualitative sense, are independent of the precise values of the QCD
parameters.  The relationships remain prominent in the more
quantitative Borel sum rule analysis as indicated in figures
\ref{rhowlambda}, \ref{rhoMw} and \ref{rhoMlambda}.  Of course, the
Borel analysis is done independent of these qualitative constraints.
The first relation, the most reliable FESR relation, is satisfied
remarkably well.

\begin{figure}[p]
\begin{center}
\epsfysize=11.7truecm
\leavevmode
\setbox\rotbox=\vbox{\epsfbox{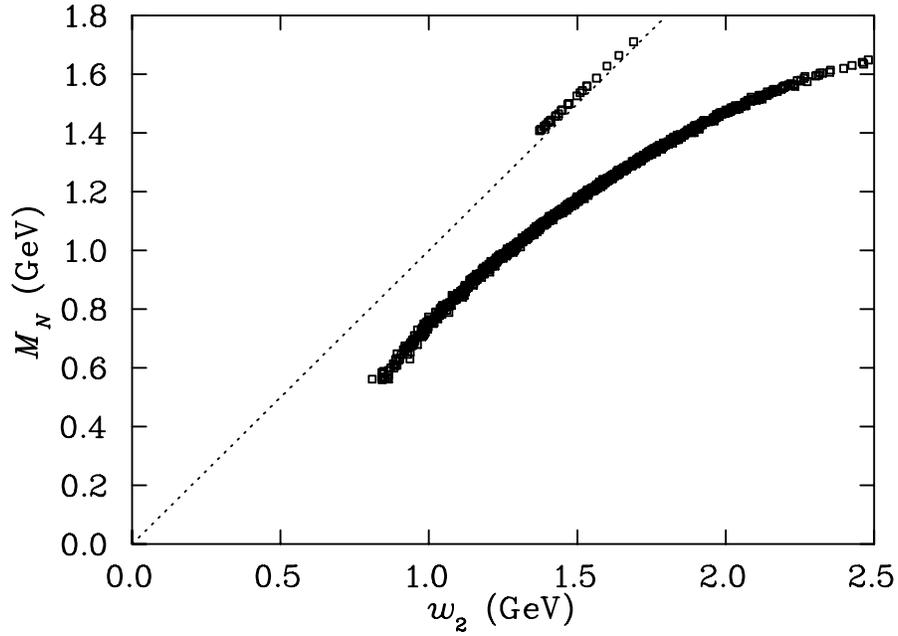}}\rotl\rotbox
\end{center}
\caption{ The association between the nucleon mass and the continuum
threshold obtained from 1000 fits of (\protect\ref{nucl11at1}) for the
optimal $\beta = -1.2$.  The dashed line indicates the equivalence of
the mass and threshold. }
\label{SR4corrMw}
\end{figure}

\begin{figure}[p]
\begin{center}
\epsfysize=11.7truecm
\leavevmode
\setbox\rotbox=\vbox{\epsfbox{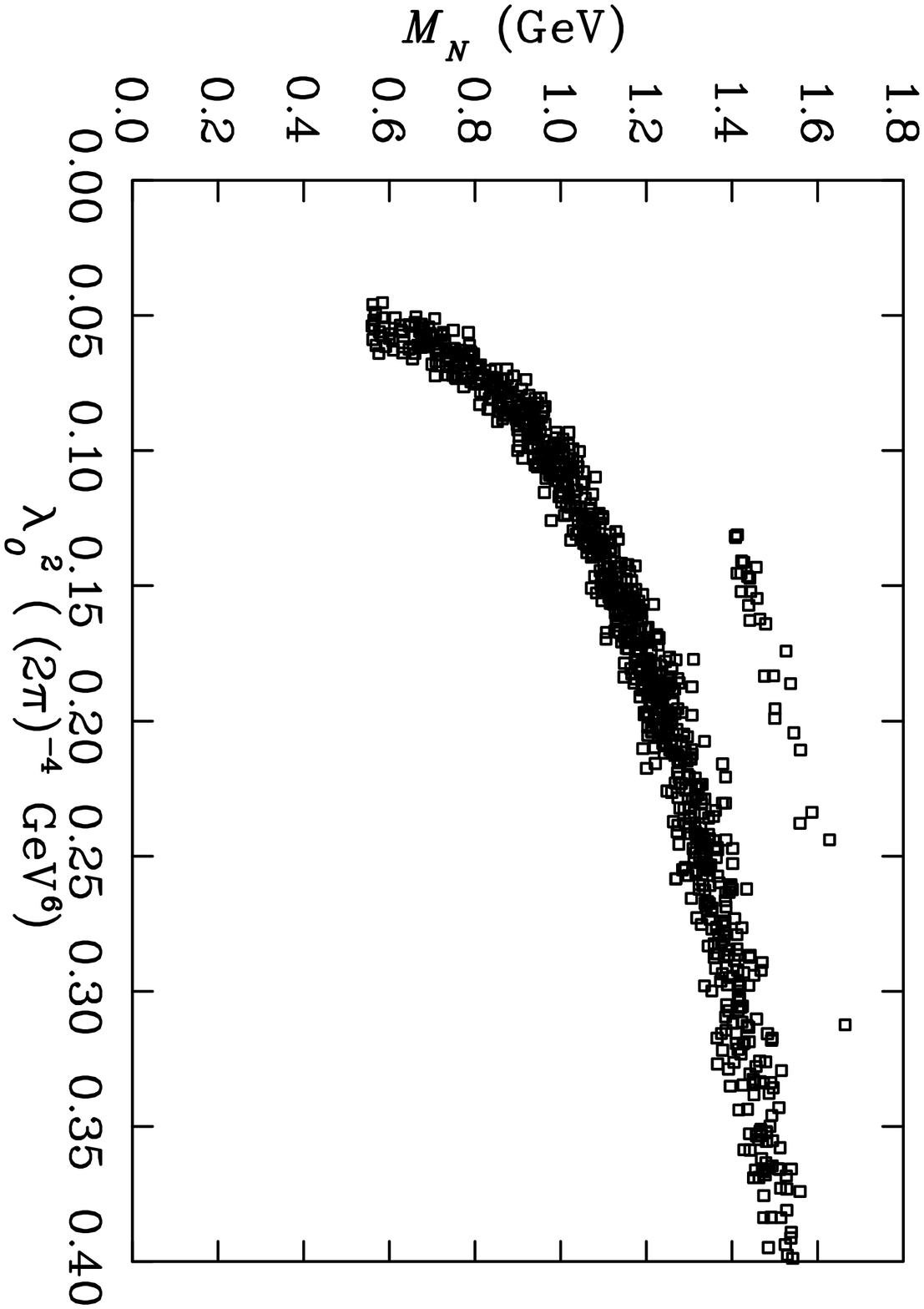}}\rotl\rotbox
\end{center}
\caption{ The association between the nucleon mass and the residue of
the pole obtained from 1000 fits of (\protect\ref{nucl11at1}) for the
optimal $\beta = -1.2$.  }
\label{SR4corrMlambda}
\end{figure}

   An even more striking correlation is illustrated in figure
\ref{SR4corrMw} for the nucleon sum rule of (\ref{nucl11at1}) with the
optimal $\beta = -1.2$.  The association between the continuum
threshold and the nucleon mass is nearly one-to-one despite the
underlying scatter of QCD parameter values.  The explanation for this
phenomena is that there is a single term in (\ref{nucl11at1}) which is
capable of resolving the pole from the continuum.  The two leading
terms of the OPE are used in deriving the continuum model.  Hence if
the third term is vanishingly small, the best fit is most certainly
pure continuum $w_2 \to 0$ as the sum rule is satisfied as a perfect
Laplace transform.  The few points lying above the curve in figure
\ref{SR4corrMw} where the mass equals the threshold are cases where
the third term in the sum rule has failed to resolve the pole from the
continuum.  Figure \ref{SR4corrMlambda} displays the corresponding
association for the mass and the residue of the pole.

   In the $\rho$-meson case there are in fact two terms independent of
the continuum model in (\ref{rhoSR}).  Both terms act together to resolve
the pole and continuum and hence give rise to some scatter in the
mass--threshold association.  Scatter in the residue of the pole
reflects the fact that all terms affect the strength of the pole.

   The strong associations among nucleon spectral parameters reflect
the underlying constraints of local duality.  Relationships analogous
to the FESR for the $\rho$ meson may be derived for the nucleon
spectral parameters \cite{krasnikov83}.  However, the larger continuum
model contributions to the nucleon correlator make the use of FESR
less desirable.  The approximate nature of the continuum model may be
too crude for a weighting function that fails to suppress excited
state strength in the spectral density.  The resultant FESR relations
are relatively crude and do not capture the essence of local duality
as well as in the $\rho$-meson case.

\begin{figure}[p]
\begin{center}
\epsfysize=11.7truecm
\leavevmode
\setbox\rotbox=\vbox{\epsfbox{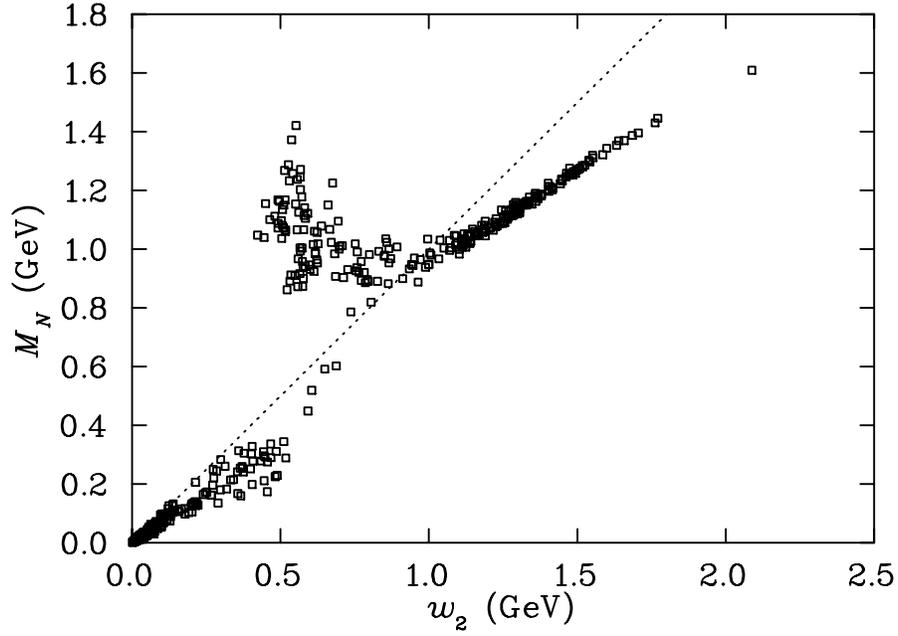}}\rotl\rotbox
\end{center}
\caption{The correlation between the nucleon mass and the continuum
threshold, $w_2$, obtained from the analysis of
(\protect\ref{nucl11at1}) for the unfavorable value of $\beta = +0.8$.
The dashed line indicates the equivalence of the mass and threshold. }
\label{SR4at0.8corrMw}
\end{figure}

\begin{figure}[p]
\begin{center}
\epsfysize=11.7truecm
\leavevmode
\setbox\rotbox=\vbox{\epsfbox{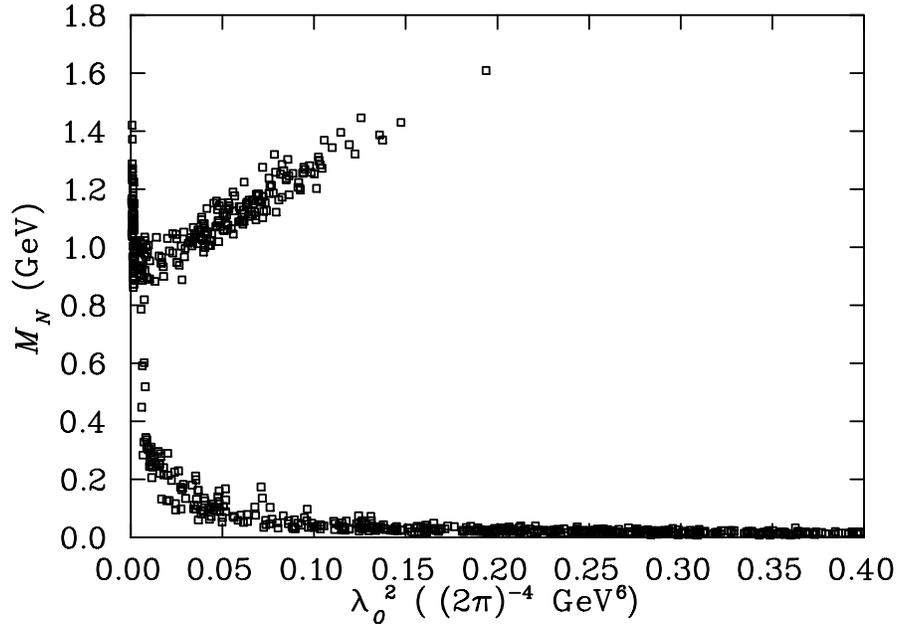}}\rotl\rotbox
\end{center}
\caption{The correlation between the nucleon mass and the residue of
the pole, $\lambda_{\cal O}$, obtained from the analysis of
(\protect\ref{nucl11at1}) for the unfavorable value of $\beta = +0.8$.
}
\label{SR4at0.8corrMlambda}
\end{figure}

   A very different picture develops for the correlations among the
nucleon fit parameters obtained from the analysis of (\ref{nucl11at1})
for the unfavorable value of $\beta = +0.8$.  Figure
\ref{SR4at0.8corrMw} displays the correlation between the mass and the
continuum threshold, $w_2$.  The corresponding correlation between the
mass and the residue is illustrated in figure
\ref{SR4at0.8corrMlambda}.

   The nature of the three regimes discussed earlier in section
\ref{SecNucl11at1.8} is resolved here.  The near-pure continuum fits
are revealed in the lower-left corner of the mass-threshold scatter
plot of figure \ref{SR4at0.8corrMw}.  The collision of the continuum
threshold and the nucleon mass gives rise to the intermediate regime
where the gap between ground and excited states has not been resolved.
The smaller residue compensating for strength from the continuum model
in these fits is illustrated in figure \ref{SR4at0.8corrMlambda}.  The
third pole-plus-continuum regime displays the same near one-to-one
correspondence between the mass and the threshold as seen in figure
\ref{SR4corrMw} for the optimal $\beta = -1.2$.

\begin{figure}[p]
\begin{center}
\epsfysize=11.7truecm
\leavevmode
\setbox\rotbox=\vbox{\epsfbox{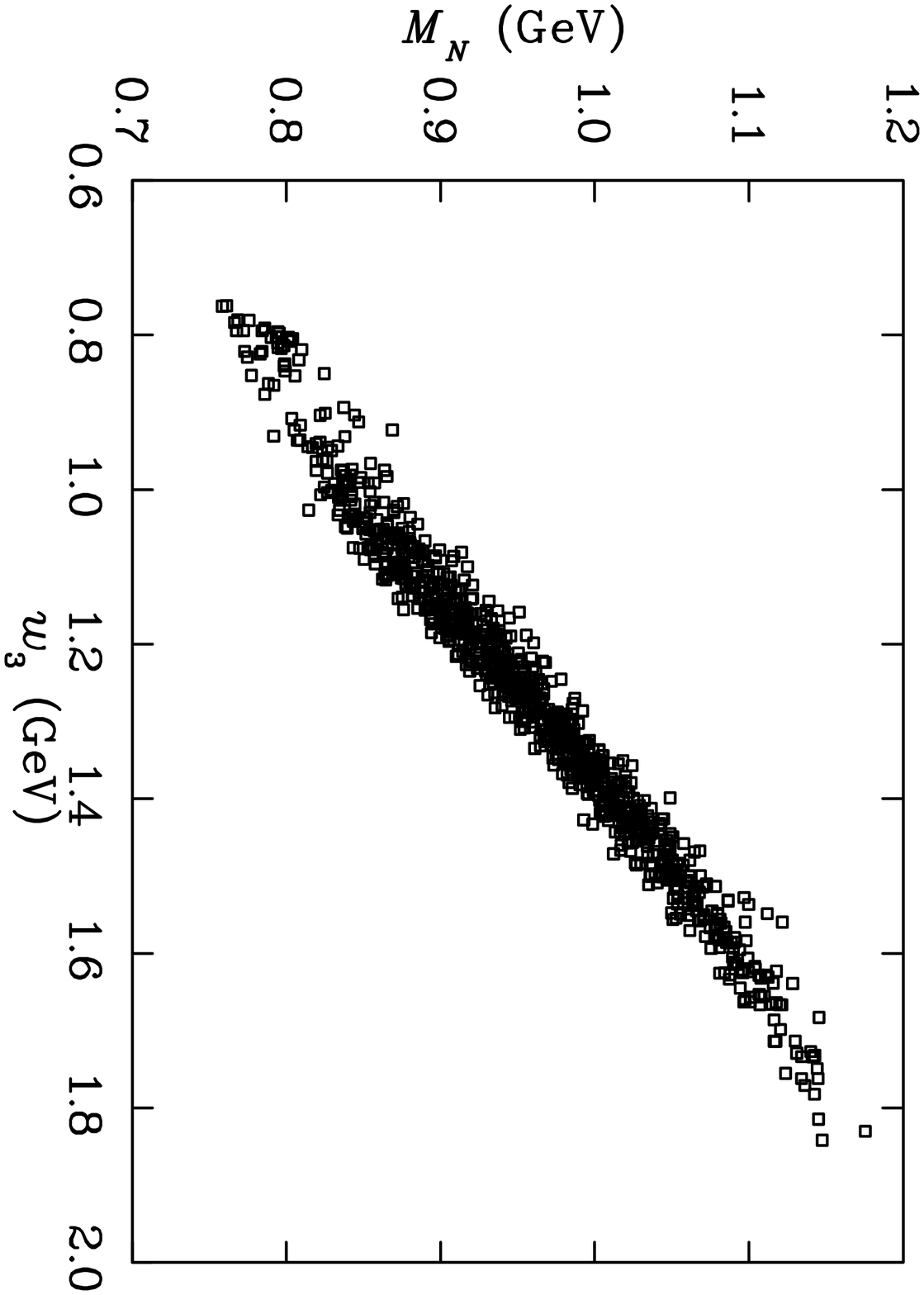}}\rotl\rotbox
\end{center}
\caption{The correlation between the nucleon mass and the continuum
threshold, $w_3$, obtained from the simultaneous analysis of sum rules
(\protect\ref{nucl13atGG}) and (\protect\ref{nucl13atGqq}) at $\beta =
0$.}
\label{SR467corrMw2}
\end{figure}

\begin{figure}[p]
\begin{center}
\epsfysize=11.7truecm
\leavevmode
\setbox\rotbox=\vbox{\epsfbox{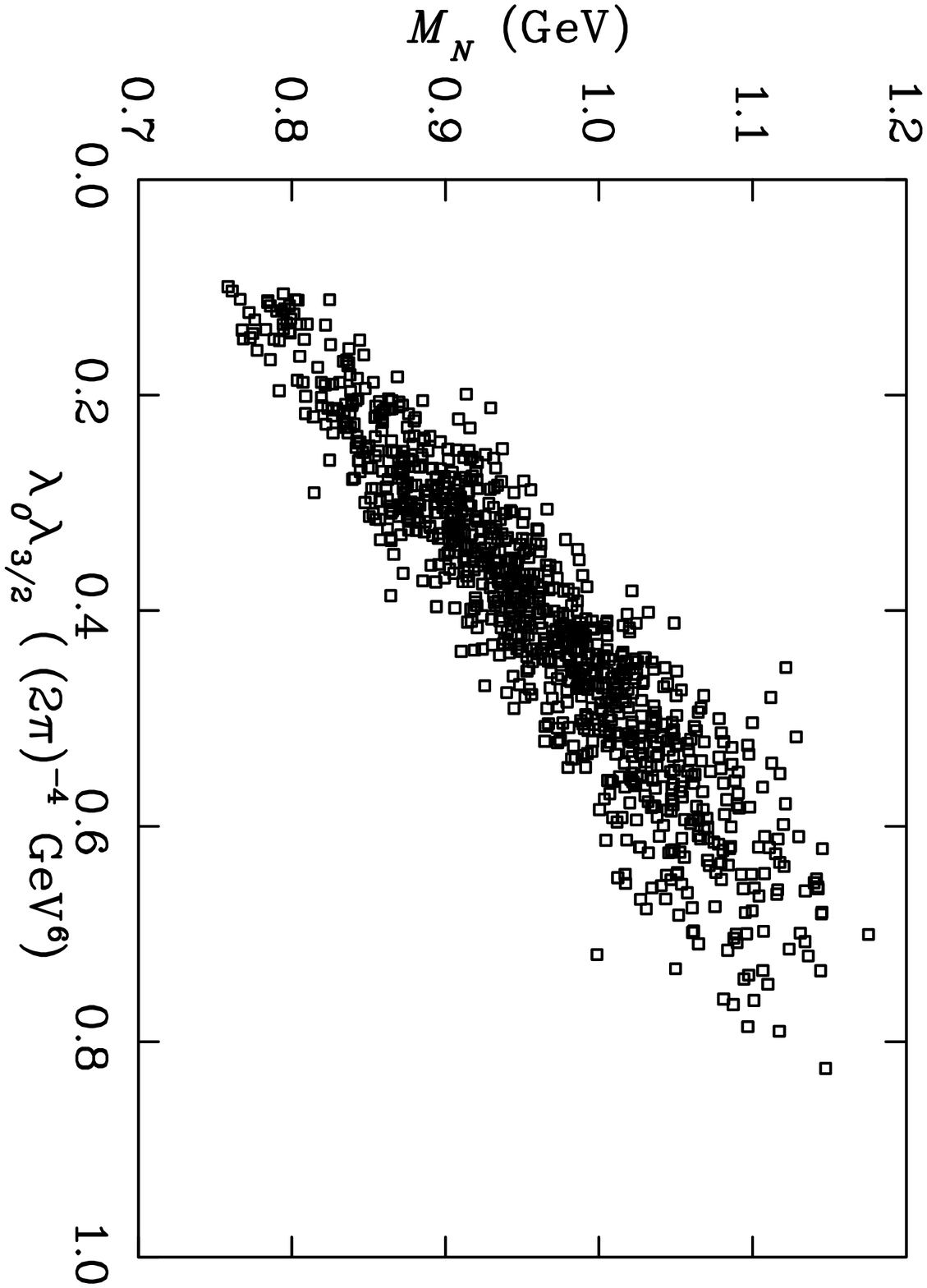}}\rotl\rotbox
\end{center}
\caption{The correlation between the nucleon mass and the residue of
the pole, $\lambda_{\cal O} \, \lambda_{3/2}$, obtained from the
simultaneous analysis of sum rules (\protect\ref{nucl13atGG}) and
(\protect\ref{nucl13atGqq}) at $\beta = 0$.  }
\label{SR467corrMlambda2}
\end{figure}

   The correlations among the spectral parameters are interesting
because they reveal the manner in which the sum rules work.
Correlations among the fit parameters from the simultaneous analysis
of (\ref{nucl13atGG}) and (\ref{nucl13atGqq}) at $\beta = 0$ are
illustrated in figures \ref{SR467corrMw2} and \ref{SR467corrMlambda2}.
These distributions are similar to those for the celebrated
$\rho$-meson sum rule, and reflect the presence of two terms
separating pole from continuum.

\section{ON THE NECESSITY OF DIRECT INSTANTONS}
\label{instantons}

   Arguments favoring the need for direct instanton contributions to
the QCD side of QCD-SRs \cite{dorokhov90,forkel93} are based on the
assumption that the average instanton size \cite{shuryak88}, $\rho_c =
0.33$ fm $\ll 1/\mu = 0.39$ fm, the separation scale of the OPE.  Even
though $\rho_c \sim 1/\mu$, proponents of direct instanton
contributions still argue that the effects of instantons are not
sufficiently accounted for in QCD condensates alone.

The analysis of direct instantons in nucleon sum rules are based on
many additional assumptions.  Here we focus on the most recent
investigation of Ref.\ \cite{forkel93}.  There, $\alpha_s$ corrections
to the identity operator are neglected.  As discussed in Section
\ref{nucleon11} these leading order corrections are 50\% and suggest
that $\alpha_s$ corrections are completely out of control.  Moreover
the possibility of a significant dimension-two power correction
arising from a summation of the perturbative series, uncertainties
surrounding the factorization of the four-quark operator, coupled with
an absence of a valid Borel regime for the sum rule of
(\ref{nucl11atP}) make any conclusions dubious.

   In addition, the OPE is truncated at dimension six.  As we saw in
Section \ref{nucleon11} the dimension six, seven, and eight operators
are crucial to separating the ground state pole from the continuum.
Without the dimension seven term, the best fit of (\ref{nucl11at1}) is
pure continuum.  The fit is perfect when $w_2 \to 0$ and $\lambda \to
0$.  Hence, the nucleon sum rule at the structure 1 in
Ref. \cite{forkel93} (equation 12) is unstable by design.  Any further
discussion of stabilizing the sum rule by adding instanton
contributions is uninteresting.  Finally, issues surrounding the
problems of double counting were handled crudely by simply eliminating
OPE contributions beyond dimension five.  Hence one should not be
prejudiced by existing analyses suggesting nucleon sum rules require
direct instanton contributions.  The issue is really not settled.

   For the spin-1/2 interpolator of (\ref{chiN1}), direct instanton
contributions to the sum rule of (\ref{nucl11at1}) are governed by the
factor \cite{forkel93}
\begin{equation}
c_2 = \left ( 13 \, \beta^2 + 10 \, \beta + 13 \right ) / 8 \, .
\end{equation}
The direct instanton contributions were argued to be of vital
importance in Ref.\ \cite{forkel93} in the vicinity of $\beta \sim
-1$.

   In the lattice QCD investigation of \cite{leinweber95b} it was
discovered that the overlap of the interpolator $\chi_2$ with the
nucleon ground state is one hundred times smaller than that for
$\chi_1$.  While the nucleon mass is certainly independent of the
interpolating field mixing parameter $\beta$, one also has the residue
of the pole, $\lambda_{\cal O}$, independent of $\beta$ at the 1\%
level.  This anticipated independence of both ground state properties
provides new opportunities for evaluating the necessity of direct
instanton effects.

   As a first test of sum rule consistency, we utilize the sum rule of
(\ref{nucl11at1}) to extract the mass as a function of the
interpolating field mixing parameter $\beta$.  As $\beta \to -1.5$ the
terms contributing to the continuum model become very small and
additional poles must be added to the phenomenological side of the sum
rule to account for strength in the correlator lying above the ground
state mass.  As a result we restrict $\beta \ge -1.3$.  Valid Borel
regimes were found over the range $-1.3 \le \beta \le -0.9$.  The
coefficient $c_2$ changes by 50\% over this range.  If direct
instanton contributions are indeed important at $\beta \sim -1$, the
absence of direct instanton contributions in this investigation might
be revealed by a significant $\beta$ dependence in the nucleon mass.

\begin{figure}[t]
\begin{center}
\epsfysize=11.7truecm
\leavevmode
\setbox\rotbox=\vbox{\epsfbox{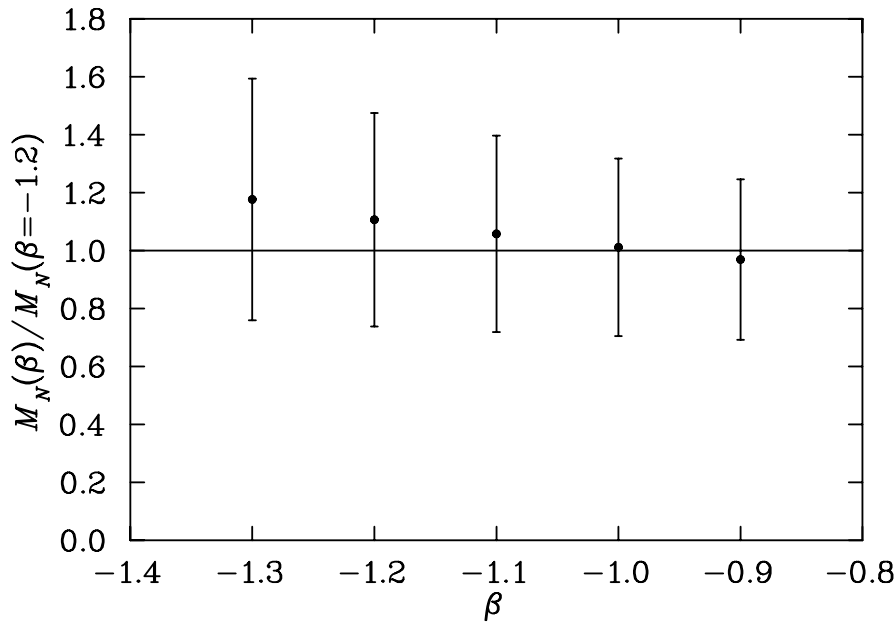}}\rotl\rotbox
\end{center}
\caption{The dependence of the nucleon mass obtained from the sum rule
of (\protect\ref{nucl11at1}) on the interpolating field mixing
parameter $\beta$.  Uncorrelated uncertainty estimates are obtained
from two samples of 200 QCD parameter sets.  }
\label{SR4MassEvol}
\end{figure}

Figure \ref{SR4MassEvol} displays the ratio of the nucleon mass
extracted at $\beta$ to the value obtained at the optimal $\beta =
-1.2$.  Uncertainties are obtained from two uncorrelated samples of
200 QCD parameter sets.  As such the central value at $\beta = -1.2$
need not be 1.  The large uncertainties associated with this sum rule
prevent a sensitive test of the necessity of direct instantons.  It is
clear that all ratios accommodate the ratio of 1.

   A more demanding test may be performed using the sum rules arising
from the consideration of the overlap of spin-1/2 and spin-3/2
interpolators.  The smaller uncertainties allow a more interesting
examination of these issues.  In addition, the continuum model terms
are independent of $\beta$ and provide a wider range of $\beta$ in
which a valid Borel regime exists.

\begin{figure}[t]
\begin{center}
\epsfysize=11.7truecm
\leavevmode
\setbox\rotbox=\vbox{\epsfbox{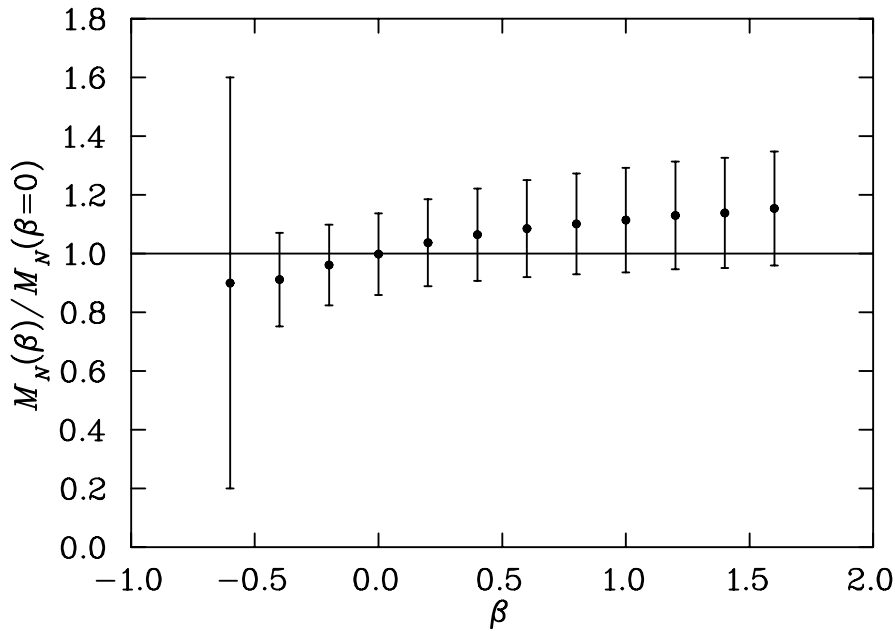}}\rotl\rotbox
\end{center}
\caption{The dependence of the nucleon mass obtained from the sum
rules of (\protect\ref{nucl13atGG}) and (\protect\ref{nucl13atGqq}) on
the interpolating field mixing parameter $\beta$.  Uncorrelated
uncertainty estimates are obtained from two samples of 200 QCD
parameter sets.  }
\label{MassEvol}
\end{figure}

   Figure \ref{MassEvol} addresses the dependence of the nucleon mass
obtained from the sum rules of (\ref{nucl13atGG}) and
(\ref{nucl13atGqq}) on the interpolating field mixing parameter
$\beta$.  Here, ratios of the mass obtained at finite $\beta$ to
$\beta = 0$ are displayed.  Sum rule consistency demands that all
ratios agree with one, and this is satisfied in figure \ref{MassEvol}
without direct instanton contributions.  The uncertainty estimates are
obtained from two uncorrelated samples of 200 QCD parameter sets.  We
note that a valid regime does not exist for $\beta \le -0.8$.  Hence
the use of the Ioffe interpolator ($\beta = -1$) is a particularly
unfavorable choice for these sum rules.  The large error bar at $\beta
= -0.6$ reflects the small valid Borel regime used in the fit.

   The corresponding dependence of $\lambda_{\cal O}$ on $\beta$ is
plagued by huge uncertainties.  Small changes in the nucleon mass
which appears squared in the exponential of the phenomenological side
of the sum rules leads to large fluctuations in $\lambda_{\cal O}$.
Hence it is difficult to gain any information on the possible size of 
direct instanton contributions.  However, they are certainly
unnecessary as all ratios encompass 1.

\begin{figure}[p]
\begin{center}
\epsfysize=11.7truecm
\leavevmode
\setbox\rotbox=\vbox{\epsfbox{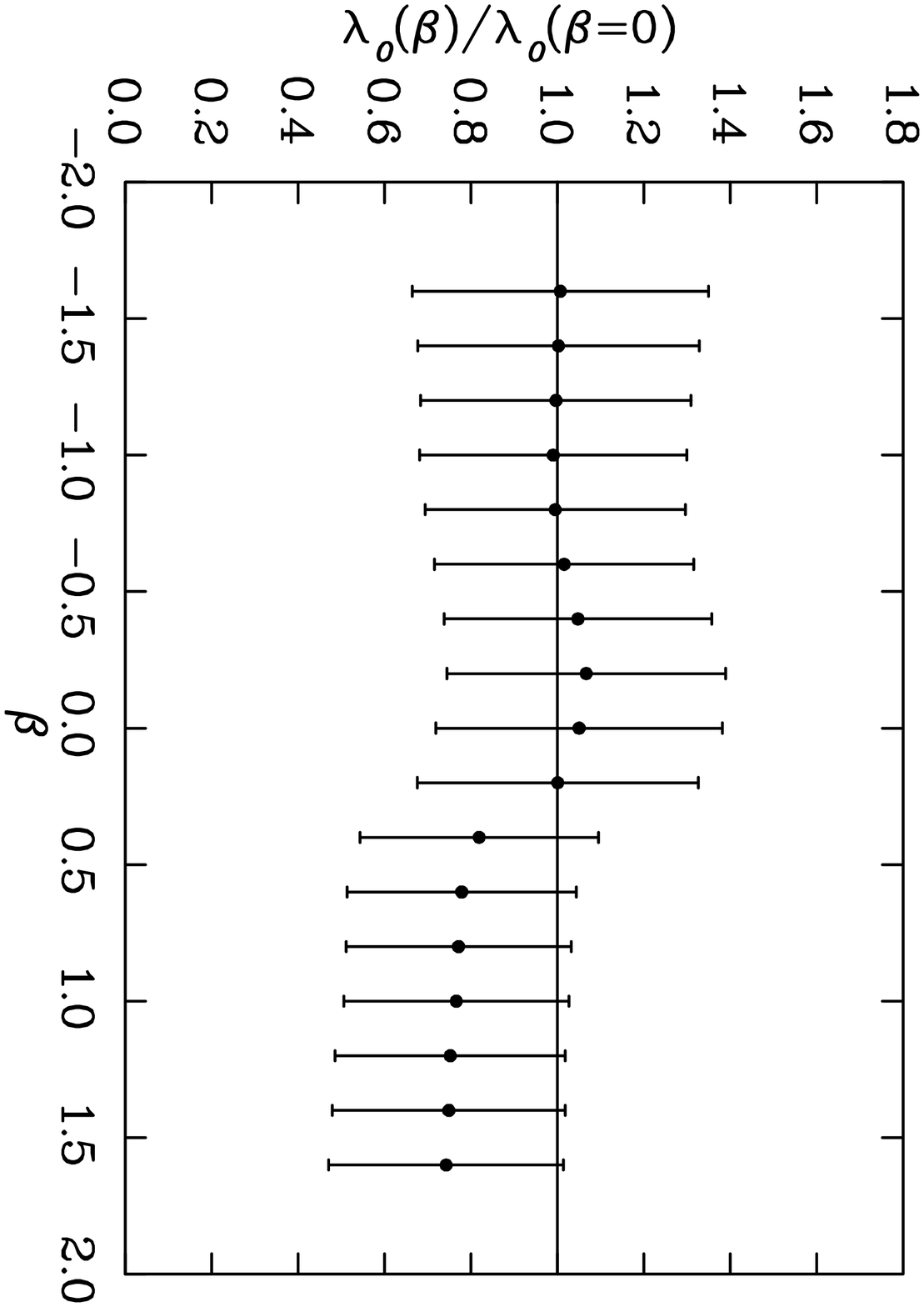}}\rotl\rotbox
\end{center}
\caption{The dependence of $\lambda_{\cal O}$ determined by the sum
rules of (\protect\ref{nucl13atGG}) and (\protect\ref{nucl13atGqq}) on
the interpolating field mixing parameter $\beta$.  Here the nucleon
mass has been fixed to the optimal value obtained at $\beta = 0$.
Uncorrelated uncertainty estimates are obtained from two samples of
200 QCD parameter sets.}
\label{LambdaEvolMfixed}
\end{figure}

\begin{figure}[p]
\begin{center}
\epsfysize=11.7truecm
\leavevmode
\setbox\rotbox=\vbox{\epsfbox{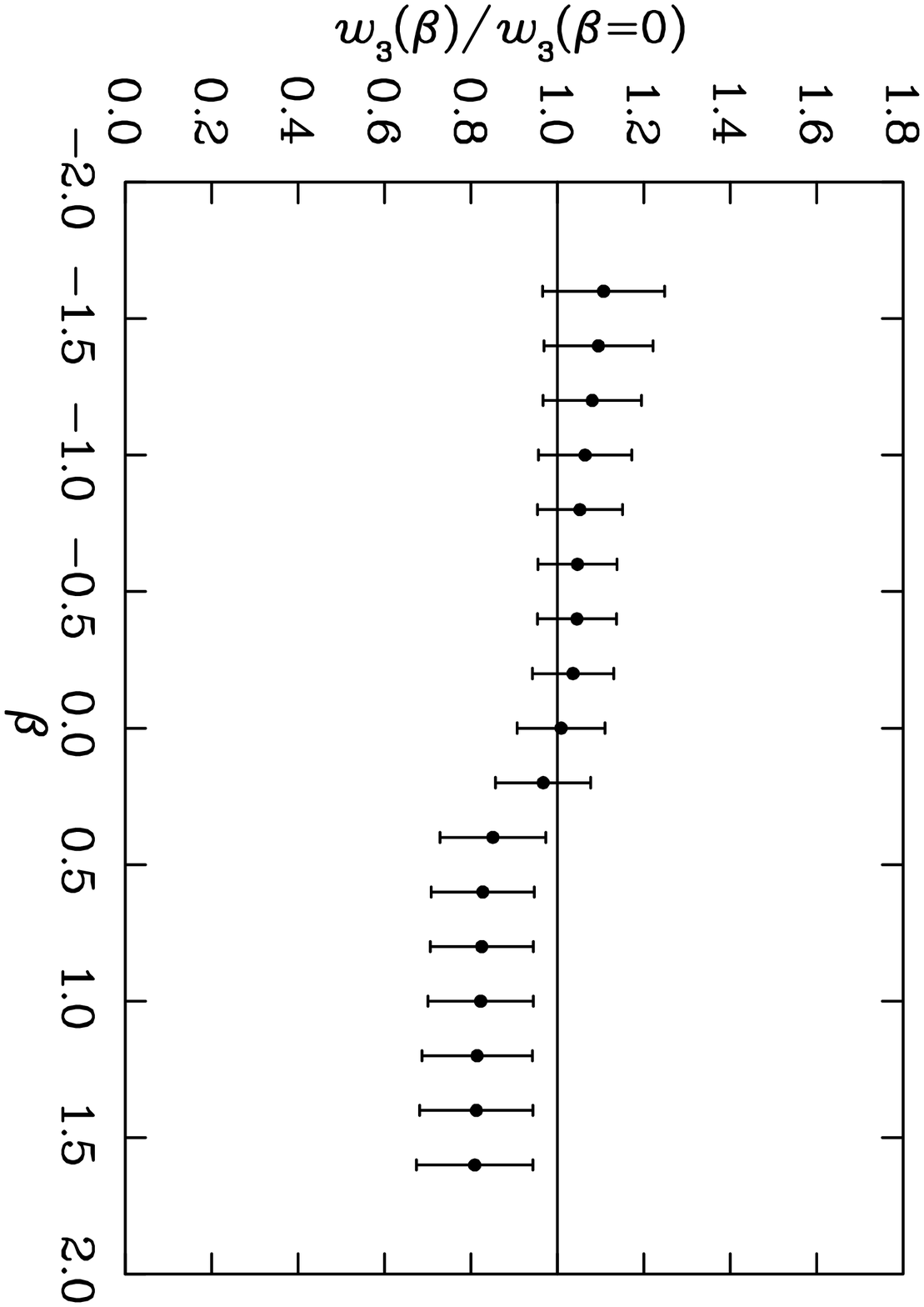}}\rotl\rotbox
\end{center}
\caption{The dependence of the continuum threshold obtained from the
sum rules of (\protect\ref{nucl13atGG}) and
(\protect\ref{nucl13atGqq}) on the interpolating field mixing
parameter $\beta$.  Here the nucleon mass has been fixed to the value
obtained at $\beta = 0$.  Uncorrelated uncertainty estimates are
obtained from two samples of 200 QCD parameter sets.  }
\label{ContEvol}
\end{figure}

   In order to proceed, we consider fixing the nucleon mass to the
value determined at $\beta = 0$ in an attempt to reduce the
uncertainty in $\lambda_{\cal O}$.  Figure \ref{LambdaEvolMfixed}
displays these ratios.  The corresponding plot for the continuum
threshold is given in figure \ref{ContEvol}.

   For $\beta < 0$ the continuum threshold tends to rise and the
distribution of the residue remains largely independent of $\beta$
without direct instanton contributions. For $\beta > 0$ the continuum
threshold drops to account for strength that would otherwise be
absorbed by the rising nucleon mass as illustrated in figure
\ref{MassEvol}.  With the threshold near the pole position, strength
which might otherwise rest in the pole has been suppressed as
discussed in section \ref{PhenomCorr}.  Hence the drop in the residue
is an artifact of our fixing the nucleon mass in an attempt to reduce
the error bar.

   In summary, the independence of the mass and residue from $|\beta|
\mathrel{\raise.3ex\hbox{$<$\kern-.75em\lower1ex\hbox{$\sim$}}} 1$ is
satisfied within uncertainties in the absence of direct instanton
contributions.  As the sum rules are improved in the future and QCD
input parameters are better determined, it will be important to repeat
this analysis to again evaluate the necessity of direct instanton
contributions or other refinements of the QCD-SR approach.

   At present, direct instanton contributions are not required.  This
is good news indeed as the amount of modeling required for instanton
estimates is relatively unpalatable when compared to that for the
standard QCD-SRs themselves.  In fact a preferable approach would be
to simply raise the OPE separation scale further into the perturbative
regime to ensure that direct instanton effects are more fully
accounted.  However, this does not yet appear to be necessary.

\section{Coordinate Space Correlator Implications}
\label{CoordImpl}

   Ground state properties have been found to be independent of the
interpolating field mixing parameter $\beta$.  On the other hand, the
leading terms of the OPE for (\ref{nucl11}) are very dependent on
$\beta$.  This situation has interesting ramifications for previous
analyses of coordinate space point-to-point correlation functions
\cite{chu93b,schafer94a,chu93a} where there has been some
confusion surrounding the physics represented in point-to-point
correlation functions.

In Ref.\ \cite{schafer94a} some conclusions are drawn from an
incorrect interpretation of the physics represented in two-point
correlators.  These authors suggest that the behavior of a ratio of the
interacting to free correlator can reveal information on clustering in
the scalar diquark channel.

   It should be emphasized that the phenomenological description of
the two-point function involves the baryon mass, the continuum
threshold describing the effective onset of excited states and the
residue of the pole which indicates the ability of the particular
interpolator to excite the baryon from the QCD vacuum.  Information on
quark configurations in the nucleon wave function is absent in
point-to-point correlation functions.

   The coordinate-space correlator may be obtained from (\ref{rhop})
or (\ref{rho1}) via 
\begin{equation}
\Pi(x) = { 1 \over (2 \pi)^4 } \int d^4p \, e^{-i \, p \cdot x} 
         \int_0^\infty { \rho(s) \over s - p^2 } \, ds \, .
\end{equation}
The important point here is that for large $x$, $\Pi(x)$ is dominated
by the contribution from the ground state pole.  As such $\Pi(x)$ is
independent of $|\beta|
\mathrel{\raise.3ex\hbox{$<$\kern-.75em\lower1ex\hbox{$\sim$}}} 1$ at
the 1\% level, or alternatively $\Pi(x)$ is independent of $|\beta|
\mathrel{\raise.3ex\hbox{$<$\kern-.75em\lower1ex\hbox{$\sim$}}} 10$ at
the 10\% level \cite{leinweber95b}.

   The coordinate space analyses focused on a ratio of $\Pi(x) /
\Pi^0(x)$ where $\Pi^0(x)$ is the free coordinate-space correlator.
For the correlator at the structure $\gamma \cdot p$ studied in
\cite{chu93b,schafer94a,chu93a} at $\beta = -1$, the generalized free
correlator is given by
\begin{equation}
\Pi^0(x) = {-3 \, i \over 4 \, x^{10} \, \pi^6} \left (5 + 2 \beta + 5
\beta^2 \right ) \, .
\end{equation}
Since the denominator is $\beta$ dependent and the numerator is
independent of $\beta$ for large $x$, the ratio is dependent on
$\beta$.  The rate at which the curve turns up as the ground state
dominates and the free correlator continues to drop off for increasing
$x$ is dependent on the interpolator mixing parameter
$\beta$.  The rate at which the ratio increases above 1 
describes the relative weightings of ground versus excited states
generated by the interpolating field.  It has nothing to do with the
possibility of scalar diquark degrees of freedom being predominant in
the nucleon wave function.  Moreover, a comparison with a
$\Delta^+$ correlation function for example has no meaning.

\section{Conclusions}
\label{conclusions}

   A rigorous procedure for extracting quantities of phenomenological
interest from QCD Sum Rules has been established.  We have performed a
comprehensive analysis of ground state $\rho$-meson and nucleon QCD
sum rules.  The inclusion of uncertainty estimates for the
phenomenological parameters has allowed us to evaluate the predictive
ability and self consistency of the QCD-SRs.  When the analysis is
done rigorously, QCD Sum Rules are predictive.  

   Some have argued that fitting sum rules appears to be more of an
art than a science.  Indeed, many of the findings here contradict the
conventional wisdom of both practitioners and skeptics alike
\cite{leinweber95h} and suggests that past arguments have been based
more on rhetoric than sound scientific arguments.

   The analysis of thousands of QCD parameter sets reveals the true
predictions of the sum rules.  For the nucleon the results may be
summarized as follows.
\begin{itemize}
\item The nucleon sum rule of (\ref{nucl11atP}), traditionally taken
to be the most reliable sum rule, fails to have a Borel regime where
both the truncated OPE is reasonably convergent and the ground state
dominates the phenomenology.
\item The nucleon sum rule of (\ref{nucl11at1}), also obtained from
spin-1/2 interpolating fields, resolves the nucleon mass to $\pm 260$
MeV.  This uncertainty is more than double the value commonly assumed
by most practitioners.
\item The nucleon sum rules of (\ref{nucl13atGG}) and
(\ref{nucl13atGqq}) obtained from the consideration of a spin-3/2
isospin-1/2 interpolator projected by the generalized spin-1/2
isospin-1/2 interpolator are much more reliable than the traditional
sum rules.  They have greater overlap with the ground state pole
relative to the continuum model, have broader valid Borel regimes and
provide phenomenological estimates of nucleon spectral properties
which are more stable than those obtained from the conventional sum
rules.
\item The nucleon mass obtained from a simultaneous analysis of
(\ref{nucl13atGG}) and (\ref{nucl13atGqq}) at the optimal $\beta = 0$
is resolved to $\pm 80$ MeV using QCD parameter estimates from the
literature.
\item The dependence of the spectral parameters on the interpolating
field mixing parameter is in accord with expectations without resort
to direct instanton contributions.  This is compelling evidence that
instanton physics is adequately accounted for in the nonperturbative
vacuum expectation values.
\end{itemize}
The latter point is truly exciting for the QCD sum rule approach.  The
amount of modeling required for instanton estimates is relatively
unpalatable when compared to that for the standard QCD-SRs themselves.

   The QCD-SR typically regarded as the most reliable or stable
nucleon sum rule has been found to be invalid.  A summary of
additional problems with this sum rule includes:
\begin{itemize}
\item The leading order $\alpha_s$ corrections are the order of 50\%.
As such, perturbative corrections to this sum rule are very likely out
of control.
\item Summation of the perturbative series may give rise to an
important dimension-two contribution \cite{brown92b,zakharov92}.  The
size of such a correction is presently unknown.
\item Contributions from the four-quark condensate are not well known
and appear to be large.  Unlike the $\rho$-meson sum rule, the
four-quark condensate is not suppressed by a factor of $\alpha_s$.
\item A regime in which the OPE is reasonably convergent while the
phenomenological description is dominated by the ground state pole is
non-existent.  If one proceeds with this sum rule, the results are
very sensitive to the model for the continuum, or inaccuracies in the
OPE lead to incorrect results as seen here.
\end{itemize}
Most QCD Sum Rule predictions of nucleon properties are based on this
sum rule.  The results are suspect, and should be reevaluated using
the new techniques introduced here.

   The ``Ioffe formula'' directly relating the nucleon mass to the
local chiral condensate has been found to be misleading at best.
While there is a strong connection between chiral symmetry breaking
and hadron properties, such detailed or direct relationships are
impossible to make and it is inappropriate to be more specific on the
origin of the nucleon mass.

   The fundamental and derivative sum rules for the $\rho$ meson have
also been examined.  Once again, the popular sum rule analysis has
been found to be undesirable.  The use of derivative sum rules for the
determination of $\rho$-meson spectral properties is unfavorable.
Increased continuum model contributions, poorer OPE convergence, and
loss of vital information in the derivative sum rule results in
discrepancies between the fundamental and derivative sum rules.

   The associations among the phenomenological fit parameters are
particularly interesting as they reveal how the sum rules resolve the
spectral properties.  We have found that the separation of the ground
state from the excited state continuum model is completely determined
by the high dimension operators of the OPE which are not used in
formulating the continuum model.  The practice of fixing the continuum
threshold to a preferred value has been exposed as strongly
determining the outcome of the sum rule analysis.  QCD sum rule
determinations of phenomenological spectral properties are necessarily
dependent on the treatment of high dimension operators where a
factorization assumption is usually invoked.  Moreover, careful
attention to OPE convergence is crucial to extracting meaningful
results.  These issues are somewhat unsettling, as we are working with
expansions which are asymptotic.

   The emphasis here has been on exploring the QCD parameter space via
Monte Carlo.  Further refinements of the parameter space might be made
by adjusting the valid Borel regime for each QCD parameter set such
that the 10\%--50\% criteria are satisfied.  This was not done here in
order to explore other criteria and reduce the sensitivity to a
reasonable yet somewhat arbitrary criteria.  In addition, the H\"older
inequalities might be applied as a further check of the validity and
consistency of the sum rules\footnote{We note that the shaded regions
satisfying the {H\"older} inequalities in Ref.\
\protect\cite{steele95} are determined using both fundamental and
derivative sum rules.  The regions are altered when determined by the
fundamental sum rule alone.} \cite{steele95}.

We are now in an excellent position to attack phenomena which is of
current interest to the nuclear physics community.  The techniques are
currently being applied to an in-medium study of vector meson
properties \cite{leinweber95f}.  There the debates over the behavior
of the vector-meson masses and the sum rules to be used in extracting
vector meson properties in nuclear matter are resolved.  Using
standard in-medium condensate estimates, the analysis leaves no doubt
that vector-meson masses decrease with increasing density.

   These techniques are also being applied to in-medium properties of
the nucleon \cite{leinweber95g}, by generalizing the vacuum sum rules
analyzed here to finite nuclear matter density.  The Monte-Carlo
techniques are resolving the predictive ability of the in-medium sum
rules.

   Another investigation utilizing these Monte-Carlo techniques
studies the QCD-SR determination of the $\rho - \omega$ meson mixing
parameter $\lambda$ \cite{leinweber95e}.  There, correlations between
the ratio of $u$ and $d$ quark masses are particularly interesting.

   The relatively large uncertainty in the residue of the pole does
not bode well for QCD Sum Rule analysis beyond two-point functions.
For example, the residue of the pole appears as a pre-factor to any
nucleon matrix element such as $g_A$ or magnetic form factors
\cite{burkardt96}.  It will not be surprising to find uncertainties in
these observables as large as the 35\% relative error in the squared
residue itself.  As the uncertainties in the additional vacuum matrix
elements relevant to the three-point functions are folded in the
uncertainties are likely to exceed 50\% and may in fact approach 100\%.
Analyses of these issues are currently underway
\cite{lee96b,hencken96}.

   When compared to uncertainty estimates previously published in the
literature, our uncertainty estimates are somewhat larger.  This is
not due to any shortcoming of the approach presented here.  Instead,
previous authors have failed to provide a reliable method for
determining realistic uncertainties.  Often, uncertainty estimates
have been conjectured without any supporting quantitative analysis.

   This in-depth examination of QCD sum rule self consistency paints a
favorable picture for further quantitative refinements of the
approach.  Of course such progress will be very difficult, as there
are many aspects of the approach which must be refined for any
particular improvement to be relevant.  The techniques introduced
here, such as maintaining independence of ground state properties from
the interpolating field mixing parameter, might be used to further
narrow the uncertainties of the approach.  It is extremely important
to refine the QCD-SR approach such that uncertainties in the QCD input
parameters may be reduced while maintaining $\chi^2/N_{\rm DF} \simeq
1$.  Research in this direction is required before meaningful
QCD-SR-based predictions for nucleon matrix elements and moments of
structure functions are possible.

\acknowledgements

   I am indebted to Kai Hencken for his careful checking of the Wilson
coefficients and his discovery of new corrections in the dimension
seven operators.  His contribution is key to resolving the discrepancy
between the sum rules presented here and modern estimates of the
vacuum condensates.  Thanks also to Xuemin Jin for numerous beneficial
discussions.  Dick Furnstahl and Xuemin Jin have provided a critical
examination of this analysis and I thank them for their insightful
comments.  Thanks also to Javed Iqbal for his contributions to the
evaluation of the finite width integrals for the $\rho$-meson
correlator.  Many years ago, Jimmy Law provided the optimization
routine used in this investigation, and I thank him for providing this
useful program.  This research was supported by the Natural Sciences
and Engineering Research Council of Canada and the U.S. Department of
Energy under grant DE-FG06-88ER40427.


\end{document}